\documentclass[useAMS,usenatbib]{mn2e}
\newcommand \kms{km s$^{-1}$}
\newcommand \zabs{z$_{\rm abs}$}
\newcommand \zem{z$_{\rm em}$}
\newcommand \url{}

\newcommand{\ion}[1]{~\textsc{#1}}

\usepackage{longtable}
\usepackage{graphicx,subcaption}
\usepackage{lscape}
\usepackage{rotating}
\usepackage{amssymb}
\usepackage{hyperref}

\def \HI{H\textsc{i}}
\def \omegaDLA{$\Omega_{\rm H \textsc{i}}$}
\def \fracOmega{$f_{\rm \Omega_{HI}}[<$N(HI)]}
\def \fNX{$f(N,X)$}
\def \lX{$\ell(X)$}
\def \nodata{. . .}

\hypersetup{draft}

\begin{document}

\title[HI contribution of XQ-100 subDLAs]{Sub-damped Lyman $\alpha$ systems in the XQ-100 survey I -- Identification and contribution to the cosmological HI budget}

\author[Berg et al.] {
\parbox[t]{\textwidth}{
Trystyn A. M. Berg$^{1,2}$,  Sara L. Ellison$^2$, Rub\'en S\'anchez-Ram\'irez$^{3,4}$, Sebasti\'an L\'opez$^{5}$, Valentina D'Odorico$^{6,7}$, George D. Becker$^{8}$, Lise Christensen$^{9}$, Guido Cupani$^{6}$, Kelly D. Denney$^{10}$, G\'abor Worseck$^{11,12}$\\\\
}
\\
$^1$European Southern Observatory, Alonso de Cordova 3107, Casilla 19001, Santiago, Chile.\\
$^2$Department of Physics and Astronomy, University of Victoria, Victoria, British Columbia, V8P 1A1, Canada.\\
$^{3}$INAF, Istituto di Astrofisica e Planetologia Spaziali, Via Fosso del Cavaliere 100, I-00133 Roma, Italy.\\
$^{4}$Instituto de Astrof\'{\i}sica de Andaluc\'{\i}a (IAA-CSIC), Glorieta de la Astronom\'{\i}a s/n, E-18008, Granada, Spain.\\ 
$^{5}$Departamento de Astronom\'{\i}a, Universidad de Chile, Casilla 36-D, Santiago, Chile.\\
$^{6}$INAF-Osservatorio Astronomico di Trieste, Via Tiepolo 11, I-34143 Trieste, Italy.\\
$^{7}$Scuola Normale Superiore Piazza dei Cavalieri, 7 I-56126 Pisa, Italy.\\
$^{8}$Department of Physics \& Astronomy, University of California, Riverside, California, 92521, USA.\\
$^{9}$DARK, Niels Bohr Institute, University of Copenhagen, Lyngbyvej 2, 2100 Copenhagen, Denmark\\
$^{10}$Department of Astronomy, The Ohio State University, 140 West 18th Avenue, Columbus, OH 43210, USA.\\
$^{11}$Max-Planck-Institut f\"{u}r Astronomie, K\"{o}nigstuhl 17, D-69117 Heidelberg, Germany.\\
$^{12}$Institut f\"{u}r Physik und Astronomie, Universit\"{a}t Potsdam, Karl-Liebknecht-Str. 24/25, D-14476 Potsdam, Germany.\\
}

\maketitle
\begin{abstract}

Sub-damped Lyman $\alpha$ systems (subDLAs; \HI{} column densities of $19.0\leq$~logN(\HI{})~$<20.3$) are rarely included in the cosmic \HI{} census performed at redshifts \zabs{}~$\gtrsim1.5$, yet are expected to contribute significantly to the overall \HI{} mass budget of the Universe. In this paper, we present a blindly selected sample of 155 subDLAs  found along 100 quasar sightlines (with a redshift path length $\Delta X=475$) in the XQ-100 legacy survey to investigate the contribution of subDLAs to the \HI{} mass density of the Universe. The impact of X-Shooter's spectral resolution on Ly$\alpha$ absorber identification is evaluated, and found to be sufficient for reliably finding absorbers down to a column density of logN(\HI{})$\geq18.9$. We compared the implications of searching for subDLAs solely using \HI{} absorption versus the use of metal lines to confirm the identification, and found that metal-selection techniques would have missed 75 subDLAs. Using a bootstrap-Monte Carlo simulation, we computed the column density distribution function (\fNX{}) and the cosmological \HI{} mass density (\omegaDLA{}) of subDLAs and compared with our previous work based on the XQ-100 damped Lyman $\alpha$ systems. We do not find any significant redshift evolution in \fNX{} or \omegaDLA{} for subDLAs. However, subDLAs contribute 10--20 per cent of the total \omegaDLA{} measured at redshifts $2<z<5$, and thus have a small but significant contribution to the \HI{} budget of the Universe.

\end{abstract}


\section{Introduction}

Studying the evolution of neutral gas reservoirs over cosmic time provides key constraints on aspects of galaxy evolution. The neutral atomic gas, primarily traced by \HI{}, is thought to be an indicator of the gas reservoirs that eventually form stars in galaxies \citep{Bird15,Somerville15}. Furthermore, the rate at which the \HI{} reservoirs change places direct constraints on the processes that can nourish or prevent future star formation \citep{Dave13,Lilly13}.

The comoving \HI{} mass density (\omegaDLA{}; measured with respect to the critical mass density of the Universe) has typically been used to quantify the amount of neutral gas in galaxies. At $z\approx 0$, \omegaDLA{} is quantified from 21-cm emission density maps \citep{Zwaan05,Martin10,Braun12}. However at redshifts $z \gtrsim 0.2$ where current radio facilities cannot easily detect the faint 21-cm emission, quasars (QSO) are used to probe sightlines through intervening \HI{} clouds in the form of Ly$\alpha$ absorption \citep[][]{Lanzetta95,Rao00,StorrieLombardi00,Ellison01,Peroux03,Prochaska05DR3,Rao06,Prochaska09,
Noterdaeme12,Zafar13,Crighton15, Neeleman16,SanchezRamirez16}.

Of the many classes of Ly$\alpha$ absorption line systems seen towards QSOs, damped Lyman $\alpha$ systems (DLAs), defined as having \HI{} column densities of logN(\HI{})$\geq 20.3$ \citep{Wolfe86,Wolfe05}, are typically used to measure \omegaDLA{}. Although few in number compared to lower \HI{} column density counterparts, DLAs dominate the \HI{} column density distribution from \zabs{}$\sim5$ to the present epoch \citep[particularly those with logN(\HI{})~$>21$;][]{Wolfe05,Prochaska05DR3,Noterdaeme12}. However, subDLAs ($19.0\leq$logN(\HI{})$<20.3$) are thought to host a modest contribution of $\approx10$--20 per cent to the \omegaDLA{} budget of the Universe \citep{Omeara07,Zafar13}. The difficulty in using subDLAs to probe the neutral gas reservoirs of the Universe is that these systems are often not fully self-shielded from the cosmic UV background, and thus do not completely trace neutral gas reservoirs \citep[e.g.][]{Zheng02,Omeara07}. Nevertheless, there has been a substantial body of work calculating the contribution of \HI{} from subDLAs to the column density distribution function (\fNX{}) and \omegaDLA{} \citep{Peroux03,Peroux03b,Peroux05,Omeara07,Guimaraes09,Zafar13}, as well as understanding whether subDLAs probe different types of gaseous systems than DLAs \citep{Peroux03b,Kulkarni07,Zavadsky09,Quiret16}.

It is expected that the observed \fNX{} can place strong constraints on feedback and ionization prescriptions in cosmological simulations of neutral gas reservoirs in the Universe \citep[e.g.][]{Rahmati13a,Bird14}. Both observational \citep{Peroux03,Peroux03b,Omeara07,Guimaraes09,Zafar13} and simulation-based studies \citep{Rahmati13b,Rahmati15,VillaescusaNavarro18} of the nature of \fNX{} for subDLA column densities have suggested that subDLAs are more common at redshifts \zabs{}$>3.5$ and contribute more than $\approx10 - 20$\% of \omegaDLA{}  traced solely by DLAs \citep{Peroux05,Guimaraes09,Noterdaeme09,Zafar13}. To date, samples of subDLAs have typically been selected using the presence of metal lines rather than using solely \HI{} absorption as is done with DLAs. Such a selection effect can potentially lead to sample incompleteness from missing metal-poor systems, and introduce biases in the comparison with DLAs. Additionally, most subDLA samples are too modest in size to detect any redshift evolution  at $>2\sigma$ confidence  \citep[][with redshift path lengths of $\Delta X=$~105, 378 and 193, respectively]{Omeara07,Guimaraes09,Zafar13}. 

This paper aims to investigate the contribution of subDLAs to the \omegaDLA{} budget at high redshifts ($2.5\leq$~\zabs{}~$\leq4.5$) using a completely blind sample of subDLAs identified in the XQ-100 survey \citep{Lopez16} solely based on their Lyman series absorption. We searched for all \HI{} absorbers in the XQ-100 spectra down to a column density threshold of logN(\HI{})$\geq18.8$ (i.e.~the subDLA threshold minus the typical logN(\HI{}) error of $\pm0.2$~dex) in order to evaluate the completeness of our absorber identification and properly account for miscalssification of absorbers as subDLAs in our computation of \fNX{} and \omegaDLA{}. However, the standard subDLA \HI{} column density limits are still adopted in all our computations. The reader is referred to \cite{SanchezRamirez16} for the HI statistics of the XQ-100 DLA sample. Through-out this paper we assume a flat $\Lambda$CDM Universe with ${\rm H_{0}=70.0~km~s^{-1}~Mpc^{-1}}$ and ${\rm \Omega_{M,0}=0.3}$.

\section{Data}
\label{secHI:Data}

\subsection{XQ-100 Survey design and data reduction}

The XQ-100 Legacy Survey (PI S.~L\'opez) observed 100 QSO sightlines at redshift \zem{}$\sim$ 3.5--4.5 with the X-Shooter Spectrograph \citep{Vernet11} on the Very Large Telescope (VLT) in Chile. The QSOs were purposefully chosen to be blind to any intervening absorption line systems, thus providing a random sample of sub-DLAs and DLAs to study the cosmological implications of these intervening absorbers  \citep{SanchezRamirez16,Berg16,Berg17,Christensen17}. For more details about the XQ-100 science cases and survey design, see \cite{Lopez16}.

For each QSO in the XQ-100 survey, the per-arm exposures were either $\sim0.5$ or $\sim1$ hour in length (depending whether the QSO was classified as `bright' or `faint'; respectively), providing median signal-noise ratios of $\approx30$ pixel$^{-1}$. The spectra for each arm were reduced using an internal IDL package developed by G.~Becker and G.~Cupani. For more details on the data reduction, see \cite{Lopez16}.

As the Ly$\alpha$ forest and intervening gas clouds along every sightline partially absorb all QSO photons below the Lyman limit, the QSO's flux is systematically absorbed for all wavelengths below the corresponding observed Lyman Limit at the redshift of the QSO ($\lambda_{LL}=(1+z_{\rm em})\times 911.7633~{\rm \AA{}}$), making both accurate continuum placement and subDLA identification difficult below this wavelength. For this reason, the wavelength regime below the Lyman limit is consequently not used for identifying the subDLAs in this work. The resulting redshift path length searched for subDLAs is $\Delta X=475$.

\subsection{Identification of intervening absorbers}
\label{sec:ID}

Starting with the manually placed QSO continuum fits outlined in \cite{Lopez16}, we visually identified candidate subDLAs by systematically scrolling through the XQ-100 spectra starting from the Ly$\alpha$ emission of the QSO to $\lambda_{LL}$.  Candidate absorbers are first identified by any saturated absorption broad enough to fit a simulated Ly$\alpha$ Voigt profile of column density logN(\HI{})$\geq18.8$ \citep[assuming a typical Doppler parameter of $b=10$~\kms{}; e.g.][]{DLAcat45}. Note that our adopted logN(\HI{})$\geq18.8$ search threshold is below the minimum subDLA threshold (logN(\HI{})~$=19.0$) by the typical \HI{} column density error (i.e.~$\pm0.2$) in order to enable a complete bootstrap error analysis (Section \ref{sec:BMC}). Simultaneously to the Ly$\alpha$ fitting, a simulated higher order Lyman series line profile (usually Ly$\beta$) was checked at the corresponding wavelength. For each of these saturated regions identified, one or more Voigt profiles were fitted by adjusting simultaneously the original continuum fits, the logN(\HI{}), and \zabs{} until the modelled Ly$\alpha$ and Ly$\beta$ Voigt profiles matched the continuum-normalized spectra. Continuum adjustments were common for large column density absorption systems on top or near QSO emission lines. In many of these adjustments, the best-fit column logN(\HI{}) was constrained by simultaneously fitting both the Ly$\alpha$ and Ly$\beta$ profiles. Errors on the logN(\HI{}) column density were estimated as the minimum and maximum possible logN(\HI{}) that would provide a visually reasonable fit to the data. For cases where the continuum was uncertain, the assigned logN(\HI{}) error was increased to incorporate this uncertainty in the continuum placement.

To ensure higher redshift Lyman series lines were not confused with lower redshift Ly$\alpha$ absorption, a full model spectrum was generated to keep track of all fitted absorption systems along each sightline. Upon identifying a new absorber, the associated Voigt profiles for several key Lyman series lines (from Ly$\alpha$ to Ly$\delta$) were added to the full model spectrum. The identification of systems was terminated when the observed-frame wavelength of additional Ly$\alpha$ absorbers was shorter than $\lambda_{LL}$.

\begin{figure*}
\begin{center}
\includegraphics[width=0.95\textwidth]{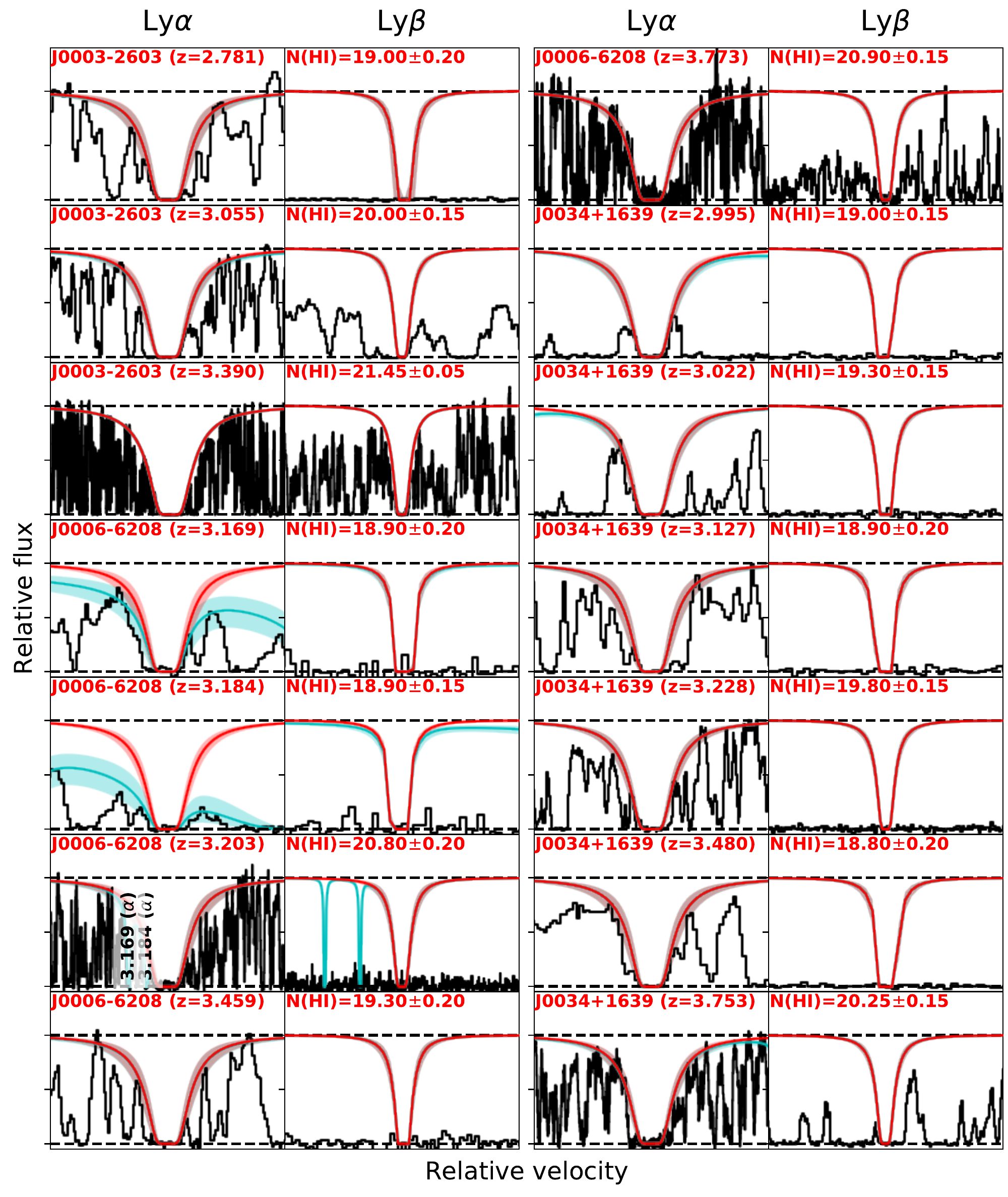}
\caption[]{Each pair of panels displays the fitted Ly$\alpha$(first and third columns) and Ly$\beta$ (second and fourth columns) profiles (dark red lines) and associated errors (dark red shaded region) for 14 of the identified candidate system. The light blue line and shaded region denote the full model spectrum and associated error combined from all absorbers within the entire sightline. The red text denotes the properties of the absorber [QSO sightline name and redshift in the corresponding Ly$\alpha$ panel, and logN(\HI{}) in Ly$\beta$ panel], while the vertical black text labels the redshift and Lyman series lines associated with additional absorption included in the model. The x-axis of each panel is scaled such that integrated region of the Ly series line's Voigt profile displayed encompasses 98 per cent of the absorbed flux. The remaining 262 systems are displayed in Figure \ref{fig:LyABapp}. }
\label{fig:LyAB1}
\end{center}
\end{figure*}

Once the absorber identification process was complete, the initial list of absorption line systems was cleaned by adjusting or removing any absorbing systems whose modeled Lyman series lines (Ly$\alpha$--Ly$\delta$, and the highest order Lyman series lines between 911\AA{} and 920\AA{}) did not fit the continuum normalized spectra. The final cleaned sample of identified candidate absorbers consists of 87 $18.8\leq$logN(\HI{})$<19.0$ Lyman limit systems, 155 sub-DLAs and 34 DLAs. Figure \ref{fig:LyAB1} shows the best-fitting Ly$\alpha$ and Ly$\beta$ profiles  for a representative sample of 14 absorbing systems, while the remaining profiles are displayed in Appendix \ref{sec:App} (Figure \ref{fig:LyABapp}). Table \ref{tab:Ncat} contains the full list of 276 candidate absorbers, including their \zabs{} and logN(\HI{}).

Given the moderate resolution of the XQ-100 spectra, it is sometimes difficult to model regions of extended saturated absorption where the damping wings of the Voigt profile are not visible into individual absorption features. The troughs could be fit reasonably by one or more subDLAs, but also by several lower column density Lyman Limit systems ($17<$~logN(\HI{})~$<19$) that are clustered together.  Unless a higher-order Lyman series line could disentangle this degeneracy (such as the absorbers seen towards J0034+1639 at \zabs{}$\approx4.2$; see Figure \ref{fig:LyABapp}), no candidate absorbers were fitted to the data in these degenerate regions. In many cases, we note that the edges of the Lyman series absorption appear to be better fit by Lyman Limit systems with high Doppler parameters rather than subDLAs. We therefore caution that our identification of subDLAs is potentially incomplete.

\subsubsection*{Comparison to \cite{SanchezRamirez16}}
In \cite{SanchezRamirez16}, we identified 41 DLAs in the XQ-100 sample using a similar absorber identification and fitting method. However, the identification of DLAs was completed over a slightly longer redshift path in \citet[$\Delta X=536$]{SanchezRamirez16} in that it was not truncated at $\lambda_{LL}$ as was done here. As a result, the initial search from \cite{SanchezRamirez16} includes eight more DLAs, which have been removed from the sample in this paper. We also identify two XQ-100 DLAs missed in the \cite{SanchezRamirez16} catalogue; one system is towards J0214$-$0517 (\zabs{}~$=3.7211$, logN(\HI{})~$=20.6\pm0.2$), and the other is along the J1111$-$0804 sightline (\zabs{}~$=3.60690$, logN(\HI{})~$=20.4\pm0.15$). 

In order to facilitate a fair comparison between the \HI{} contributions of subDLAs and DLAs in XQ-100, it is important that we use the same continuum fits for both the DLA and subDLA identification; thus we re-measured the N(\HI{}) values for all the previously identified XQ-100 DLAs. All of the \cite{SanchezRamirez16} DLAs within the common redshift range of the two studies were recovered, and their column densities are consistent (within the \HI{} column density errors) with those that we derived in our subDLA search. The largest discrepancies occurred in regions where the continuum is difficult to constrain, such as systems that are nearby the background QSO's Ly$\alpha$ emission line. For this paper, we use the newly measured logN(\HI{}) values of the XQ-100 DLAs for consistency.

\subsection{Completeness analysis}
\label{sec:Complete}
\subsubsection*{Recovery test}
To assess the completeness of the visual identification process and ensure that the modest XQ-100 spectral resolution (FWHM resolution R$\sim$5400--8900) did not affect the determination of column densities, an absorption line recovery test was necessary. To ensure the Ly$\alpha$ forest statistics and X-Shooter instrumental effects are reproduced while maintaining the same continuum fits used in the original identification process, artificial absorption systems were randomly inserted into the XQ-100 spectra. 

Prior to insertion, a list of artificial systems was first generated to ensure that an approximately uniform distribution of column densities was used. In bins of column density from logN(\HI{})$=18.0$ to 20.3\footnote{The recovery rate for DLAs in the XQ-100 dataset was previously measured at 100\% in \cite{SanchezRamirez16}. Given that we recovered all the previously identified DLAs in an identical recovery test in \cite{SanchezRamirez16}, we did not repeat the experiment for DLAs.} (of bin width 0.1 dex), $15\pm3$ systems were generated with column densities uniformly selected within the bin's limits. Therefore the total number of systems (between 276 and 414 systems) was randomly generated, preventing any subconscious counting during the recovery process. For each of these fake systems, a model Lyman series (down to Ly$\delta$) spectrum was generated, and randomly inserted into one of the XQ-100 QSO spectra at a random redshift. To simulate noise in the model spectrum, the flux of the model spectra was then varied randomly using the XQ-100 error spectrum flux as the standard deviation of a Gaussian distribution. Between 0 and 7 fake absorption systems were assigned to each XQ-100 sightline.

\begin{figure}
\begin{center}
\includegraphics[width=0.5\textwidth]{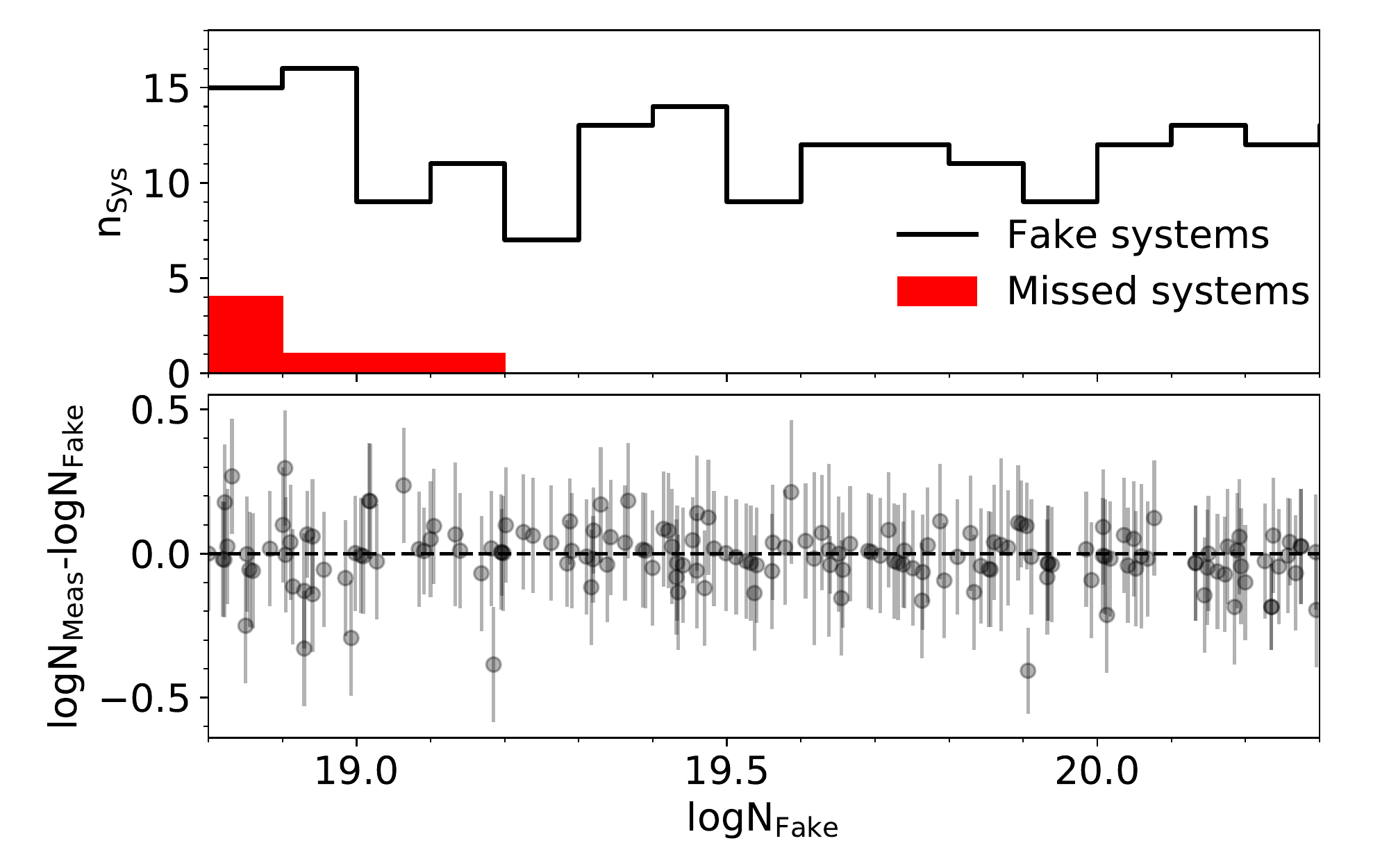}
\caption[XQ-100 N(\HI{}) recovery statistics ]{\emph{Top panel:} Distribution of the number of fake systems of a given \HI{} column density (logN$_{\rm Fake}$) used for the recovery test (black line) and those systems that were missed in the recovery process (solid red histogram). Note that only systems whose Ly$\alpha$ is beyond $\geq \lambda_{LL}$ are included in this plot. \emph{Bottom panel:} The difference in measured and fake logN(\HI{}) column densities (i.e. logN$_{\rm Meas}$-logN$_{\rm Fake}$) as a function of logN$_{\rm Fake}$. }
\label{fig:Recovery}
\end{center}
\end{figure}

The identification process described in Section \ref{sec:ID} was repeated on the spectra with fake absorbers inserted, starting with the locations of the previously-identified real absorbers flagged in the model spectra. The top panel of Figure \ref{fig:Recovery} shows the column density distribution of fake systems inserted into the XQ-100 spectra (with observed-frame Ly$\alpha$ absorption at wavelengths larger than $\lambda_{LL}$; black line). The solid red bars indicate that we missed seven, logN(\HI{})$\leq19.2$ systems in our recovery analysis.  However, the three missed systems with the higher column densities (logN(\HI{})$\geq18.9$) happened to be randomly inserted on top of already identified, higher column density Ly$\alpha$ absorbers. Since the \HI{} contribution from real overlapping systems would have already be accounted in the original fit to the larger \HI{} systems, we do not include these three missed systems in the recovery rate derivation. Therefore only $\approx 27$ percent of systems with logN(\HI{})$=18.8$ were missed in the recovery analysis.

For all the recovered systems, the bottom panel of Figure \ref{fig:Recovery} shows the difference in logN(\HI{}) between the measured column density of the fake absorber (N$_{\rm Meas}$) and the randomly generated value (N$_{\rm Fake}$). All but ten of the recovered systems ($\approx 3$\% of the total number of systems) have consistent column densities with N$_{\rm Fake}$. For these ten systems, the column densities are only inconsistent by  $\sim0.1$ dex, which is statistically consistent with the random flux errors inserted into the fake model spectra.

In summary, the \HI{} recovery test suggests that our identification of XQ-100 absorbers is 100\% complete for logN(\HI{})$\geq18.9$, and 73\% complete for logN(\HI{})$=18.8$.

\subsubsection*{Effects from continuum placement}
The location of the continuum becomes more uncertain towards bluer wavelengths as there is more absorption from additional Lyman series lines (i.e.~Ly$\beta$ and Ly$\gamma$) removing regions of clean continuum. To test whether or not our continuum placements introduce significant errors in our column density determinations, we artificially inserted a fake absorption line in the \emph{un-normalized} spectrum and fitted the line using our continuum placements (without continuum adjustment). This experiment was done in three regions within the rest-frame of the QSO, near the wavelengths: $\lambda=$~1120.7, 999.1, and 961.1  \AA{}. These three regions correspond to locations in the QSO spectrum where there is only Ly$\alpha$, Ly$\alpha+$Ly$\beta$, and Ly$\alpha+$Ly$\beta+$Ly$\gamma$ absorption (respectively). In order to remove the effects of individual sightlines on the experiment, five QSO spectra were selected at random for each region. However, we required that each of the selected QSOs had no previously identified Ly$\alpha$ absorption within the region where the line was inserted to avoid any effects from previous continuum placement modifications during the Ly$\alpha$ fit. The column density of the fake absorption line started at logN(\HI{})~$=18.8$, and was increased in steps of 0.2 dex up to a column density of logN(\HI{})~$=20.0$ (inclusive) to test the effects as a function of column density. For each fake absorption line, the wavelength was held fixed. For all but one QSO, we were able to recover the column density to within 0.2 dex (the typical quoted error on the \HI{} columns). These errors were not systematically offset in one particular direction (especially in the Ly$\alpha+$Ly$\beta$ and Ly$\alpha+$Ly$\beta+$Ly$\gamma$ region), suggesting there is no significant effect from the increased absorption (and thus more uncertain continuum) on our column density determinations. We note that for the one QSO sightline which provided inconsistent fits, the identification of the line was difficult due to significant blending with Ly$\alpha$ forest for the inserted Ly$\alpha$ profiles with column densities logN(\HI{})~$=19.0$ and 19.2, and the offsets were measured to be $+0.3$ dex each (i.e. the column densities were over-estimated), and is likely not strongly influenced from the continuum placement.

\subsubsection*{N(\HI{}) comparison to higher resolution data}
\label{secHI:HiRes}
Although the resolution of the XQ-100 data (FWHM resolution R$\sim$5400--8900) is sufficient to resolve the absorbing systems, it may be inadequate to resolve close blends of multiple lines. In order to test the absorber recovery and fit accuracy, we compared the fits derived from our X-Shooter data with spectra obtained with high resolution (R$\approx 80 000$) observations with the UVES instrument.  Using all available archival UVES spectra\footnote{Data accessed from \url{http://archive.eso.org/wdb/wdb/eso/uves/form}.} of the XQ-100 sightlines (19 sightlines total; with signal-noise of $\approx5-15$ pixel$^{-1}$), we repeated the Ly$\alpha$ profile fitting at the already identified redshifts of 62 \HI{} absorbers with UVES coverage.  Comparing the obtained logN(\HI{}) from both the UVES and XQ-100 fits, all column densities are consistent within the errors (at most 0.2 dex difference), with two exceptions. For these particular exceptions, the XQ-100 absorbers (J0134$+$0400, \zabs{}=3.999; J0137$-$4224, \zabs{}=3.665) each appear to be better fit by two, smaller component systems using the UVES data. However, the total \HI{} column measured contained within the UVES absorbers remains consistent with the single \HI{} profile fit measured in the XQ-100 data.

\begin{figure}
\begin{center}
\includegraphics[width=0.5\textwidth]{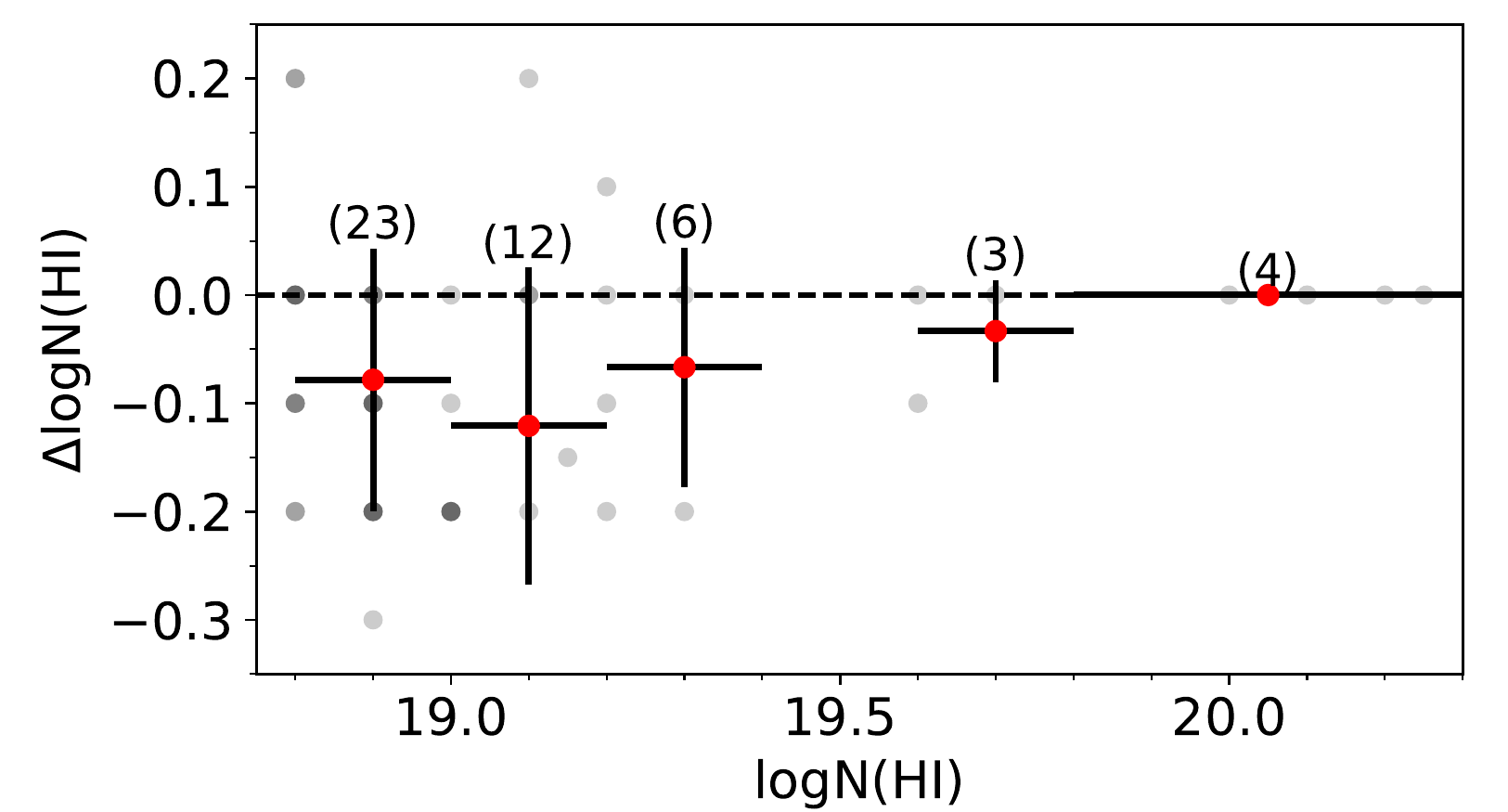}
\caption[]{The red points show the mean offset of logN(\HI{}) measured between UVES and XQ-100 ($\Delta$logN(\HI{}); positive values imply a higher UVES-derived column density) in bins of column density. Error bars show the standard deviation in each bin, while the number in parenthesis above each data point indicate the number of absorbers within the bin. Offsets measured for individual absorbers are displayed as the grey circles, where a darker grey represent a higher density of systems with the same logN(\HI{}) and $\Delta$logN(\HI{}).}
\label{fig:UVESComparison}
\end{center}
\end{figure}

Despite the overall consistency between the UVES and XQ-100 data, we note a systematic difference in logN(\HI{}) exists between the two datasets. Figure \ref{fig:UVESComparison} shows the mean N(\HI{}) offset between the UVES and XQ-100 ($\Delta$logN(\HI{}); measured relative to the XQ-100 value) as a function of logN(\HI{}), and indicates that the UVES data tend to provide lower logN(\HI{}) fits than the XQ-100 data for absorbers with logN(\HI{})$<19.8$ by up to 0.1 dex on average. All the offsets are tabulated in Table \ref{tab:UVESdN}. Table \ref{tab:UVESdN} also includes the fraction of system which have a non-zero  $\Delta$logN(\HI{}) (n$_{\rm off}$) relative to the total number of absorbers (n$_{\rm tot}$) within each column density bin. As both the VIS (R$\approx$8900) and UVB (R$\approx$5400) arms of X-Shooter have different spectral resolutions, we checked for a dependence of the measured $\Delta$logN(\HI{}) for absorbers identified in the respective arms. Absorbers detected in the VIS arm have a slightly smaller $\Delta$logN(\HI{}) than those found in the lower resolution UVB arm by at most $\approx0.05$ dex, so the effects of the differing arm resolutions are minimal.

\begin{table}
\begin{center}
\caption{Mean logN(\HI{}) offsets between UVES and XQ-100 data}
\label{tab:UVESdN}
\begin{tabular}{rcc}
XQ-100 logN(\HI{}) range & $\Delta$logN(\HI{}) & n$_{\rm off}$/n$_{\rm tot}$ \\
\hline
$18.8\leq$logN(\HI{})$<19.0$ & $-0.08$ & 16/23 \\
$19.0\leq$logN(\HI{})$<19.2$ & $-0.10$ & 8/13\\
$19.2\leq$logN(\HI{})$<19.4$ & $-0.06$ & 4/6\\
$19.4\leq$logN(\HI{})$<19.6$ & \nodata{} & 0/0\\
$19.6\leq$logN(\HI{})$<19.8$ & $-0.03$ & 1/3\\
$19.8\leq$logN(\HI{})$<20.3$ & $0.0$ & 0/4\\
logN(\HI{})$\geq20.3$ & $0.0$ & 0/5\\
\end{tabular}
\end{center}
\end{table}

\subsubsection*{False positives and metal lines}
\label{sec:FalsePos}
In addition to ensuring that all the subDLAs and DLAs were identified in the XQ-100 dataset, it is possible that two closely-separated Lyman limit systems or other Lyman series absorption can mimic a low column density subDLA, leading to falsely identified subDLA systems.  For low-to-modest resolution spectra or low redshift absorption systems, the Lyman series lines are difficult to detect within the blanket Lyman limit absorption, or separate from the Ly$\beta$ forest. One method to determine the false positive rate is to use higher resolution data where it is easier to resolve distinct absorption lines. In Section \ref{secHI:HiRes}, we found only two cases of the 63 systems where two Ly$\alpha$ lines were a better fit to the higher resolution UVES data, suggesting these false positives are not a common occurrence in the XQ-100 data. However this is not a complete assessment of the false positive rate as it is still possible that lines can be blended at even closer separations.  We caution that our comparison with the UVES data does not uniformly probe the range in wavelength (with respect to the rest frame of the quasar) to properly assess the frequency of mis-identification of systems due to blending.

\begin{figure}
\begin{center}
\includegraphics[width=0.48\textwidth]{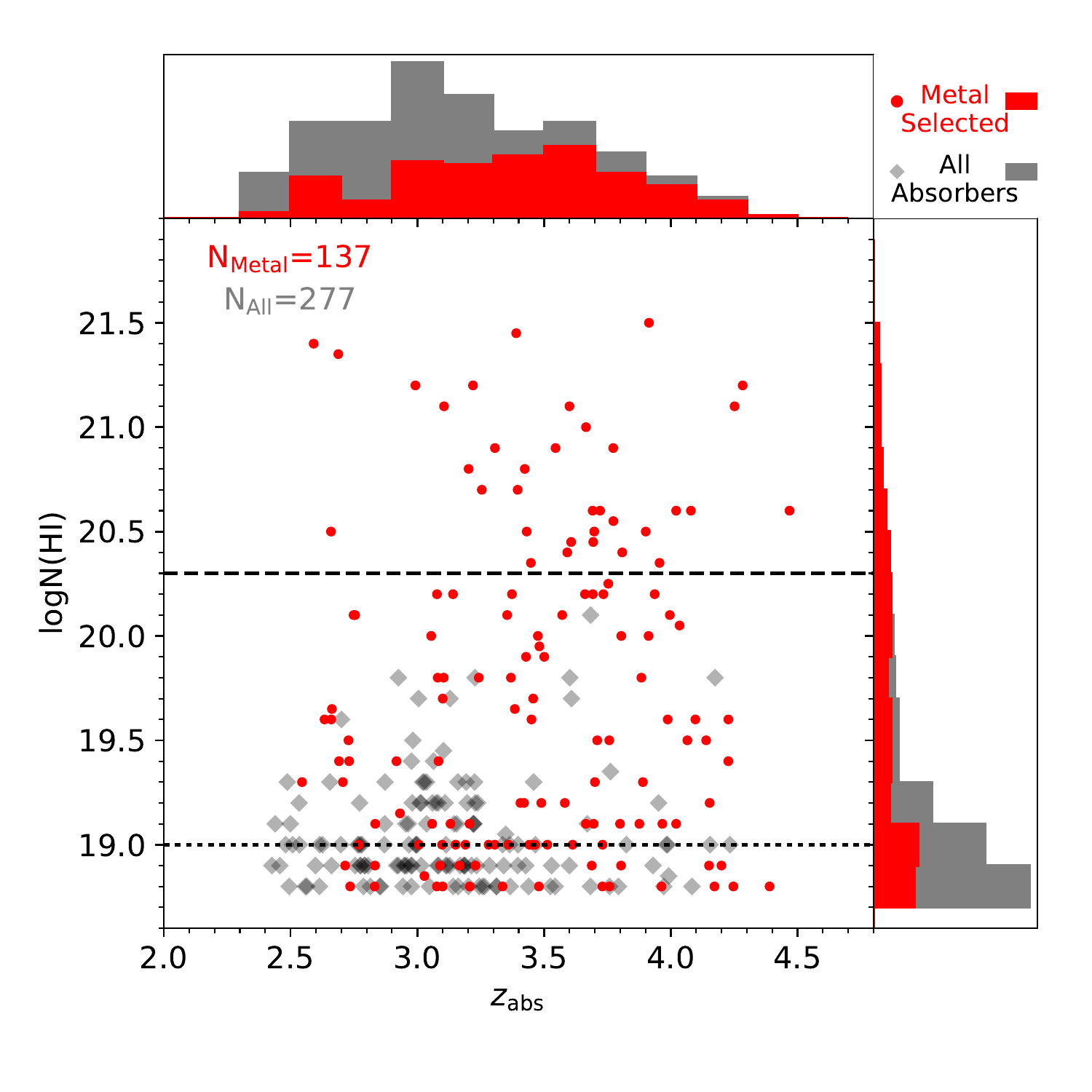}
\caption[N(\HI{}) vs. \zabs{} for all XQ-100 absorber]{The \HI{} column density (logN(\HI{})) of all candidate XQ-100 absorbers (light grey diamonds) as a function of their redshift (\zabs{}). The red circles indicate the XQ-100  metal-selected absorbers. The total number of absorbers (N) for each sample are given in the top left corner of the main panel. The distributions of \zabs{} and logN(\HI{}) are given in the top and right-most panels, using the same colour scheme for the two subsamples. For reference, the dotted and dashed horizontal lines denote the column density thresholds for subDLAs and DLAs (respectively).}
\label{fig:Nz}
\end{center}
\end{figure}

Another common method to remove these `false positive' absorbers from a sample is to check for the presence of higher order Lyman series lines or metal lines associated with the absorber, particularly using high resolution data \citep{Peroux03b,Omeara07,Zafar13}. The typical metal lines used are: C\ion{ii}, C\ion{iv}, Si\ion{ii}, Si\ion{iv}, Mg\ion{ii}, or Fe\ion{ii}. However, using metal lines for confirmation of absorbers may either: \emph{(i)} artificially remove systems with weak or no metal lines that are not easily identified (i.e.~metal-poor or systems with large dust or ionization corrections to neutral species), or \emph{(ii)} select blends of lower N(\HI{}) absorbers which still have strong metal lines. Either of these two cases can introduce different biases into the sample.  

From the full sample of candidate absorption systems with logN(\HI{})$\geq18.8$, we identified a subsample of metal-selected absorbers, where at least one of the following ionic species is detected: C\ion{iv} ($\lambda\lambda$ 1548~\AA{} \& 1550~\AA{}), Si\ion{iv} ($\lambda\lambda$ 1393~\AA{} \& 1402~\AA{}), Mg\ion{ii} ($\lambda\lambda$ 2796~\AA{} \& 2803~\AA{}), or Fe\ion{ii} ($\lambda$ 2344~\AA{}, 2382~\AA{}, \& 2600~\AA{}). The systems with observed metal lines are identified by a `Y' in the `Metal absorption detected?' column of Table \ref{tab:Ncat}. Figure \ref{fig:Nz} shows the distribution of logN(\HI{}) and \zabs{} for all the absorbers with logN(\HI{})$\geq18.8$ (grey diamonds) and metal-selected absorbers (red circles). We point out that all 34 DLAs have metal lines associated.  It is clear that the metal selection process preferentially removes low redshift, low \HI{}-column density (logN(\HI{})$\leq19.4$) candidate absorbers from the full list of absorbers; these removed systems tend to be found where higher-order Lyman series blending is more likely to produce false positive Ly$\alpha$ absorption.

\begin{figure*}
\begin{center}
\includegraphics[width=0.98\textwidth]{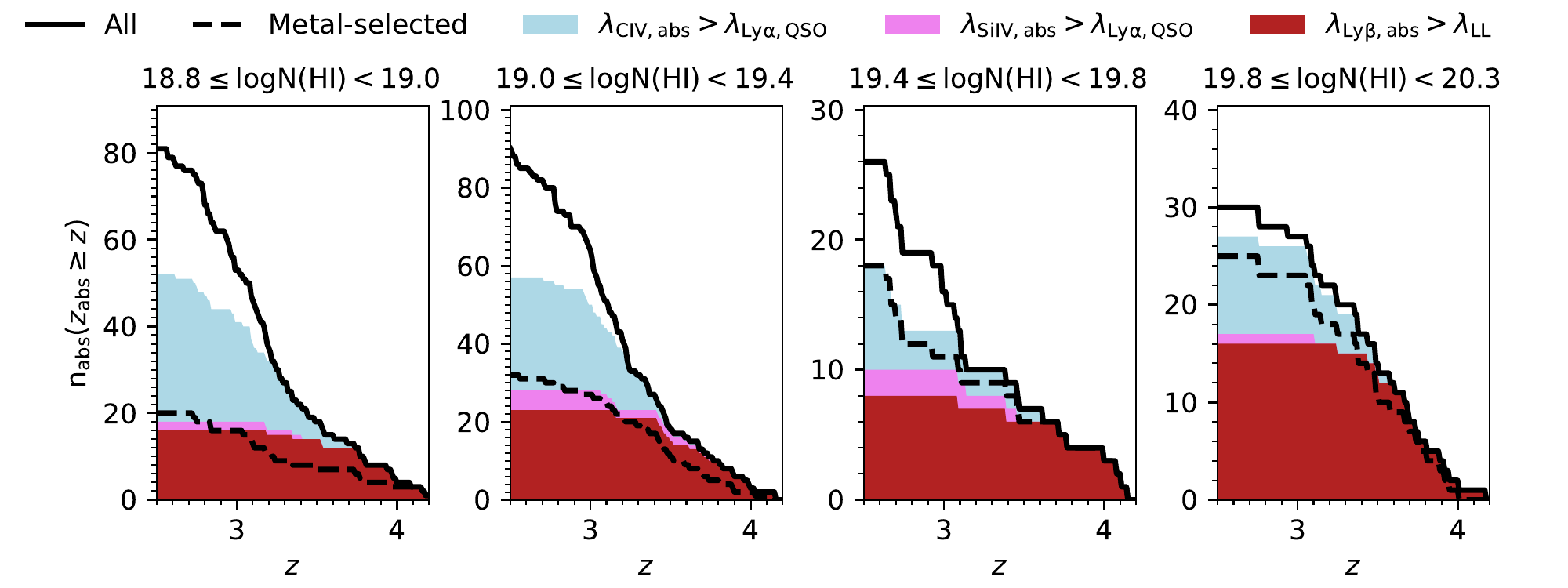}
\caption{The cumulative distribution of XQ-100 absorbers (with $18.8\leq$~logN(\HI{})~$<20.3$) above a given redshift $z$ ($n_{\rm abs}(z_{\rm abs}\geq z)$). The four panels split the absorbers into different logN(\HI{}) bins ([18.8,19.0), [19.0,19.4), [19.4,19.8), [19.8,20.3); from left to right). The solid and dashed black line denote the curves using all the identified absorbers and metal-selected systems (respectively). The shaded regions show the distribution of absorbers whose key diagnostic lines (Ly$\beta$, C\ion{iv}, and Si\ion{iv}; dark red, light blue, and magenta, respectively) are in uncontaminated regions of the spectrum (i.e. Ly$\beta$ redward of the QSO's Lyman Limit; metal lines outside of the Ly$\alpha$ forest). Unshaded regions below the solid line denote the number of absorbers that cannot be confirmed as metal-selected using either C\ion{iv} or Si\ion{iv}, and contains $\approx40$ per cent of all the absorbers identified (mostly below \zabs{}~$\lesssim3.2$ and logN(\HI{})~$\lesssim19.4$). The shaded region above the dashed line estimates the inability to detect the given metal line (due to low signal-noise, blending, or being metal-poor) combined with false-positives. This regions is significantly populated by logN(\HI{})$<19.4$ absorbers.}
\label{fig:zHdist}
\end{center}
\end{figure*}

To understand what is driving the large discrepancy between the number of absorbers between metal-selected sample and all identified absorbers at low \zabs{} and \HI{}, Figure \ref{fig:zHdist} shows the cumulative number of subDLAs as a function of redshift for all the absorbers (solid black lines) and the metal-selected sample (dashed black lines). The distributions are split up into three panels for three different column density ranges. For comparison, the shaded regions represent the number of absorbers whose commonly-detected metal diagnostic lines (i.e.~C\ion{iv} 1548\AA{} and Si\ion{iv} 1393 \AA{}; light blue and magenta, respectively) are outside of the Ly$\alpha$ forest. Although only C\ion{iv} and Si\ion{iv} doublets have been displayed, other commonly-observed low-ionization species (e.g.~Si\ion{ii} 1260\AA{}, C\ion{ii} 1334\AA{}) are bluer than the Si\ion{iv} doublet, or are often blended with telluric lines at these redshifts (e.g.~Fe\ion{ii} 2586\AA{}, and the Mg\ion{ii} doublet at 2796\AA{} and 2803\AA{}); and thus these lines are used less frequently for metal line identification in this work. Eleven of the absorbers with confirmed metal lines ($19.3\leq$~logN(\HI{})~$\leq20.2$) have been confirmed using only low ionization metal lines; eight of these systems have blended C\ion{iv} or Si\ion{iv}, and one subDLA (logN(\HI{})~$=19.8$) has no associated C\ion{iv} and Si\ion{iv} absorption.

The dark red shaded regions of Figure \ref{fig:zHdist} show the number of absorbers whose Ly$\beta$ is redward of $\lambda_{\rm LL}$ to summarize the frequency in which can confirm the presence of \HI{} absorption through Ly$\beta$. The redshift at which the metal-selected dashed line is greater than the red shaded region (\zabs{}~$\lesssim3.5$) denotes the highest redshift where it is no longer possible to remove false-positives in a metal-blind fashion using the Lyman series. This method for removing false-positives appears to be successful for the XQ-100 subDLAs, as the solid line for all logN(\HI{})~$\geq 19$ absorbers are very similar above \zabs{}~$\gtrsim3.5$. However, the large red-shaded region between the solid and dashed lines in the leftmost panels suggests that using Ly$\beta$ absorption to remove false-positives becomes challenging for systems with low \HI{} column densities (logN(\HI{})~$\lesssim19.4$) up to \zabs{}~$\approx4$ in the XQ-100 data. Upon investigation, this difficulty in removing false-positive using Ly$\beta$ is mostly due to blending of the weaker Ly$\beta$ line of these absorbers with the Ly$\alpha$ forest. As a result it is more difficult to confirm the observed absorption is primarily from Ly$\beta$ for logN(\HI{})~$<19$ XQ-100 absorbers at all redshifts. 

Of the 86 systems without detected metal lines, 77 of these systems have the corresponding C\ion{iv} and Si\ion{iv}  absorption in the Ly$\alpha$ forest; this is represented by the unshaded regions below the solid black line.  Thus for nearly 40 per cent of all identified subDLAs, \emph{we cannot determine if the commonly detected metal lines are present and estimate the false positive rate}. The subDLAs which populate the unshaded region are typically found at \zabs{}~$\lesssim3$ as a result of the XQ-100 QSO redshifts. We point out that this percentage is much less for subDLAs with logN(\HI{})$\geq19.8$ as neutral metal species typically outside the forest and not in telluric lines (e.g. Fe\ion{ii} 1608\AA{} and Al\ion{ii} 1670\AA{}) are more often detected due to larger column densities of gas.

In Figure \ref{fig:zHdist}, signatures of false positives would manifest as shaded regions above the black dashed lines (i.e.~no diagnostic absorption lines are detected). We caution that this shaded regime can also be driven by many different factors other than false-positives such as regions where the line is blended (particularly for Ly$\beta$; red shaded region), systems which are metal-poor (blue and magenta regions only), or low signal-noise data. For the 14 absorbers which have undetected C\ion{iv} or Si\ion{iv} in uncontaminated regions of the spectrum, we measured detection thresholds of logN(C\ion{iv})$\leq12.97$ and logN(Si\ion{iv})~$\leq12.76$. For C\ion{iv} (which should be easily identified in absorbers within the shaded regime), the column density detection threshold is higher than the expected column densities observed in subDLAs from the literature \citep[logN(C\ion{iv})~$\gtrsim12.75$;][]{Fox07}, and thus the XQ-100 sample can be limited by insufficient signal-noise to detect C\ion{iv} for absorbers with logN(\HI{})~$<19.4$ at lower redshift ($z<3$). However, we note that generally higher metallicities are observed in low redshift subDLAs compared to DLAs \citep{Pettini99,Rafelski14,Quiret16}.

To better understand whether the XQ-100 spectra have the capability of detecting low N(\HI{}), metal-poor absorbers, we computed the signal-noise ratio ($SNR$) required to detect a neutral metal line absorption line (with rest-frame wavelength $\lambda$ and oscillator strength $f$), at 3$\sigma$ significance, for an absorber with an observed metallicity [M/H] (i.e.~no dust or ionization corrections) using 
\begin{equation}
\label{eq:snr}
SNR = \frac{3 m_{e} c ~ FWHM}{\pi e_{c}^{2} f \lambda^{2}}\left(N(\HI)10^{[M/H]+log\left(\frac{n_{\rm M_{\odot}}}{n_{\rm H_{\odot}}}\right)}\right)^{-1};
\end{equation}
where $m_{e}$ and $e_{c}$ are the electron mass and charge, and $\frac{n_{\rm M_{\odot}}}{n_{\rm H_{\odot}}}$ is the observed solar abundance of the species from \cite{Asplund09}. Equation \ref{eq:snr} assumes the metal line is detected at the corresponding full width half-maximum ($FWHM$; measured in the observed frame) of the instrumental resolution of the NIR arm ($R\approx 5600$).

\begin{figure}
\begin{center}
\includegraphics[width=0.5\textwidth]{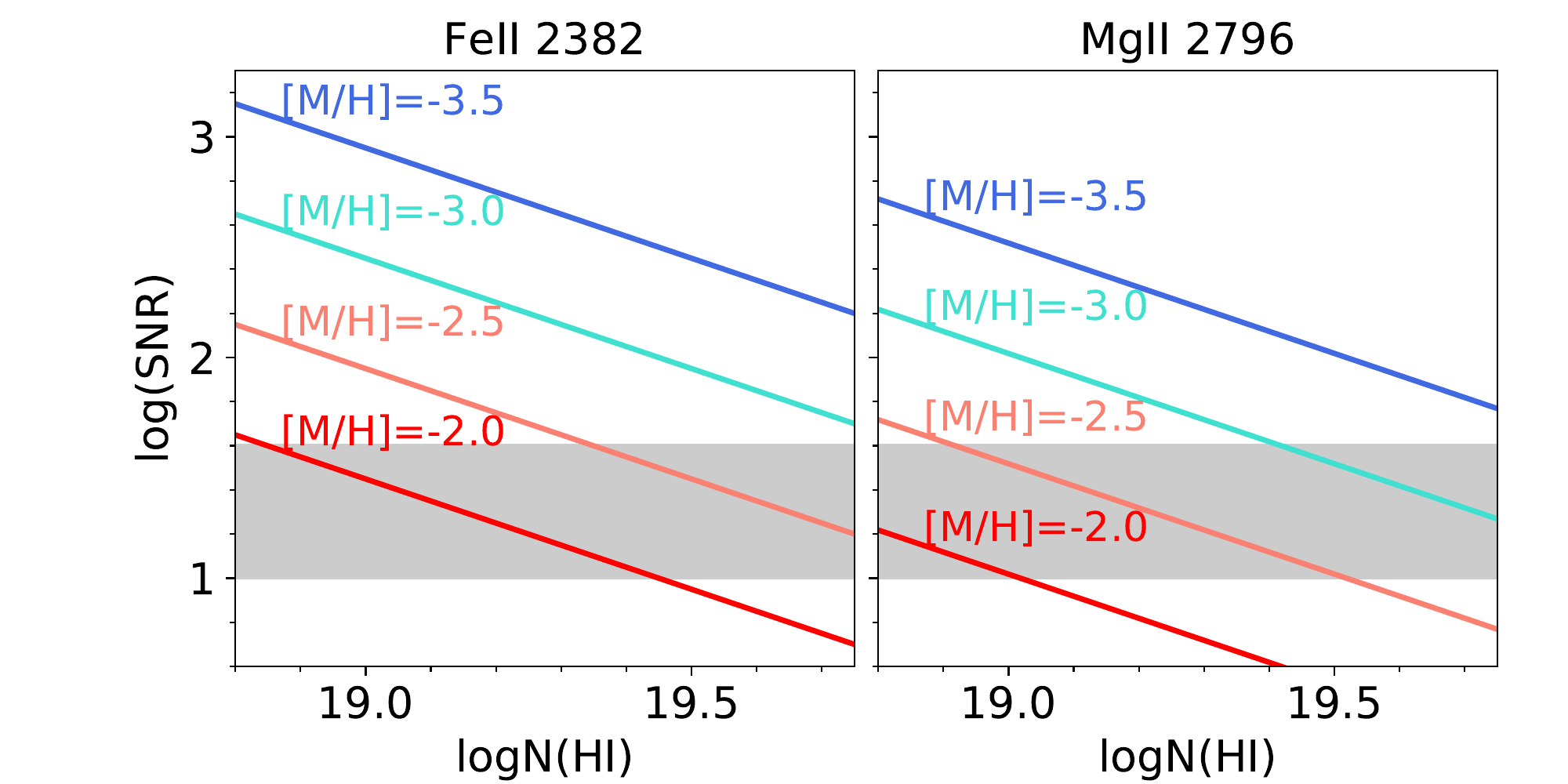}
\caption{The predicted spectral $SNR$ required to observe the metal lines Fe\ion{ii} $\lambda$ 2382~\AA{} (left panel) and Mg\ion{ii} $\lambda$ 2796~\AA{} (right panel) as a function of absorber logN(\HI{}).  A range in metallicities is represented by the different coloured lines. The typical range in signal-noise of the XQ-100 spectra is denoted by the light grey region. }
\label{fig:SNR}
\end{center}
\end{figure}

Figure \ref{fig:SNR} shows the range of signal-noise required to detect the strong neutral metal lines Fe\ion{ii} $\lambda$ 2382~\AA{} (left panel) and Mg\ion{ii} $\lambda$ 2796~\AA{} (right panel), as a function of logN(\HI{}) and observed metallicity ([M/H]; coloured lines). The typical range of $SNR$ in the XQ-100 spectra is shown by the grey shaded regions. Figure \ref{fig:SNR} demonstrates that subDLAs should have detectable Fe\ion{ii} 2382~\AA{} in our X-Shooter data for [Fe/H]~$>-2.0$. Few subDLAs have previously been detected with an observed [M/H]$<-2$ \citep{Quiret16,Fumagalli16}. Therefore, the ability of our X-Shooter data to detect metal lines for typical subDLA metallicities suggests that up to $\approx46$ per cent (128 systems which do not have confirmed metal lines) of the candidate subDLAs might be lower column density or blended systems. However the lack of low-metallicity subDLAs in the literature could be a result of metal-line selection, particularly as DLAs can probe down to [M/H]$\approx -3.0$ \citep{Cooke15,Quiret16}. A detailed analysis of the metal abundances of the XQ-100 subDLAs will be presented in a future paper (Berg et al., in prep).

In summary, the region between the solid (all identified absorbers with logN(\HI{})$\geq18.8$) and dashed lines (metal-selected absorbers) in Figure \ref{fig:zHdist} represents a combination of the false positives in our full sample of absorbers, metal-poor absorbers ([M/H]$\lesssim-2$), and absorbers whose metal lines lie in regions suffering from blending or insufficient signal-noise to classify into the metal-selected sample. As these two curves encompass the two extremes in the number of subDLAs identified, we use all the identified absorbers and metal-selected sample as bounds to the true number of subDLAs within the XQ-100 sample for our analysis. It is worth emphasizing that the discrepancy in the number of absorbers identified that do not have confirmed metal lines is strongest at low column densities (logN(\HI{})~$\leq19.4$) and for systems at low redshifts where metal lines are shifted into the Ly$\alpha$ forest (\zabs{}~$\lesssim3.2$). Although the false-positive rate may be significant at low column densities ($19.0\leq$~logN(\HI{})~$\leq19.4$), the difference in their contribution to the total \HI{} mass budget of subDLAs and DLAs is only a couple of per cent (see Section \ref{sec:Mass}).

\section{Results}
\label{secHI:Analysis}

In \cite{SanchezRamirez16}, we computed both \fNX{} and \omegaDLA{} for the XQ-100 DLA sample and analyzed the evolution of \omegaDLA{}. In this section, we will present the added information obtained from calculating \fNX{} and \omegaDLA{} from the XQ-100 subDLA sample. For consistency with \cite{SanchezRamirez16}, we restricted the XQ-100 sample to exclude systems with an absorption redshift within 5000~\kms{} of the QSO emission redshift. These proximate systems likely trace a different population of systems compared to their intervening counterparts, \citep{Ellison02,Ellison10,Berg16,Perrotta16} and are typically ignored when computing \omegaDLA{}.  To properly measure \fNX{} for DLAs, a sample size much larger than the 33 DLAs identified in the XQ-100 sample must be used in order to statistically probe the high column density end of \fNX{} (logN(\HI{})$>21.5$). We thus combined the XQ-100 DLA sample with literature sample compiled in \cite{SanchezRamirez16}, which consists of 1019 DLAs measured in various literature surveys across a redshift range of $1.6<$\zabs{}$<5.1$.  It is important to note that this supplementary DLAs sample does not include any subDLAs, and probes a very different redshift path than the XQ-100 subDLAs.  For this reason, we divide the entire XQ-100 absorber sample into DLAs and subDLAs in the rest of this analysis.

\subsection{Bootstrap Monte Carlo re-sampling}
\label{sec:BMC}
To estimate the errors in deriving \fNX{} and \omegaDLA{} from the XQ-100 dataset, the bootstrap Monte Carlo (further abbreviated B-MC) resampling technique presented in \cite{SanchezRamirez16} was used. This technique provides a robust method of combining the random errors quantified for measurements (such as logN(\HI{})) with the sampling errors inherent to a relatively small sample size. In essence, the B-MC method takes the original absorber N(\HI{}) distribution and duplicates the sample a large number of times to create N$_{\rm samp}$ simulated mock surveys. For each mock survey, the logN(\HI{}) of each absorbers is randomly varied within its error-bar (assuming a Gaussian error distribution) to create a `new' set of logN(\HI{}) measurements. From this new set of measurements, absorbers are randomly drawn (with replacements) until a new mock survey is generated with the same number of absorbers. As the drawing of values is done with replacement, this mimics the effect of sampling errors by randomly excluding or duplicating absorbers in the sample. The choice of N$_{\rm samp}$ is arbitrary, but needs to be very large ($\gtrsim 10000$) in order to provide a robust statistical sample.

We also experimented with statistically incorporating the effects of using the modest resolution of X-Shooter to measure \HI{} column densities over high resolution datasets (e.g.~UVES; see Section \ref{secHI:HiRes}). Using the frequency and average measured offset seen between the X-Shooter and UVES column densities (as presented in Table \ref{tab:UVESdN}), we varied a random fraction of the column densities in the B-MC simulation by the average measured N(\HI{}) offset. After implementing these offsets, we did not detect a significant difference in the results of this paper. In addition to not having a sufficient sample size of absorbers with UVES coverage to provide robust statistics for different redshifts and column densities (see Section \ref{secHI:HiRes}), we did not implement this resolution experiment into the final B-MC simulation.

As the XQ-100 absorber sample is split into two subsamples based on column density (subDLAs and DLAs), the B-MC technique also needs to take into account the effects of removing/including systems that could be misclassified into the wrong \HI{} absorber type (e.g.~a subDLA with a measured logN(\HI{})$=20.1\pm0.2$ could be a DLA in reality, or conversely a logN(\HI{})$=18.8\pm0.2$ absorber could be a subDLA). By design, this misclassification effect is naturally taken into account by the random variation in logN(\HI{}) during each resampling of the identified absorbers down to a column density threshold of logN(\HI{})$=18.8$ (Section \ref{sec:ID}).  However, as the supplementary literature DLA sample from \cite{SanchezRamirez16} does not contain subDLAs and covers a different redshift path; only absorbers along the associated redshift path of the XQ-100 sightlines are used in the subDLA computation, even if the supplementary literature DLA would be classified as a subDLA. We emphasize that even though all absorbers identified with logN(\HI{})$\geq18.8$ are used in the B-MC simulation, absorbers with a randomly-varied logN(\HI{}) are still classified as subDLAs and DLAs based on the classical logN(\HI{}) cuts of $19.0\leq$~logN(\HI{})~$<20.3$ and logN(\HI{})~$\geq20.3$, respectively, within each mock generated survey. As already stated, the B-MC simulations were repeated using all identified absorbers with logN(\HI{})$\geq18.8$ and those that were metal-selected to provide bounds on the quantities computed in this section. The subDLAs generated from these two sets of B-MC simulations will be further referred to as the full and metal-selected samples (abbreviated FS and MS), respectively.

\subsection{The frequency of subDLAs}
\label{sec:Freq}
\subsubsection{\fNX{}}
Using the B-MC resampling of the XQ-100 and supplementary DLA samples, \fNX{} was computed using:  
\begin{equation}
\label{eq:fNX}
f(N,X) dN dX=  \frac{m_{\rm abs}}{\Delta N \times \sum_{i}^{n_{\rm QSO}} \Delta X_{i}} dN dX,
\end{equation}
where $m_{\rm abs}$ is the total number of absorbers with column densities between $N-\Delta N/2$ and $N+\Delta N/2$ along the observed absorption distance $\Delta X_{i}$ of the $i^{th}$ QSO sightline, and $n_{\rm QSO}$ is the number of  QSO sightlines observed. The total redshift path covered by the QSO sightline ($\Delta X_{i}$) is computed as
\begin{equation}
\label{eq:zpath}
\Delta X(z)=\int_{z_{min}}^{z_{max}} (1+z)^2 [\Omega_M(1+z)^3 + \Omega_\Lambda]^{-0.5} dz,
\end{equation}
where $z_{min}$ and $z_{max}$ are the minimum and maximum redshift observed along the QSO path, while $\Omega_M$ and $\Omega_\Lambda$ are the matter and dark energy densities of the Universe observed at the current epoch. $z_{min}$ is set by the Lyman limit of the QSO, while $z_{max}$ is set as \zem{}~$-5000$~\kms{} to remove proximate systems that can be associated with the QSO \citep{Ellison02,Ellison10,Berg16,Perrotta16}.

Rather than using the standard bins of column density, which can hide or amplify features depending on the choice of bins, we elected to use a sliding logN(\HI{}) bin to quantify the median \fNX{} in each column density bin. The sliding bin keeps a fixed bin width, but the centre of the bin shifts by a small amount in each step to capture the effects of individual data being included and excluded from the bin.  Using a logN(\HI{}) bin width of 0.4 dex, we computed the median, 1$\sigma$ and 2$\sigma$ percentiles of the  underlying \fNX{} distribution of the 10 000 B-MC samples. For the subDLAs, the centre of the bin starts at logN(\HI{})$=19.2$ (such that bin edge includes the minimum subDLA threshold of logN(\HI{})$=19.0$), and is then shifted by 0.05 dex until the the centre of the bin reaches logN(\HI{})$=20.1$ (such that bin edge includes logN(\HI{})$=20.3$). The same is done for the DLAs, but using logN(\HI{}) bin centres from 20.5 to 21.8. Note that the sliding bins of both the DLA and subDLA samples do not pass the logN(\HI{})$=20.3$ threshold due to the different redshift path coverage of the two samples. The B-MC computation of \fNX{} was repeated using both the FS and MS absorbers from XQ-100.

\begin{figure*}
\begin{center}
\includegraphics[width=0.95\textwidth]{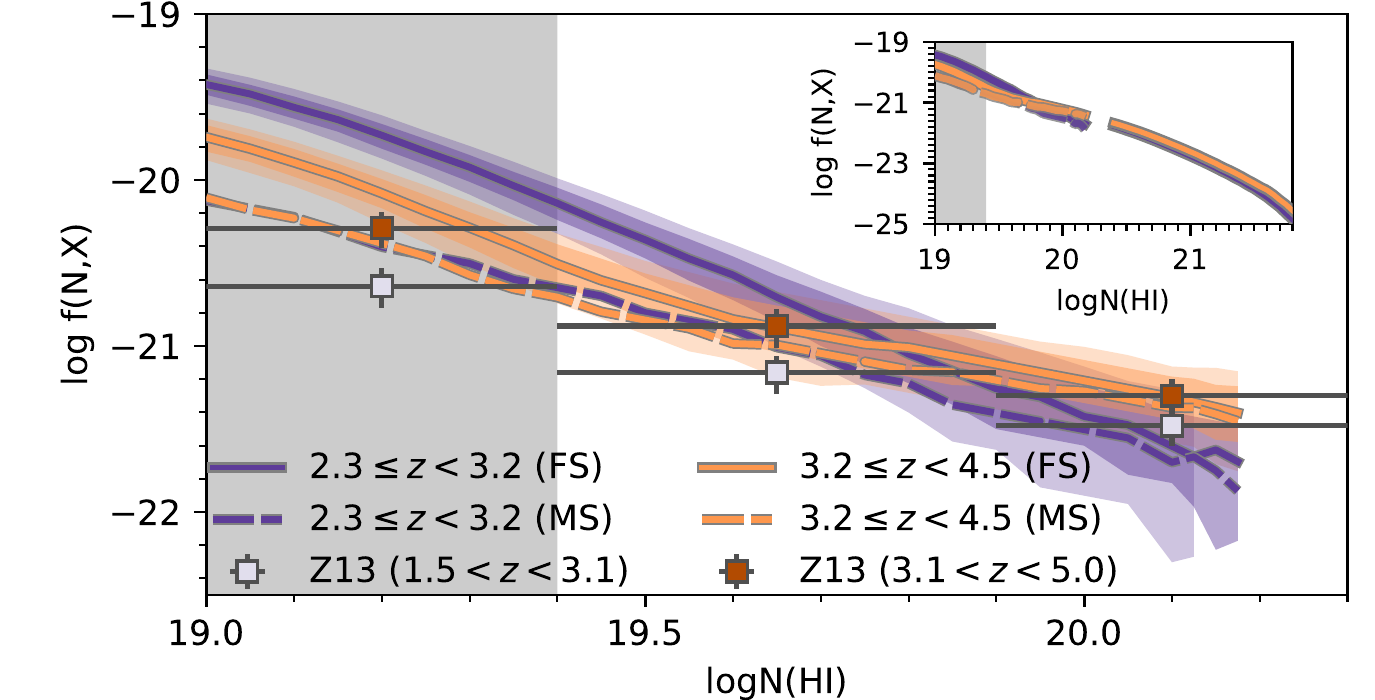}
\caption[Column density distribution of subDLAs and DLAs]{The column density distribution of subDLAs in the XQ-100 sample. The median \fNX{}  for the full sample of subDLAs (FS) is given by the solid line, while the dark and light shaded regions represent the $1\sigma$ and $2\sigma$ percentiles (respectively). The dashed line shows the median \fNX{} for the metal-selected sample of subDLAs (MS). The confidence intervals for the MS selected sample are similar to those of the FS, but are not shown for clarity. Note that the dashed lines from both of the MS redshift bins overlap below logN(\HI{})$\approx19.8$.  Both subDLA samples were split by the median redshift of the XQ-100 absorbers (\zabs$=3.2$), such that those in the lower redshift bin ($2.3\leq$~\zabs{}~$<3.2$) are in dark purple while those at higher redshifts ($3.2\leq$~\zabs{}~$<4.5$) are in light orange. The greyed region below logN(\HI{})$=19.4$ represents the regime where incompleteness and false-positive effects could  significantly influence the determined \fNX{} (Section \ref{sec:Complete}). The square points show the data for \citet[Z13]{Zafar13} subDLAs in two similar redshifts ($1.5\leq$~\zabs{}~$<3.1$ and $3.1\leq$~\zabs{}~$<5$; light purple and dark orange squares, respectively). The inset panel show the median \fNX{} curves of both the subDLA and DLA samples for reference.}
\label{fig:fNX}
\end{center}
\end{figure*}

Figure \ref{fig:fNX} shows the resulting median column density distribution functions of the FS and MS subDLA samples (solid and dashed lines; respectively). The $1\sigma$ and $2\sigma$ errors are shown as the dark and light shaded regions (respectively) for the FS absorbers. Each sample was split into two redshift bins by the median redshift of the XQ-100 absorbers (\zabs$=3.2$); \fNX{} for the lower and higher redshift bins are denoted by the purple and orange colours (respectively). The numerical values of the FS and MS \fNX{} curves are tabulated in the appendix (Tables \ref{tab:fNXFS} and \ref{tab:fNXMS}; respectively). For reference, the full \fNX{} range for subDLAs and DLAs is shown in the inset panel of Figure \ref{fig:fNX} with the same notation.

There is a clear difference in the shape of \fNX{} between all candidate subDLAs (FS) and those absorbers which have metal lines (MS), resulting in up to a $\approx0.4-0.6$~dex difference in \fNX{}  at low column densities. It is interesting to note that the difference between the FS and MS curves changes with logN(\HI{}), and is strongest at the lowest column densities. This discrepancy could result from either incompleteness in the MS (i.e.~lack of ability to detect metal lines at low column densities) or a high false positive rate in the FS [i.e.~from blending of low logN(\HI{})]. We emphasize again that using both the FS and MS likely bounds the true nature of \fNX{} at these column densities. The derived \fNX{} curves were compared with the values computed by \citet[squares in Figure \ref{fig:fNX}]{Zafar13} for subDLAs over approximately the same redshift range. The \cite{Zafar13} sample consisted of a search for subDLAs and DLAs in archival high resolution spectra taken with UVES in addition to a literature compilation. The identification of absorbers presented in \cite{Zafar13} started with an automated search for candidate subDLAs and DLAs characterized by saturated absorption over a large range of wavelength. The absorber was confirmed as a (sub)DLA if higher order Lyman series or metal lines were present; which is analogous to the methods used in determining the XQ-100 MS absorbers. It is therefore both encouraging and not surprising that \fNX{} for the MS absorbers is roughly consistent with \cite{Zafar13} over the similar high redshift bin (3.1~$\leq z<$~5.0; orange squares in Figure \ref{fig:fNX}). The discrepancy seen between the low redshift data is likely a result of different redshift coverage, as nearly two thirds of the absorbers in the low redshift bin of \citet[1.5~$\leq z<$~3.1; purple squares in Figure \ref{fig:fNX}]{Zafar13} are at a redshift \zabs{}~$<2.5$. We note that requiring coverage Ly$\beta$ redward of $\lambda_{LL}$ to remove potential false-positives (see Section \ref{sec:FalsePos}) does not significantly affect the resulting \fNX{} curve generated for the 3.2~$\leq z<$~4.5 redshift bin. Given the redshift range of the QSOs of the XQ-100 sample, there is insufficient spectral coverage of Ly$\beta$ to perform the same experiment within the lower redshift bin (2.3~$\leq z<$~3.2).

To better quantify the significance of the difference in \fNX{} between the MS and FS subDLAs, as well as search for any global evolution in \fNX{} with redshift, we computed the best-fit power law to \fNX{} (i.e.~\fNX{}$=K$N$^\alpha$; where $K$ and $\alpha$ are the parameters of the fit). For each mock sample generated in the B-MC simulation, an unweighted least-squares fit to \fNX{} was computed using the sliding bin data. The median and 1$\sigma$ percentile errors of $\alpha$ and $K$ of all 10 000 samples least squares fits are presented in Table \ref{tab:PowLaw}. We point out that a single power law does not appear to be a good fit for \fNX{} for the $3.2\leq$~\zabs{}~$<4.5$ FS subDLAs. Comparing the power law fit parameters $\alpha$ and $K$ between the MS and FS datasets demonstrates that the power law fits between the two subDLA ($19.0\leq$~logN(\HI{})~$<20.3$) samples are inconsistent at the $\approx2\sigma$ level, demonstrating that using metal selection could have a profound effect on the interpreted \fNX{}.  

Comparing the power law indices from each redshift bin, an insignificant evolution of $\alpha$ with redshift is detected at $\approx1.3\sigma$ and $\approx0.6\sigma$ significance for FS and MS subDLAs (respectively), and slight evolution for DLAs at $\approx1.5\sigma$. The increased number of logN(\HI{})~$\gtrsim19.7$ subDLAs at higher redshifts  is consistent with previous observational studies \citep[][]{Peroux03,Zafar13}. However, we note that the simulations predict a consistently higher \fNX{} across all column densities at higher redshifts \citep{Rahmati13a,VillaescusaNavarro18}, whereas both XQ-100 datasets suggest this is not the case at logN(\HI{})~$\lesssim 19.7$.  In the FS subDLAs, the increased number of low redshift, low column density subDLAs could partly be explained by a high false-positive rate due to blending of low redshift Ly$\alpha$ with the Ly$\beta$ forest. However the MS also shows no redshift evolution in \fNX{} for logN(\HI{})~$\lesssim 19.7$, which is approximately consistent with the lack of \fNX{} redshift evolution seen in simulations when local ionizing sources are included \citep{Rahmati13b}.

Previous analyses comparing the power law indices of DLAs and subDLAs have indicated a flattening in \fNX{} in the subDLA regime \citep{Peroux05,Omeara07,Guimaraes09,Zafar13}, which has been attributed to the loss of Ly$\alpha$ self-shielding in low N(\HI{}) subDLA systems \citep{Zheng02}. \cite{Omeara07} noted that this flattening in \fNX{} at low column densities is dominated by logN(\HI{})$\leq19.3$ systems; an effect which was tentatively seen in their power law fits to \fNX{} when only logN(\HI{})$\geq19.3$ absorbers were included. We attempted to search for this flattening as well by restricting the subDLA sample only to absorbers with $19.3\leq$logN(\HI{})$<20.3$ and recomputing the power law fit to limited column density range (see Table \ref{tab:PowLaw} for fit parameters). The power law indices between the two logN(\HI{}) regimes are inconsistent at $1.3\sigma$ and $0.6\sigma$ significance for the FS and MS (respectively), and thus we do not find significant evidence of this flattening.

\begin{table*}
\small
\begin{center}
\caption{Power law fit parameters to \fNX{}$=KN^{\alpha}$}
\label{tab:PowLaw}
\begin{tabular}{l|c|cc|cc}
logN(\HI{}) & \zabs{} & \multicolumn{2}{c|}{$\alpha$} & \multicolumn{2}{c}{log($K$)}\\
 & & FS absorbers & MS absorbers& FS absorbers & MS absorbers\\
 
 \hline
\rule{0pt}{3ex}$\in$[19.0,20.3) & $\in$[2.3,3.2) & $-2.00_{-0.176}^{+0.176}$ & $-1.46_{-0.196}^{+0.276}$ & $18.64_{-3.369}^{+3.381}$ & $7.74_{-5.293}^{+3.847}$ \\
\rule{0pt}{4ex} & $\in$[3.2,4.5) & $-1.37_{-0.119}^{+0.083}$ & $-1.12_{-0.198}^{+0.186}$ & $6.18_{-1.602}^{+2.294}$ & $1.15_{-3.688}^{+3.752}$ \\
\hline
\rule{0pt}{3ex}$\in$[19.3,20.3) & $\in$[2.3,3.2) & $-1.94_{-0.399}^{+0.230}$ & $-1.41_{-0.305}^{+0.337}$ & $17.46_{-4.457}^{+7.831}$ & $6.72_{-6.660}^{+6.050}$ \\
\rule{0pt}{4ex} & $\in$[3.2,4.5) & $-1.14_{-0.160}^{+0.260}$ & $-0.90_{-0.293}^{+0.373}$ & $1.68_{-5.255}^{+3.077}$ & $-3.30_{-7.461}^{+5.769}$ \\
\hline
\rule{0pt}{3ex}$\in$[20.3,22.0) & $\in$[2.3,3.2) & $-2.10_{-0.098}^{+0.114}$ & $-2.08_{-0.084}^{+0.062}$ & $21.23_{-2.392}^{+2.058}$ & $20.88_{-1.319}^{+1.713}$ \\
\rule{0pt}{4ex} & $\in$[3.2,4.5) & $-1.96_{-0.082}^{+0.131}$ & $-1.95_{-0.116}^{+0.111}$ & $18.52_{-2.729}^{+1.690}$ & $18.20_{-2.308}^{+2.354}$ \\
\hline

\end{tabular}
\end{center}
\end{table*}

\subsubsection{The line density of subDLAs}

\begin{figure}
\begin{center}
\includegraphics[width=0.5\textwidth]{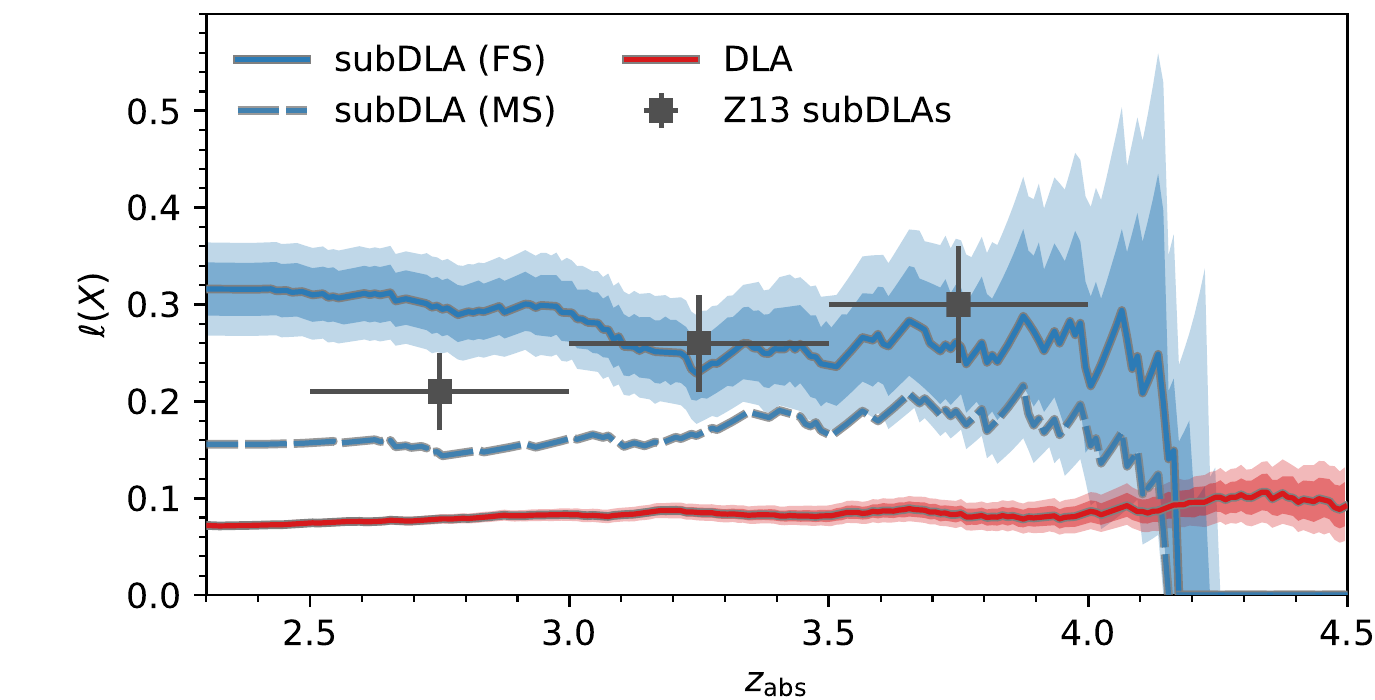}
\caption{The redshift evolution of \lX{} as probed by DLAs (red) and subDLAs (blue) in the XQ-100 survey.  The solid line shows the median \lX{} evolution for the full sample of subDLAs (FS) while the dashed line represents when only the metal-selected subDLAs (MS) is used. The shaded regions show the $1\sigma$ and $2\sigma$ confidence intervals for the FS; the confidence intervals for the MS are approximately the same size, but are not shown for clarity. The values of \lX{} computed for the \citet[Z13]{Zafar13} subDLAs are shown as the grey squares. }
\label{fig:lX}
\end{center}
\end{figure}

In addition to \fNX{}, the sightline density of absorbers (\lX{}) provides additional constraints on the nature of the absorbers. \lX{} is defined as:
\begin{equation}
\ell(X) = \int_{N_{min}}^{N_{max}} f(N,X) dN = \frac{m_{\rm abs}}{\sum_{i}^{n_{\rm QSO}} \Delta X_{i}}.
\end{equation}
Using the same sliding bin technique described above, we computed \lX{} in bins of $\Delta X=2.5$ with redshift steps of 0.01. Figure \ref{fig:lX} shows the resulting redshift evolution curves of \lX{} for the subDLAs (both FS and MS, blue solid and dashed lines) and DLAs \citep[red line; from][]{SanchezRamirez16}. The curves presented in Figure \ref{fig:lX} are available in Table \ref{tab:lX} (online only). Previous studies have disagreed on whether there is indeed any redshift evolution in \lX{} for subDLAs \citep{Omeara07,Zafar13}. However, any redshift evolution present in \lX{}  can physically manifests as a change in either (or both) the \HI{} cross-sections of the absorbers or their number density \citep[e.g.][]{Prochaska05DR3}.

After applying a unweighted least squares linear fit to the redshift evolution of \lX{} for each of the 10 000 iterations in the MC-B simulation, we find the median slope of the redshift evolution is almost negligible; $-0.02_{-0.013}^{+0.011}$ for the FS subDLAs, $0.01_{-0.009}^{+0.006}$ for the MS subDLAs, and $0.01_{-0.009}^{+0.009}$ for both DLA samples. We point out that the evolution of \lX{} in the MS is shallower than the evolution detected by \citet[grey points in Figure \ref{fig:lX}]{Zafar13}. The difference in the evolution of the \lX{} curves of the MS and FS (both in terms of the slope and relative offset) results from the increased number of low column density FS absorbers seen in Figure \ref{fig:fNX}. 

The lack of evolution in \lX{} seen for DLAs has been suggested to be due to a balance between the decreasing density of the Universe and a decrease in the strength of ionizing UV background in order to maintain an approximately constant \lX{} \citep{Prochaska05DR3,Omeara07}. Despite subDLAs being about two to four times more common than DLAs along a QSO sightline, the redshift evolution of both the subDLAs frequency and cross-sections appear to follow the evolution of the respective DLA quantities over these redshifts.

\subsection{The HI mass contribution of subDLAs}
\label{sec:Mass}
\omegaDLA{} was estimated using:
\begin{equation}
\label{eq:OmegaInt}
\Omega_{\rm H \textsc{i}} (X) = \frac{H_{0} m_{H}}{c \rho_{0}} \int_{N_{min}}^{N_{max}} N f(N,X) dN,
\end{equation}
where $H_{0}$ is the Hubble constant at the current epoch, $c$ is the speed of light, $m_{H}$ is the atomic mass of hydrogen, and $\rho_{0}$ is the critical density of the Universe.

Following \cite{SanchezRamirez16}, \omegaDLA{} is computed in sliding bins of redshift. Rather than holding the bin width fixed in redshift space, \cite{SanchezRamirez16} elected to use a fixed redshift path length $\Delta$X$=2.5$, with redshift bin steps of 0.01. For consistency, we elected to use the same sliding bin setup to reproduce the observed \omegaDLA{} for the XQ-100 DLAs. For the derivation of \omegaDLA{} (Equation \ref{eq:OmegaFrac}), we start with the same $z_{min}$ and $z_{max}$ used in \cite{SanchezRamirez16} for all the XQ-100 and supplementary QSO sightlines. However, due to the incompleteness of subDLA absorbers below $\lambda_{LL}$ (Section \ref{sec:ID}), we increased $z_{min}$ when necessary to remove all XQ-100 absorbers whose Ly$\alpha$ fell below $\lambda_{LL}$. 

\begin{figure*}
\begin{center}
\includegraphics[width=0.9\textwidth]{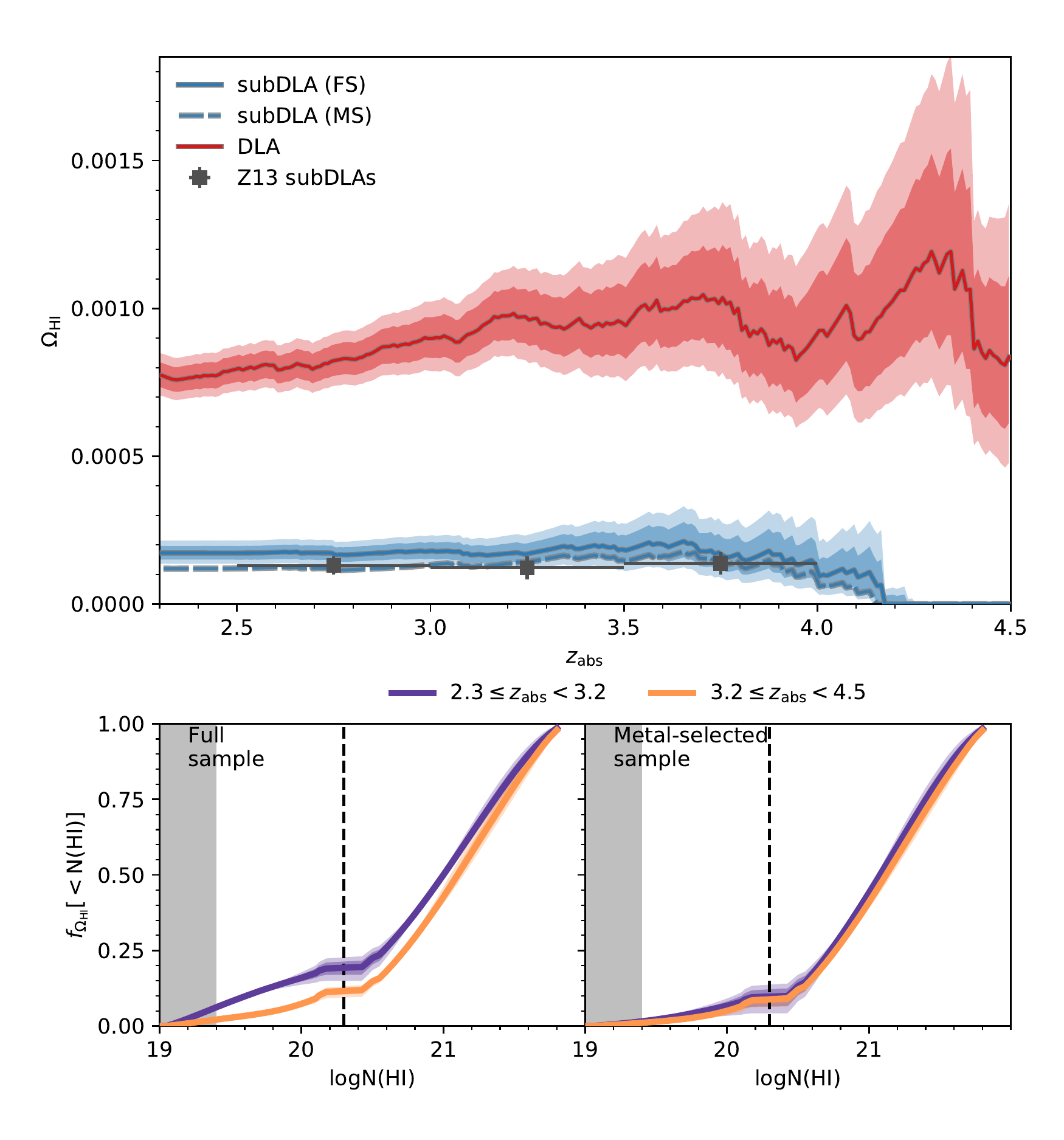}
\caption[\omegaDLA{} of XQ-100 subDLAs and DLAs]{\emph{Top panel:} The redshift evolution of \omegaDLA{} as probed by DLAs (red) and subDLAs (blue) in the XQ-100 survey.  The notation for the DLA and subDLA samples are the same as Figure \ref{fig:lX}. The values of \omegaDLA{} computed for subDLAs in \citet[Z13]{Zafar13} are shown as the grey squares. \emph{Bottom panels:} The cumulative contribution of absorbers to the total \omegaDLA{} for a given maximum column density (\fracOmega). \fracOmega{} is defined to be zero when only including absorbers with logN(\HI)$<19.0$, and has the value one when including all absorbers with logN(\HI{})$\leq22.0$. The left and right panels in the bottom row show \fracOmega{} for the respective full (FS) and metal-selected (MS) subDLAs, while the colours show the two redshift ranges $2.3\leq$~\zabs{}~$<3.2$ (purple) and $3.2\leq$~\zabs{}~$<4.5$ (orange). The vertical dashed line indicates the logN(\HI{})$=20.3$ threshold that separates DLAs and subDLAs, while the greyed region below logN(\HI{})$=19.4$ represents the regime where incompleteness and false positives could significantly affect the calculation (Section \ref{sec:Complete}).}
\label{fig:omega}
\end{center}
\end{figure*}

Similar to Figure \ref{fig:fNX}, the top panel of Figure \ref{fig:omega} shows the redshift evolution of \omegaDLA{} for both the DLAs (red) and subDLAs (blue), using both the FS (solid line) and MS absorbers (dashed line; subDLAs only). The \omegaDLA{} curves presented in Figure \ref{fig:omega} are available in the Appendix (Table \ref{tab:omega}). Although there is a slight but significant difference in the magnitude of \omegaDLA{} for the subDLAs between the FS and MS (detected at the $\approx1\sigma$ level), no significant evolution in \omegaDLA{} with redshift is detected for subDLAs in the range $2.5\lesssim$~\zabs{}~$\lesssim4.2$. We remind the reader that beyond \zabs{}~$\gtrsim4.2$  the redshift path coverage of the XQ-100 survey significantly drops \citep[Figure 5 of][]{SanchezRamirez16}, so it is unclear if the drop in \omegaDLA{} beyond \zabs{}~$\gtrsim4.2$ for subDLAs is real. 

The relatively constant \omegaDLA{} as a function of redshifts as traced by subDLAs is remarkably similar to the gradual decrease in \omegaDLA{} measured in DLAs over the same redshift range ($2.5\lesssim$~\zabs{}~$\lesssim4.2$). The canonical picture that has emerged to explain the gradual evolution in \omegaDLA{} in DLAs is a constant replenishment of gas in star-forming galaxies to counteract the consumption of gas from star formation and ejection from galactic outflows \citep{Noterdaeme09,Noterdaeme12,Zafar13}. The similar lack of significant redshift evolution in the \HI{} statistics of both DLAs and subDLAs (\fNX{}, \lX{} and \omegaDLA{}) is suggestive that the same astrophysical processes are responsible in the host galaxies of these absorbers.

Previous cosmological simulations have demonstrated that only higher mass galaxies contribute significantly to \omegaDLA{} below \zabs{}~$\lesssim 3$ while lower mass galaxies dominate the contribution at \zabs{}~$\gtrsim 3$, which leads to a relatively constant \omegaDLA{} with redshift \citep[][]{Dave13,Lagos14}. We speculate that the lack of significant evolution in \omegaDLA{} for both subDLAs and DLAs for $2.5\lesssim$~\zabs{}~$\lesssim4.2$ in Figure \ref{fig:omega} could potentially suggest that the average subDLA and DLA absorbers are probing the same range of host galaxy halo masses. As such, this potential equality in halo mass range for subDLA and DLA host galaxies would be in tension with previous works which use the observed higher metallicity of subDLAs (relative to DLAs) as a proxy for a higher host galaxy mass \citep[e.g.][]{Kulkarni07,Kulkarni10,Quiret16}.

To estimate the relative mass contribution of subDLAs to the \HI{} budget, we computed the relative ratio of \omegaDLA{} from subDLAs to the total \omegaDLA{} (i.e. DLAs and subDLAs). To reduce the effects of the vastly different redshift sampling of the DLA and subDLA samples, we elected to use the median \fNX{} curves (Figure \ref{fig:fNX}) for the two redshift bins using the discrete limit of Equation \ref{eq:OmegaInt}, namely
\begin{equation}
\label{eq:OmegaFrac}
\Omega_{\rm H \textsc{i}} (X) = \frac{H_{0}  m_{H}}{c \rho_{0}} \frac{\sum_{j}^{m_{\rm abs}} N_{i}}{\sum_{j}^{n_{\rm QSO}} \Delta X_{i}}.
\end{equation}
The bottom two panels of Figure \ref{fig:omega} shows the cumulative fraction contribution to \omegaDLA{} from absorbers of a given \HI{} column density or lower (\fracOmega{}). Since we are only focussing on a range of N(\HI{}) column densities, we restrict \fracOmega{} to have the value 0 for all absorbers that have logN(\HI)$<19.0$, and is 1 for all absorbers with logN(\HI{})~$<22.0$.

The bottom left and right panels of Figure \ref{fig:omega} show \fracOmega{} for both the FS and MS (respectively), with the different colours representing the two redshift bins: $2.3\leq$~\zabs{}~$<3.2$ (purple) and $3.2\leq$~\zabs{}~$<4.5$ (orange). We note that the discontinuity in \fracOmega{} at N(\HI{})~$\approx20.3$ is a result from the switch between the subDLA and DLA \fNX{} curves. We remind the reader that this discontinuity is a result of the different redshift paths covered by the DLA and subDLA samples, resulting in a truncated \fNX{} at logN(\HI{})~$=20.3$. Overall, the relative contribution of mass from subDLAs appears to be between $\approx12$--22\% (for the MS and FS, respectively), which is consistent with estimates in the literature \citep{Peroux05,Guimaraes09,Noterdaeme09,Zafar13}.  Despite the potentially high false-positive rate discussed in Section \ref{sec:FalsePos} for subDLAs with logN(\HI{})$\leq19.4$ at \zabs{}~$\lesssim3$, there is at most a 3 per cent contribution to the \HI{} mass contribution (i.e.~the difference between the FS and MS curves in the left panel) from the potential false psoitives identified. We also note the lack of a redshift evolution in \fracOmega{} for the MS, as previously seen by \cite{Zafar13}. Despite the increase in the number of logN(\HI{})$\gtrsim19.7$ absorbers at higher redshifts (Figure \ref{fig:fNX}), there is little effect on the \omegaDLA{} contribution from these systems.

\section{Conclusion}

We have presented a metal-blind catalogue of 242 candidate absorbers with \HI{} column densities of $18.8\leq$~logN(\HI{})~$<20.3$, identified solely using their \HI{} absorption, in the XQ-100 legacy survey. By comparing to archival UVES data, we demonstrated that X-Shooter has sufficient spectral resolution (FWHM resolution R$\sim$5400--8900) to identify potential subDLAs and measure their \HI{} column density accurately to within the typically quoted errors ($\pm0.2$ dex), which are primarily due to continuum placement. Using associated metal absorption lines as a method of removing false positive systems, we note that the 128 absorbers without confirmed metal lines were preferentially found to have logN(\HI{})~$\lesssim19.4$ and are at lower redshifts (\zabs{}~$\lesssim 3.0$). At these low column densities and redshifts, the corresponding prominent metal lines (C\ion{iv} and Si\ion{iv}) are shifted into the Ly$\alpha$ forest, Ly$\beta$ shifts beyond the Lyman limit of the QSO, and blending of the Ly$\alpha$ and Ly$\beta$ forests can mimic low column density subDLAs. Therefore there is a competing effect between incompleteness from using a metal-selected subsample and the true false-positive identification rate of $\lesssim49$ per cent of the sample. The false positive rate in XQ-100 is likely significant at low redshifts (\zabs{}~$\lesssim 3.5$) and low \HI{} column densities (logN(\HI{})~$\lesssim19.4$).

Using a Monte Carlo-Bootstrap simulation applied to the candidate absorber sample, we computed the column density distribution function (\fNX{}; Figure \ref{fig:fNX}), line density (\lX{}, Figure \ref{fig:lX}) and cosmic \HI{} mass density (\omegaDLA{}; Figure \ref{fig:omega}) for subDLAs selected using the standard column density range $19.0\leq$~logN(\HI{})~$<20.3$. We compared the \fNX{} computed for two different redshift bins ($2.3\leq$~\zabs{}$<3.2$ and $3.2\leq$~\zabs{}$<4.5$; split by the median absorber redshift of XQ-100) for both the metal-selected and full catalogue of subDLAs (dashed and solid lines in Figure \ref{fig:fNX}, respectively). Irrespective of which subDLA catalogue is used, we find that solely high column density subDLAs (logN(\HI{})$\gtrsim19.7$) are more frequently found at higher redshifts (\zabs{}$\geq3.2$), whereas all column densities are expected to increase in frequency in simulations \citep{Rahmati13a,VillaescusaNavarro18}. However, the power law fits to the entire subDLA \fNX{} curves suggest there is no  significant redshift evolution in \fNX{} ($\lesssim1.3\sigma$ significance). 

Using the full and metal-selected samples of subDLAs to bound the true value of \omegaDLA{} and \lX{},  we demonstrated that there is little evolution in \lX{} and \omegaDLA{} as a function of redshift for DLAs; a result that is also seen for DLAs \citep{Prochaska05DR3,Noterdaeme12,SanchezRamirez16}. In comparison with simulations \citep{Dave13,Lagos14} where the contribution from varying galaxy mass influences the nature of the redshift evolution  of \omegaDLA{}, the lack of significant redshift evolution in \lX{} and \omegaDLA{} is suggestive that the subDLAs and DLAs could potentially trace host galaxies with roughly the same range of halo masses and that the evolution of their absorption cross-sections trace each other. Despite subDLAs being about two to four times more common than DLAs (Figure \ref{fig:lX}), subDLAs contribute a modest amount (between $\approx$10--20 per cent) to the total \omegaDLA{} budget over redshifts $2.3\leq$~\zabs{}$<4.5$ (Figure \ref{fig:omega}).

\section*{Acknowledgments}
We thank the anonymous referee for their comments on improving the clarity of this paper, and C\'eline P\'eroux for providing reduced 1D UVES data for comparison. SL was partially funded by UCh/VID project ENL18/18.
\bibliography{bibref}

\appendix

\section{Ly$\alpha$ fits to XQ-100 absorbers}
\label{sec:App}

\onecolumn
\begin{center}
\begin{longtable}{ccccc}
\caption{XQ-100 \HI{} absorbers}
\label{tab:Ncat}\\
\hline
QSO& \zem{}& \zabs{}& logN(\HI{})& Metal absorption confirmed?\\
\hline
\endfirsthead
\multicolumn{5}{l}{{\bfseries \tablename\ \thetable{} -- continued from previous page}}\\
\hline
QSO& \zem{}& \zabs{}& logN(\HI{})& Metal absorption confirmed?\\
\hline
\endhead
\hline \multicolumn{5}{l}{\bfseries Continued on next page}\\
\hline
\endfoot
\endlastfoot
\\
J0003-2603& 4.010& 2.78140& 19.00$\pm$0.20& N\\
J0003-2603& 4.010& 3.05490& 20.00$\pm$0.15& Y\\
J0003-2603& 4.010& 3.39020& 21.45$\pm$0.05& Y\\
J0006-6208& 4.460& 3.16890& 18.90$\pm$0.20& N\\
J0006-6208& 4.460& 3.18430& 18.90$\pm$0.15& N\\
J0006-6208& 4.460& 3.20250& 20.80$\pm$0.20& Y\\
J0006-6208& 4.460& 3.45890& 19.30$\pm$0.20& N\\
J0006-6208& 4.460& 3.77300& 20.90$\pm$0.15& Y\\
J0034+1639& 4.290& 2.99540& 19.00$\pm$0.15& N\\
J0034+1639& 4.290& 3.02170& 19.30$\pm$0.15& N\\
J0034+1639& 4.290& 3.12750& 18.90$\pm$0.20& N\\
J0034+1639& 4.290& 3.22750& 19.80$\pm$0.15& N\\
J0034+1639& 4.290& 3.47970& 18.80$\pm$0.20& Y\\
J0034+1639& 4.290& 3.75350& 20.25$\pm$0.15& Y\\
J0034+1639& 4.290& 3.82470& 19.00$\pm$0.20& N\\
J0034+1639& 4.290& 4.22715& 19.60$\pm$0.20& Y\\
J0034+1639& 4.290& 4.25163& 21.10$\pm$0.20& Y\\
J0034+1639& 4.290& 4.28360& 21.20$\pm$0.10& Y\\
J0042-1020& 3.880& 2.75482& 20.10$\pm$0.15& Y\\
J0048-2442& 4.150& 3.75735& 19.50$\pm$0.25& Y\\
J0056-2808& 3.620& 2.72863& 19.50$\pm$0.15& Y\\
J0100-2708& 3.520& 3.24270& 19.80$\pm$0.10& Y\\
J0113-2803& 4.300& 3.07400& 19.20$\pm$0.20& N\\
J0113-2803& 4.300& 3.10550& 21.10$\pm$0.10& Y\\
J0113-2803& 4.300& 3.39726& 19.00$\pm$0.15& N\\
J0113-2803& 4.300& 3.89000& 19.30$\pm$0.20& Y\\
J0121+0347& 4.130& 2.87250& 19.30$\pm$0.20& N\\
J0121+0347& 4.130& 2.95176& 18.90$\pm$0.15& N\\
J0121+0347& 4.130& 2.97700& 19.40$\pm$0.15& N\\
J0121+0347& 4.130& 3.36021& 19.00$\pm$0.20& Y\\
J0121+0347& 4.130& 3.79347& 18.80$\pm$0.15& N\\
J0124+0044& 3.840& 2.83320& 18.90$\pm$0.20& Y\\
J0124+0044& 3.840& 3.07770& 20.20$\pm$0.15& Y\\
J0132+1341& 4.160& 2.94856& 18.90$\pm$0.20& N\\
J0132+1341& 4.160& 3.05980& 19.20$\pm$0.20& N\\
J0132+1341& 4.160& 3.08204& 19.20$\pm$0.25& N\\
J0132+1341& 4.160& 3.09983& 18.80$\pm$0.20& Y\\
J0132+1341& 4.160& 3.28489& 18.90$\pm$0.20& N\\
J0132+1341& 4.160& 3.93640& 20.20$\pm$0.20& Y\\
J0134+0400& 4.150& 2.94367& 18.80$\pm$0.20& N\\
J0134+0400& 4.150& 3.08042& 18.90$\pm$0.20& N\\
J0134+0400& 4.150& 3.23010& 18.90$\pm$0.20& Y\\
J0134+0400& 4.150& 3.69159& 20.60$\pm$0.15& Y\\
J0134+0400& 4.150& 3.75850& 18.80$\pm$0.20& N\\
J0134+0400& 4.150& 3.77345& 20.55$\pm$0.20& Y\\
J0134+0400& 4.150& 3.99630& 20.10$\pm$0.15& Y\\
J0134+0400& 4.150& 4.02108& 19.10$\pm$0.15& Y\\
J0137-4224& 3.970& 3.10030& 19.70$\pm$0.20& Y\\
J0137-4224& 3.970& 3.39627& 18.90$\pm$0.20& N\\
J0137-4224& 3.970& 3.66516& 19.10$\pm$0.20& Y\\
J0153-0011& 4.190& 2.99368& 19.00$\pm$0.25& N\\
J0153-0011& 4.190& 3.01484& 19.20$\pm$0.20& N\\
J0153-0011& 4.190& 3.03616& 19.30$\pm$0.20& N\\
J0153-0011& 4.190& 3.87608& 19.10$\pm$0.30& Y\\
J0211+1107& 3.980& 3.11780& 18.90$\pm$0.20& N\\
J0211+1107& 3.980& 3.14080& 20.20$\pm$0.20& Y\\
J0211+1107& 3.980& 3.50080& 19.90$\pm$0.20& Y\\
J0214-0517& 3.990& 2.76992& 19.00$\pm$0.20& N\\
J0214-0517& 3.990& 2.95904& 18.90$\pm$0.15& N\\
J0214-0517& 3.990& 3.08439& 18.90$\pm$0.15& N\\
J0214-0517& 3.990& 3.08966& 18.90$\pm$0.15& Y\\
J0214-0517& 3.990& 3.22189& 19.10$\pm$0.20& N\\
J0214-0517& 3.990& 3.72110& 20.60$\pm$0.20& Y\\
J0234-1806& 4.310& 3.00500& 19.70$\pm$0.25& N\\
J0234-1806& 4.310& 3.25732& 18.80$\pm$0.20& N\\
J0234-1806& 4.310& 3.34870& 19.05$\pm$0.20& N\\
J0234-1806& 4.310& 3.36450& 19.00$\pm$0.20& N\\
J0234-1806& 4.310& 3.69403& 20.45$\pm$0.20& Y\\
J0234-1806& 4.310& 4.22700& 19.40$\pm$0.20& Y\\
J0244-0134& 4.050& 2.85146& 18.80$\pm$0.20& N\\
J0244-0134& 4.050& 2.91910& 18.90$\pm$0.20& N\\
J0244-0134& 4.050& 2.93176& 19.15$\pm$0.20& Y\\
J0244-0134& 4.050& 3.92982& 18.90$\pm$0.20& N\\
J0244-0134& 4.050& 3.96680& 19.10$\pm$0.15& Y\\
J0247-0556& 4.240& 3.03668& 19.10$\pm$0.20& N\\
J0247-0556& 4.240& 3.10946& 19.20$\pm$0.20& N\\
J0247-0556& 4.240& 3.11344& 19.00$\pm$0.20& N\\
J0247-0556& 4.240& 3.33643& 19.00$\pm$0.20& N\\
J0247-0556& 4.240& 3.99141& 18.85$\pm$0.20& N\\
J0247-0556& 4.240& 4.13958& 19.50$\pm$0.20& Y\\
J0255+0048& 4.010& 2.79119& 18.90$\pm$0.20& N\\
J0255+0048& 4.010& 2.87008& 19.00$\pm$0.20& N\\
J0255+0048& 4.010& 3.25453& 20.70$\pm$0.15& Y\\
J0255+0048& 4.010& 3.45038& 19.60$\pm$0.15& Y\\
J0255+0048& 4.010& 3.91400& 21.50$\pm$0.20& Y\\
J0307-4945& 4.720& 3.35450& 20.10$\pm$0.20& Y\\
J0307-4945& 4.720& 3.59170& 20.40$\pm$0.25& Y\\
J0307-4945& 4.720& 3.60826& 19.70$\pm$0.25& N\\
J0307-4945& 4.720& 4.46840& 20.60$\pm$0.10& Y\\
J0311-1722& 4.040& 3.73420& 20.20$\pm$0.20& Y\\
J0403-1703& 4.230& 3.20617& 19.10$\pm$0.25& Y\\
J0403-1703& 4.230& 3.54373& 18.80$\pm$0.20& N\\
J0415-4357& 4.070& 2.81455& 18.80$\pm$0.20& N\\
J0415-4357& 4.070& 3.02721& 19.30$\pm$0.20& N\\
J0415-4357& 4.070& 3.80850& 20.40$\pm$0.20& Y\\
J0415-4357& 4.070& 4.03473& 20.05$\pm$0.30& Y\\
J0424-2209& 4.320& 2.99250& 21.20$\pm$0.20& Y\\
J0424-2209& 4.320& 3.04835& 18.80$\pm$0.20& N\\
J0424-2209& 4.320& 3.14142& 18.80$\pm$0.20& N\\
J0424-2209& 4.320& 3.24419& 18.80$\pm$0.20& N\\
J0424-2209& 4.320& 3.36663& 18.80$\pm$0.20& N\\
J0525-3343& 4.410& 3.20174& 18.80$\pm$0.15& N\\
J0529-3526& 4.410& 3.06433& 19.40$\pm$0.20& N\\
J0529-3526& 4.410& 3.19297& 19.30$\pm$0.20& N\\
J0529-3526& 4.410& 3.31146& 18.80$\pm$0.20& N\\
J0529-3526& 4.410& 3.57170& 20.10$\pm$0.15& Y\\
J0529-3526& 4.410& 3.76200& 19.35$\pm$0.20& N\\
J0529-3552& 4.170& 2.97530& 18.90$\pm$0.20& N\\
J0529-3552& 4.170& 3.01493& 18.90$\pm$0.20& N\\
J0529-3552& 4.170& 3.68360& 20.10$\pm$0.20& N\\
J0529-3552& 4.170& 3.70972& 19.50$\pm$0.20& Y\\
J0529-3552& 4.170& 4.06530& 19.50$\pm$0.15& Y\\
J0714-6455& 4.460& 3.10313& 19.45$\pm$0.20& N\\
J0714-6455& 4.460& 3.21970& 19.10$\pm$0.20& N\\
J0714-6455& 4.460& 3.26472& 18.80$\pm$0.15& N\\
J0714-6455& 4.460& 3.31292& 18.80$\pm$0.20& N\\
J0714-6455& 4.460& 4.38971& 18.80$\pm$0.20& Y\\
J0747+2739& 4.170& 2.91793& 19.40$\pm$0.20& Y\\
J0747+2739& 4.170& 2.99956& 19.00$\pm$0.20& N\\
J0747+2739& 4.170& 3.42400& 20.80$\pm$0.15& Y\\
J0747+2739& 4.170& 3.90134& 20.50$\pm$0.20& Y\\
J0755+1345& 3.670& 3.09806& 19.00$\pm$0.20& Y\\
J0800+1920& 3.960& 3.42900& 19.90$\pm$0.20& Y\\
J0800+1920& 3.960& 3.73076& 19.00$\pm$0.20& Y\\
J0818+0958& 3.670& 2.56424& 18.80$\pm$0.20& N\\
J0818+0958& 3.670& 3.02811& 18.85$\pm$0.20& Y\\
J0818+0958& 3.670& 3.30631& 20.90$\pm$0.20& Y\\
J0818+0958& 3.670& 3.45449& 19.00$\pm$0.20& Y\\
J0833+0959& 3.750& 2.80570& 18.90$\pm$0.20& N\\
J0835+0650& 3.990& 2.77229& 19.20$\pm$0.20& N\\
J0835+0650& 3.990& 2.77647& 18.90$\pm$0.20& N\\
J0835+0650& 3.990& 2.79340& 18.90$\pm$0.20& N\\
J0835+0650& 3.990& 2.87150& 19.10$\pm$0.20& N\\
J0835+0650& 3.990& 2.96430& 19.10$\pm$0.20& N\\
J0835+0650& 3.990& 2.97858& 18.90$\pm$0.20& N\\
J0835+0650& 3.990& 3.19004& 19.00$\pm$0.20& Y\\
J0835+0650& 3.990& 3.51317& 19.00$\pm$0.20& Y\\
J0835+0650& 3.990& 3.60160& 19.80$\pm$0.20& N\\
J0835+0650& 3.990& 3.95570& 20.35$\pm$0.20& Y\\
J0839+0318& 4.250& 3.43922& 18.80$\pm$0.20& N\\
J0839+0318& 4.250& 4.09690& 19.60$\pm$0.20& Y\\
J0920+0725& 3.640& 3.05873& 19.10$\pm$0.20& Y\\
J0937+0828& 3.700& 2.71534& 18.90$\pm$0.20& Y\\
J0937+0828& 3.700& 3.12919& 19.70$\pm$0.30& N\\
J0955-0130& 4.430& 3.16128& 18.80$\pm$0.20& N\\
J0955-0130& 4.430& 3.23765& 19.20$\pm$0.20& N\\
J0955-0130& 4.430& 3.47570& 20.00$\pm$0.20& Y\\
J0955-0130& 4.430& 3.95186& 19.20$\pm$0.20& N\\
J0955-0130& 4.430& 3.98709& 19.00$\pm$0.20& N\\
J0955-0130& 4.430& 4.02114& 20.60$\pm$0.20& Y\\
J0955-0130& 4.430& 4.17243& 18.80$\pm$0.20& Y\\
J0959+1312& 4.060& 2.92544& 18.90$\pm$0.20& N\\
J0959+1312& 4.060& 3.14507& 19.10$\pm$0.20& N\\
J0959+1312& 4.060& 3.91230& 20.00$\pm$0.10& Y\\
J1013+0650& 3.790& 2.78778& 18.80$\pm$0.20& N\\
J1013+0650& 3.790& 3.01137& 19.20$\pm$0.20& N\\
J1013+0650& 3.790& 3.48930& 19.20$\pm$0.20& Y\\
J1018+0548& 3.520& 2.50832& 19.00$\pm$0.20& N\\
J1018+0548& 3.520& 2.83192& 18.80$\pm$0.20& Y\\
J1018+0548& 3.520& 3.38470& 19.65$\pm$0.20& Y\\
J1020+0922& 3.640& 2.59127& 21.40$\pm$0.20& Y\\
J1020+0922& 3.640& 2.74873& 20.10$\pm$0.20& Y\\
J1024+1819& 3.530& 2.53488& 19.00$\pm$0.20& N\\
J1024+1819& 3.530& 2.69809& 19.00$\pm$0.20& N\\
J1024+1819& 3.530& 3.18451& 18.90$\pm$0.20& N\\
J1032+0927& 3.990& 3.80480& 20.00$\pm$0.15& Y\\
J1036-0343& 4.510& 3.46553& 19.00$\pm$0.20& N\\
J1036-0343& 4.510& 3.67068& 19.10$\pm$0.20& N\\
J1036-0343& 4.510& 4.15340& 19.20$\pm$0.20& Y\\
J1036-0343& 4.510& 4.17470& 19.80$\pm$0.20& N\\
J1037+0704& 4.100& 3.28155& 19.00$\pm$0.20& Y\\
J1057+1910& 4.100& 3.22329& 19.10$\pm$0.20& N\\
J1057+1910& 4.100& 3.37340& 20.20$\pm$0.15& Y\\
J1057+1910& 4.100& 3.79985& 19.10$\pm$0.20& Y\\
J1058+1245& 4.330& 3.20690& 18.80$\pm$0.20& Y\\
J1058+1245& 4.330& 3.22850& 19.20$\pm$0.20& N\\
J1058+1245& 4.330& 3.43160& 20.50$\pm$0.20& Y\\
J1058+1245& 4.330& 3.59987& 18.90$\pm$0.20& N\\
J1103+1004& 3.610& 2.48702& 19.30$\pm$0.20& N\\
J1103+1004& 3.610& 3.13101& 19.10$\pm$0.20& Y\\
J1108+1209& 3.670& 2.53374& 19.20$\pm$0.20& N\\
J1108+1209& 3.670& 2.65507& 19.30$\pm$0.20& N\\
J1108+1209& 3.670& 2.66164& 18.90$\pm$0.20& N\\
J1108+1209& 3.670& 2.73522& 18.80$\pm$0.20& Y\\
J1108+1209& 3.670& 3.39594& 20.70$\pm$0.15& Y\\
J1108+1209& 3.670& 3.54530& 20.90$\pm$0.15& Y\\
J1111-0804& 3.920& 2.70126& 19.60$\pm$0.20& N\\
J1111-0804& 3.920& 2.98005& 19.20$\pm$0.20& N\\
J1111-0804& 3.920& 3.48180& 19.95$\pm$0.15& Y\\
J1111-0804& 3.920& 3.60690& 20.45$\pm$0.20& Y\\
J1111-0804& 3.920& 3.75820& 18.80$\pm$0.20& Y\\
J1126-0124& 3.740& 2.61480& 19.00$\pm$0.20& N\\
J1202-0054& 3.590& 2.66046& 19.60$\pm$0.15& Y\\
J1202-0054& 3.590& 2.76953& 19.00$\pm$0.15& Y\\
J1202-0054& 3.590& 3.15122& 19.00$\pm$0.20& Y\\
J1248+1304& 3.720& 2.62589& 19.00$\pm$0.20& N\\
J1248+1304& 3.720& 3.40689& 19.20$\pm$0.15& Y\\
J1249-0159& 3.630& 2.48080& 19.00$\pm$0.15& N\\
J1249-0159& 3.630& 2.49471& 18.80$\pm$0.20& N\\
J1249-0159& 3.630& 3.52476& 18.80$\pm$0.20& N\\
J1249-0159& 3.630& 3.52963& 18.90$\pm$0.15& N\\
J1304+0239& 3.650& 2.75477& 18.90$\pm$0.15& N\\
J1304+0239& 3.650& 3.33631& 18.80$\pm$0.20& Y\\
J1312+0841& 3.740& 2.65950& 20.50$\pm$0.20& Y\\
J1320-0523& 3.700& 2.83411& 19.10$\pm$0.20& Y\\
J1323+1405& 4.040& 2.92550& 19.80$\pm$0.20& N\\
J1330-2522& 3.950& 2.77513& 18.90$\pm$0.20& N\\
J1330-2522& 3.950& 3.08060& 19.80$\pm$0.20& Y\\
J1332+0052& 3.510& 3.08365& 19.40$\pm$0.20& Y\\
J1332+0052& 3.510& 3.42120& 19.20$\pm$0.20& Y\\
J1336+0243& 3.800& 2.69140& 19.40$\pm$0.20& Y\\
J1352+1303& 3.700& 3.16676& 18.90$\pm$0.20& Y\\
J1401+0244& 4.440& 3.34057& 18.90$\pm$0.20& N\\
J1416+1811& 3.590& 2.59808& 18.90$\pm$0.20& N\\
J1416+1811& 3.590& 2.66330& 19.65$\pm$0.20& Y\\
J1421-0643& 3.690& 2.55859& 18.80$\pm$0.20& N\\
J1421-0643& 3.690& 3.16020& 19.30$\pm$0.20& N\\
J1421-0643& 3.690& 3.44802& 20.35$\pm$0.15& Y\\
J1503+0419& 3.660& 2.49915& 19.10$\pm$0.20& N\\
J1517+0511& 3.560& 2.42657& 18.90$\pm$0.20& N\\
J1517+0511& 3.560& 2.43960& 19.10$\pm$0.20& N\\
J1517+0511& 3.560& 2.45773& 18.90$\pm$0.20& N\\
J1517+0511& 3.560& 2.61383& 18.80$\pm$0.20& N\\
J1517+0511& 3.560& 2.68838& 21.35$\pm$0.15& Y\\
J1524+2123& 3.610& 2.54540& 19.30$\pm$0.20& Y\\
J1524+2123& 3.610& 2.63417& 19.60$\pm$0.15& Y\\
J1524+2123& 3.610& 2.73116& 19.40$\pm$0.15& Y\\
J1542+0955& 3.990& 2.76560& 19.00$\pm$0.20& N\\
J1542+0955& 3.990& 3.61287& 19.00$\pm$0.25& Y\\
J1552+1005& 3.730& 2.97674& 18.80$\pm$0.20& N\\
J1552+1005& 3.730& 3.07712& 18.80$\pm$0.20& Y\\
J1552+1005& 3.730& 3.44240& 19.00$\pm$0.20& Y\\
J1552+1005& 3.730& 3.60030& 21.10$\pm$0.20& Y\\
J1552+1005& 3.730& 3.66550& 21.00$\pm$0.20& Y\\
J1621-0042& 3.700& 3.10423& 19.80$\pm$0.15& Y\\
J1633+1411& 4.330& 4.15083& 18.90$\pm$0.20& Y\\
J1658-0739& 3.740& 2.70650& 19.30$\pm$0.20& Y\\
J1658-0739& 3.740& 2.95470& 19.10$\pm$0.20& N\\
J1658-0739& 3.740& 3.68826& 18.90$\pm$0.15& Y\\
J1658-0739& 3.740& 3.69580& 19.10$\pm$0.15& Y\\
J1723+2243& 4.520& 3.19667& 19.20$\pm$0.20& N\\
J1723+2243& 4.520& 3.42745& 18.90$\pm$0.20& N\\
J1723+2243& 4.520& 3.69800& 20.50$\pm$0.15& Y\\
J1723+2243& 4.520& 3.97228& 18.80$\pm$0.15& N\\
J1723+2243& 4.520& 3.98250& 19.00$\pm$0.15& N\\
J1723+2243& 4.520& 4.08350& 18.80$\pm$0.15& N\\
J1723+2243& 4.520& 4.15470& 19.00$\pm$0.20& N\\
J1723+2243& 4.520& 4.23341& 19.00$\pm$0.20& N\\
J1723+2243& 4.520& 4.24700& 18.80$\pm$0.15& Y\\
J2215-1611& 3.990& 2.77623& 19.00$\pm$0.20& N\\
J2215-1611& 3.990& 2.85521& 18.80$\pm$0.20& N\\
J2215-1611& 3.990& 2.96694& 19.00$\pm$0.20& N\\
J2215-1611& 3.990& 3.00628& 19.00$\pm$0.20& Y\\
J2215-1611& 3.990& 3.22585& 19.30$\pm$0.20& N\\
J2215-1611& 3.990& 3.30557& 19.00$\pm$0.20& Y\\
J2215-1611& 3.990& 3.46525& 19.00$\pm$0.15& Y\\
J2215-1611& 3.990& 3.66150& 20.20$\pm$0.15& Y\\
J2215-1611& 3.990& 3.70080& 19.30$\pm$0.15& Y\\
J2216-6714& 4.470& 3.11219& 18.90$\pm$0.20& N\\
J2216-6714& 4.470& 3.15509& 19.10$\pm$0.20& N\\
J2216-6714& 4.470& 3.21393& 18.90$\pm$0.15& N\\
J2216-6714& 4.470& 3.36940& 19.80$\pm$0.15& Y\\
J2216-6714& 4.470& 3.68304& 18.80$\pm$0.20& N\\
J2216-6714& 4.470& 3.73025& 18.80$\pm$0.20& Y\\
J2239-0552& 4.560& 3.18862& 18.90$\pm$0.15& N\\
J2239-0552& 4.560& 4.07929& 20.60$\pm$0.10& Y\\
J2239-0552& 4.560& 4.19986& 18.90$\pm$0.15& Y\\
J2251-1227& 4.160& 2.98253& 19.50$\pm$0.30& N\\
J2251-1227& 4.160& 3.45717& 19.70$\pm$0.20& Y\\
J2251-1227& 4.160& 3.98793& 19.60$\pm$0.15& Y\\
J2344+0342& 4.240& 3.21980& 21.20$\pm$0.20& Y\\
J2344+0342& 4.240& 3.88400& 19.80$\pm$0.15& Y\\
J2349-3712& 4.210& 3.58203& 19.20$\pm$0.20& Y\\
J2349-3712& 4.210& 3.69250& 20.20$\pm$0.10& Y\\
J2349-3712& 4.210& 3.80387& 18.90$\pm$0.15& Y\\
J2349-3712& 4.210& 3.96340& 18.80$\pm$0.15& Y\\
\end{longtable}
\end{center}

\begin{figure*}
\begin{center}
\begin{subfigure}{\textwidth}
\includegraphics[width=0.95\textwidth]{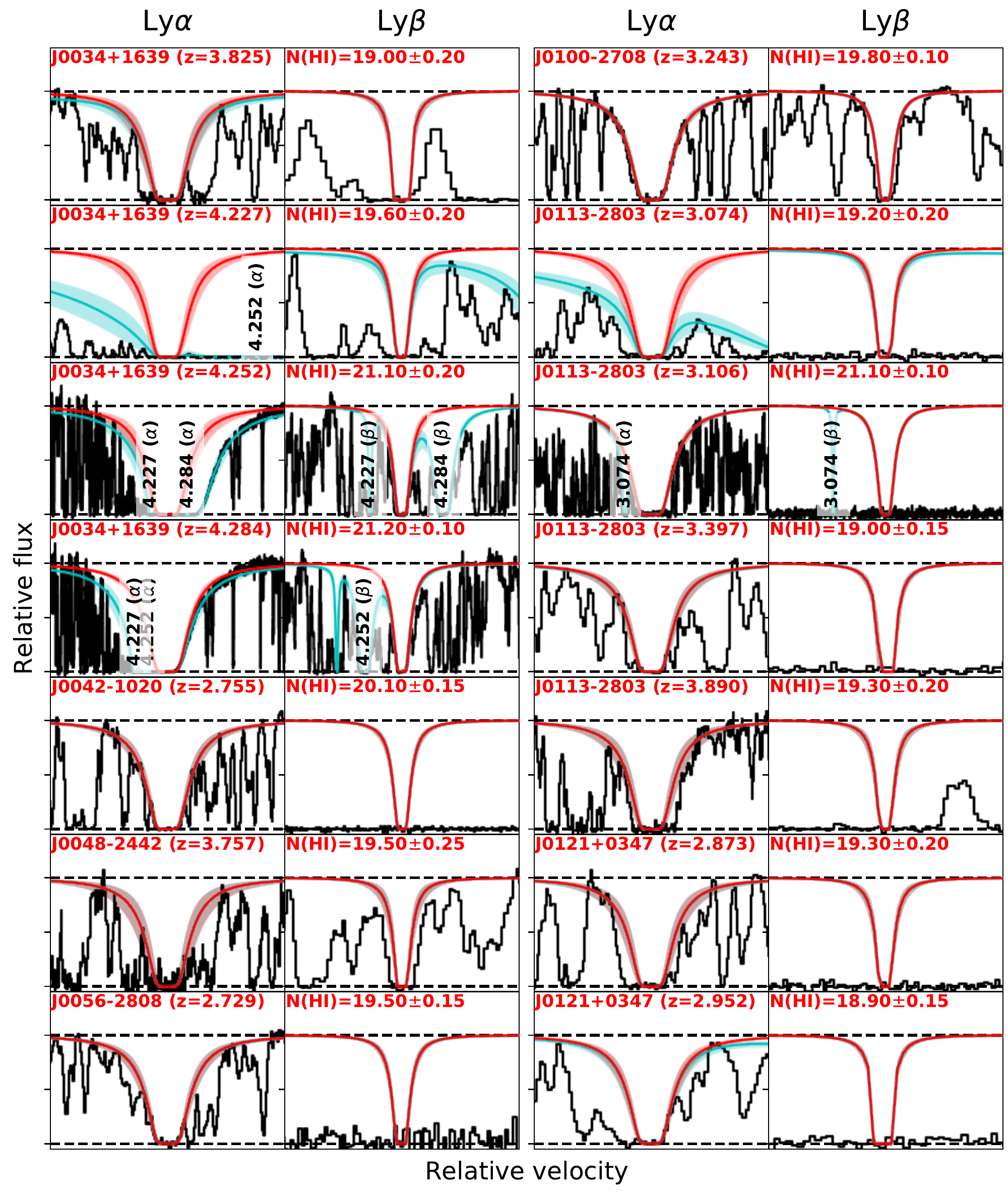}
\end{subfigure}
\caption[]{Each pair of panels displays the fitted Ly$\alpha$(first and third columns) and Ly$\beta$ (second and fourth columns) profiles (dark red lines) and associated errors (dark red shaded region) for 14 of the identified candidate system. The light blue line and shaded region denote the full model spectrum and associated error combined from all absorbers within the entire sightline. The red text denotes the properties of the absorber [QSO sightline name and redshift in the corresponding Ly$\alpha$ panel, and logN(\HI{}) in Ly$\beta$ panel], while the vertical black text labels the redshift and Lyman series lines associated with additional absorption included in the model. The x-axis of each panel is scaled such that integrated region of the Ly series line's Voigt profile displayed encompasses 98 per cent of the absorbed flux.}
\label{fig:LyABapp}
\end{center}
\end{figure*}

\clearpage

\begin{figure*}
\ContinuedFloat
\begin{center}
\begin{subfigure}{\textwidth}
\includegraphics[width=0.95\textwidth]{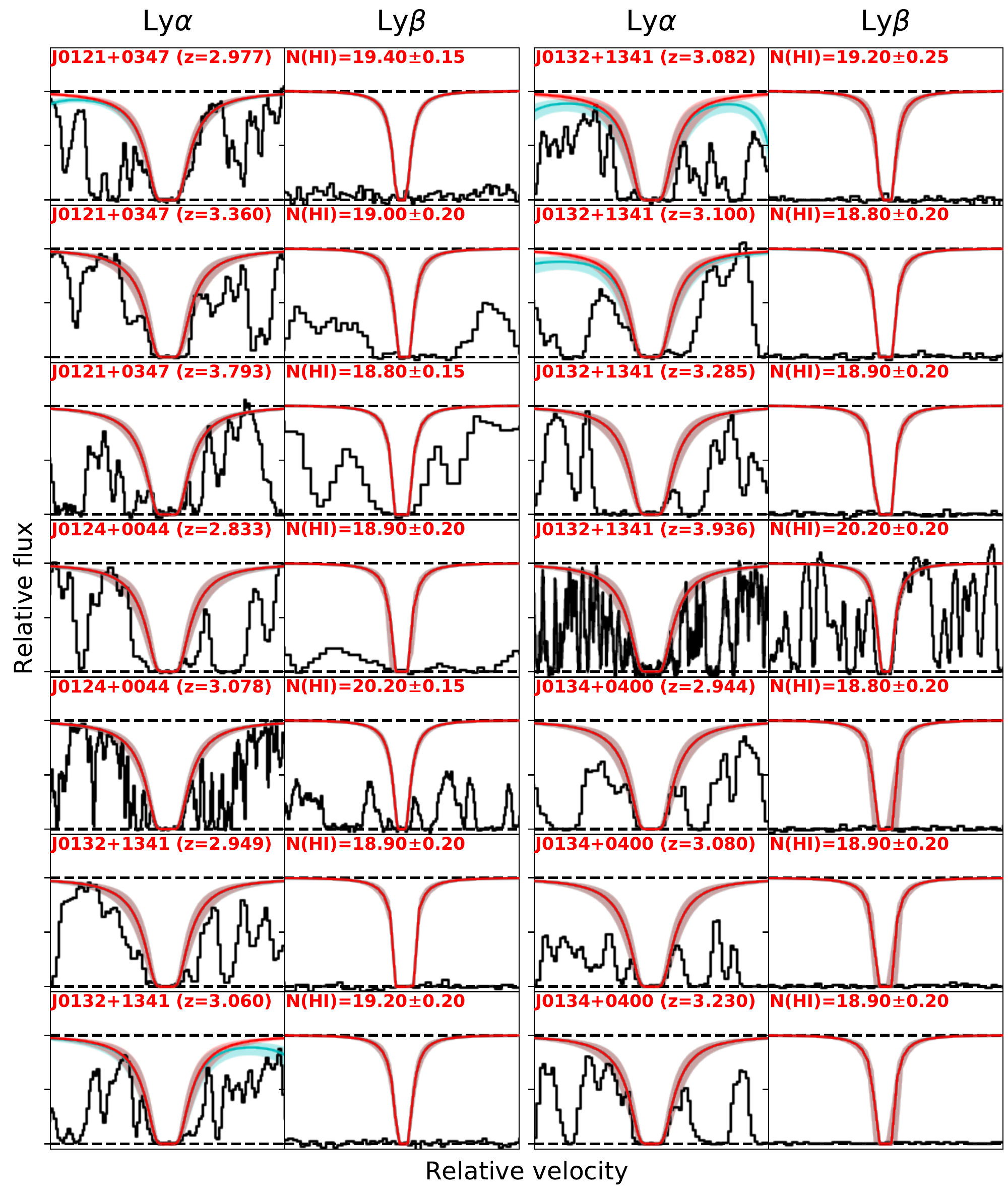}
\end{subfigure}
\caption{(cont'd)}
\end{center}
\end{figure*}

\clearpage

\begin{figure*}
\ContinuedFloat
\begin{center}

\begin{subfigure}{\textwidth}
\includegraphics[width=0.95\textwidth]{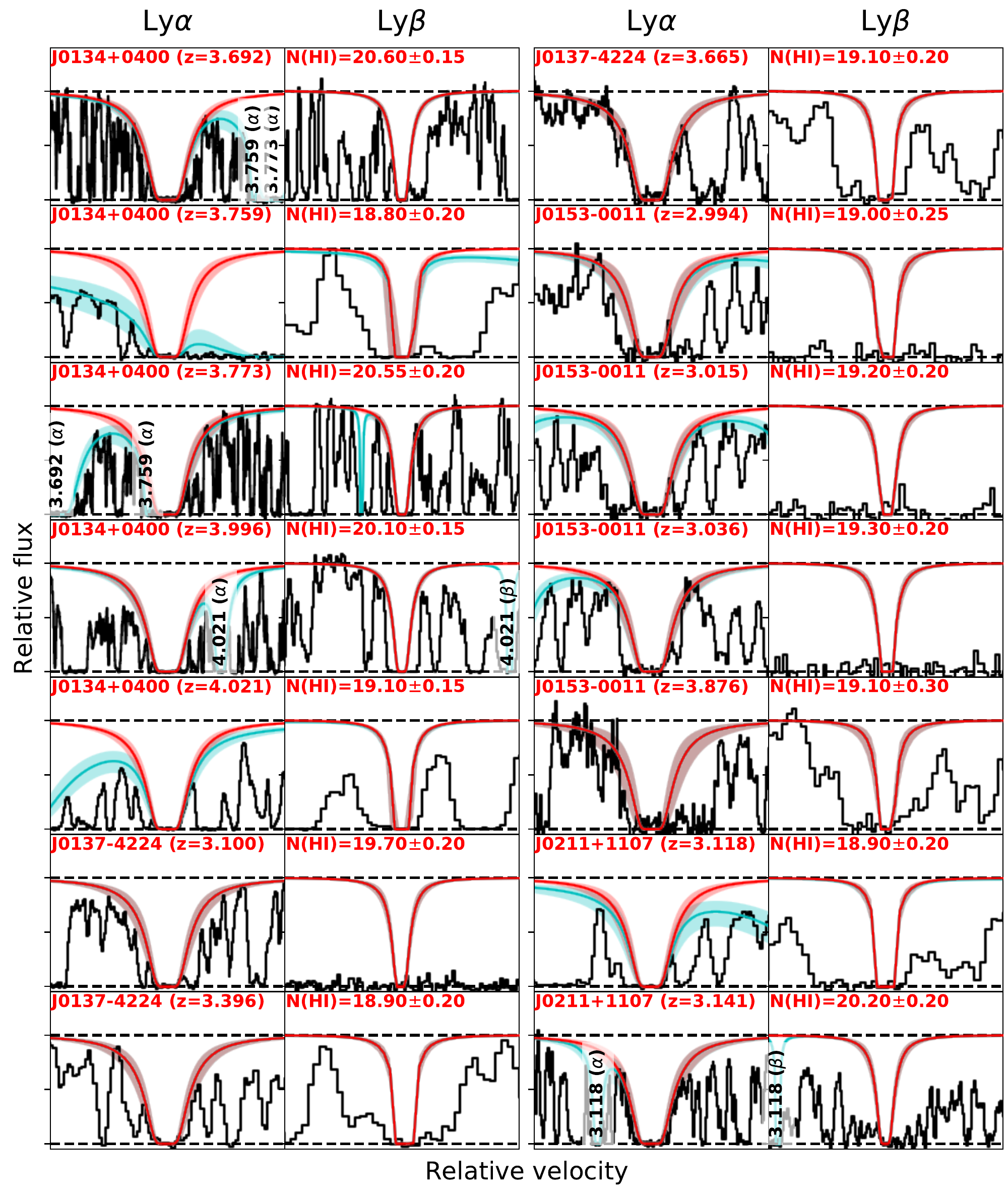}
\end{subfigure}
\caption{(cont'd)}
\end{center}
\end{figure*}

\clearpage

\begin{figure*}
\ContinuedFloat
\begin{center}
\begin{subfigure}{\textwidth}
\includegraphics[width=0.95\textwidth]{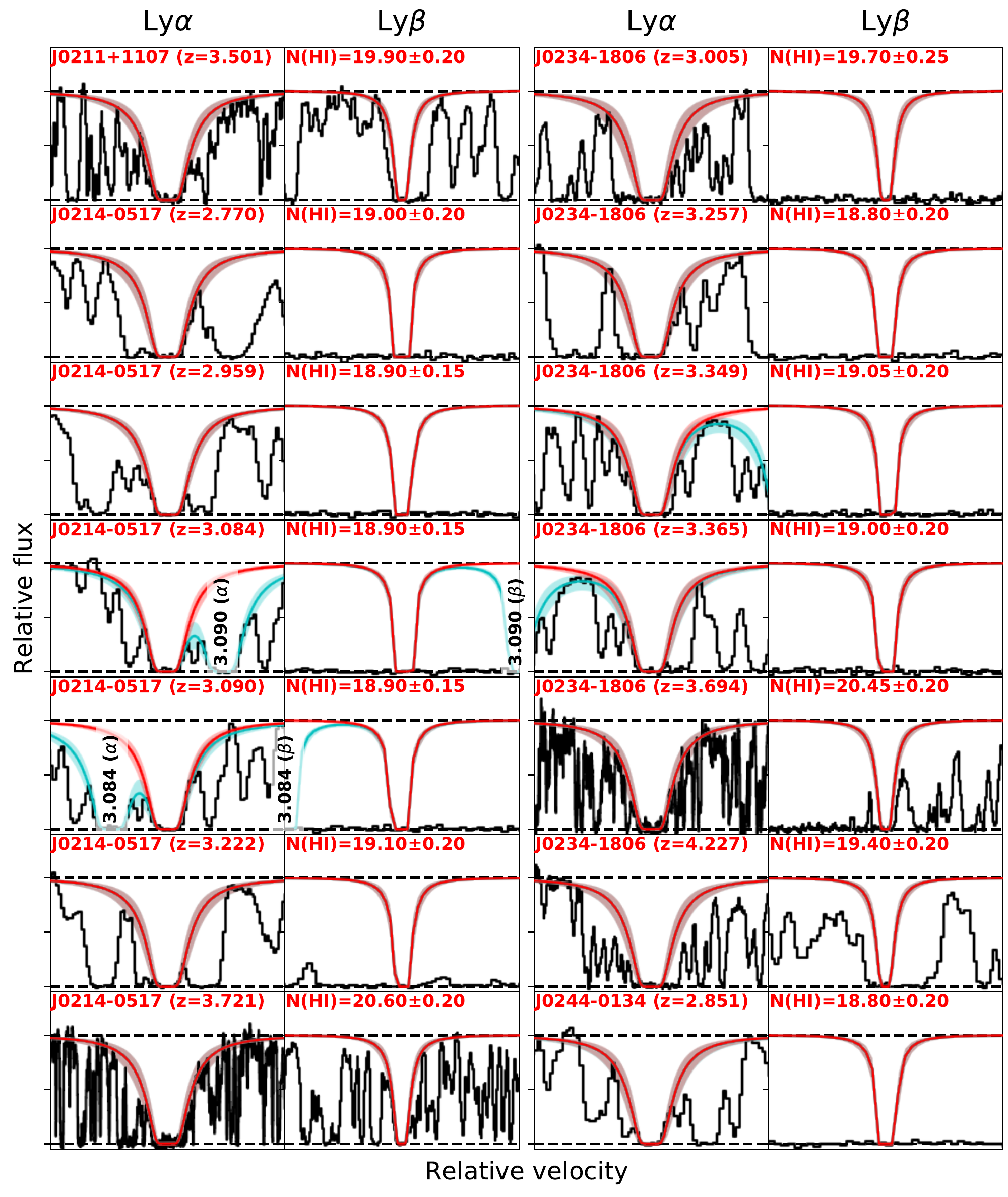}
\end{subfigure}
\caption{(cont'd)}
\end{center}
\end{figure*}

\clearpage

\begin{figure*}
\ContinuedFloat
\begin{center}
\begin{subfigure}{\textwidth}
\includegraphics[width=0.95\textwidth]{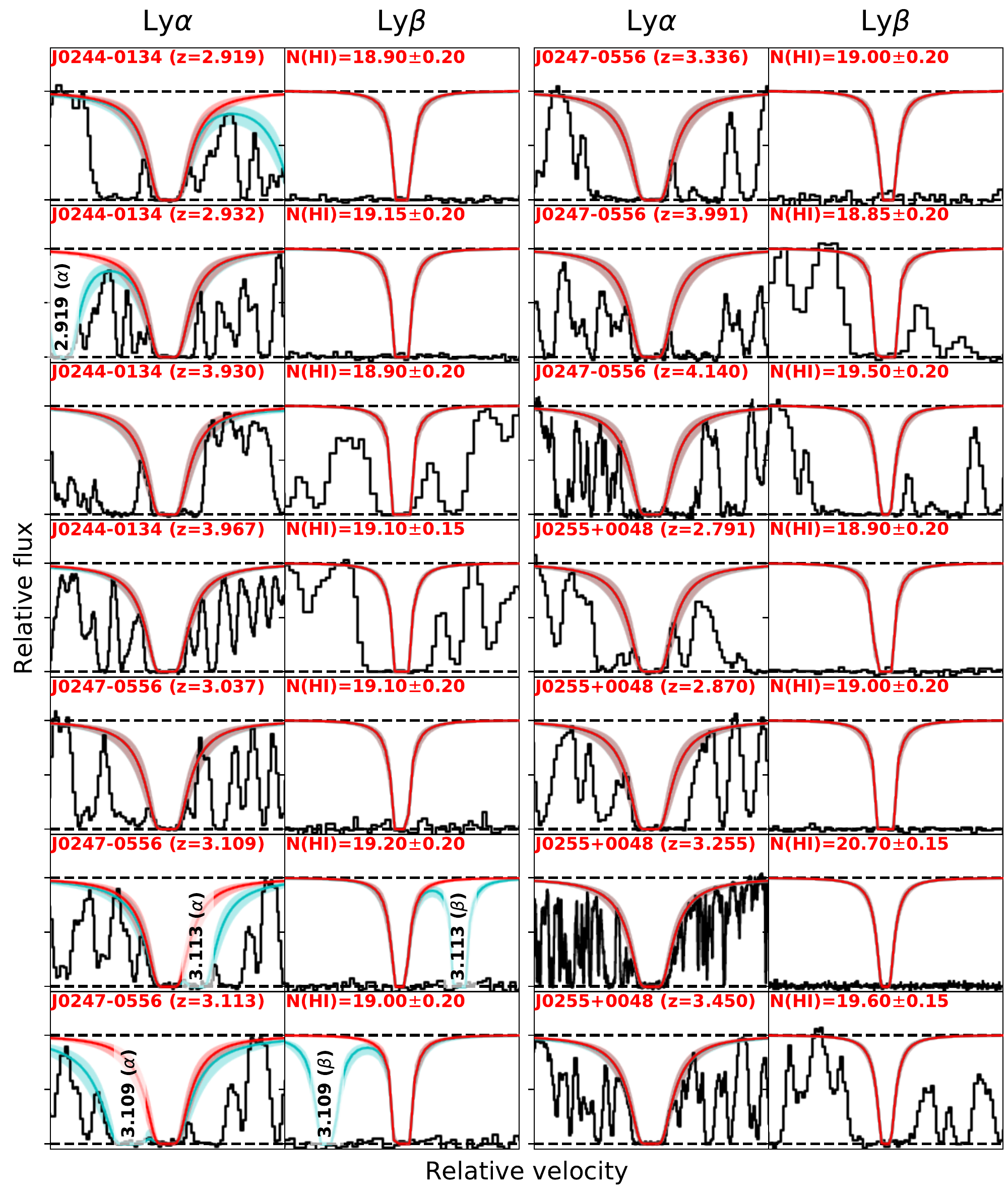}
\end{subfigure}
\caption{(cont'd)}
\end{center}
\end{figure*}

\clearpage

\begin{figure*}
\ContinuedFloat
\begin{center}
\begin{subfigure}{\textwidth}
\includegraphics[width=0.95\textwidth]{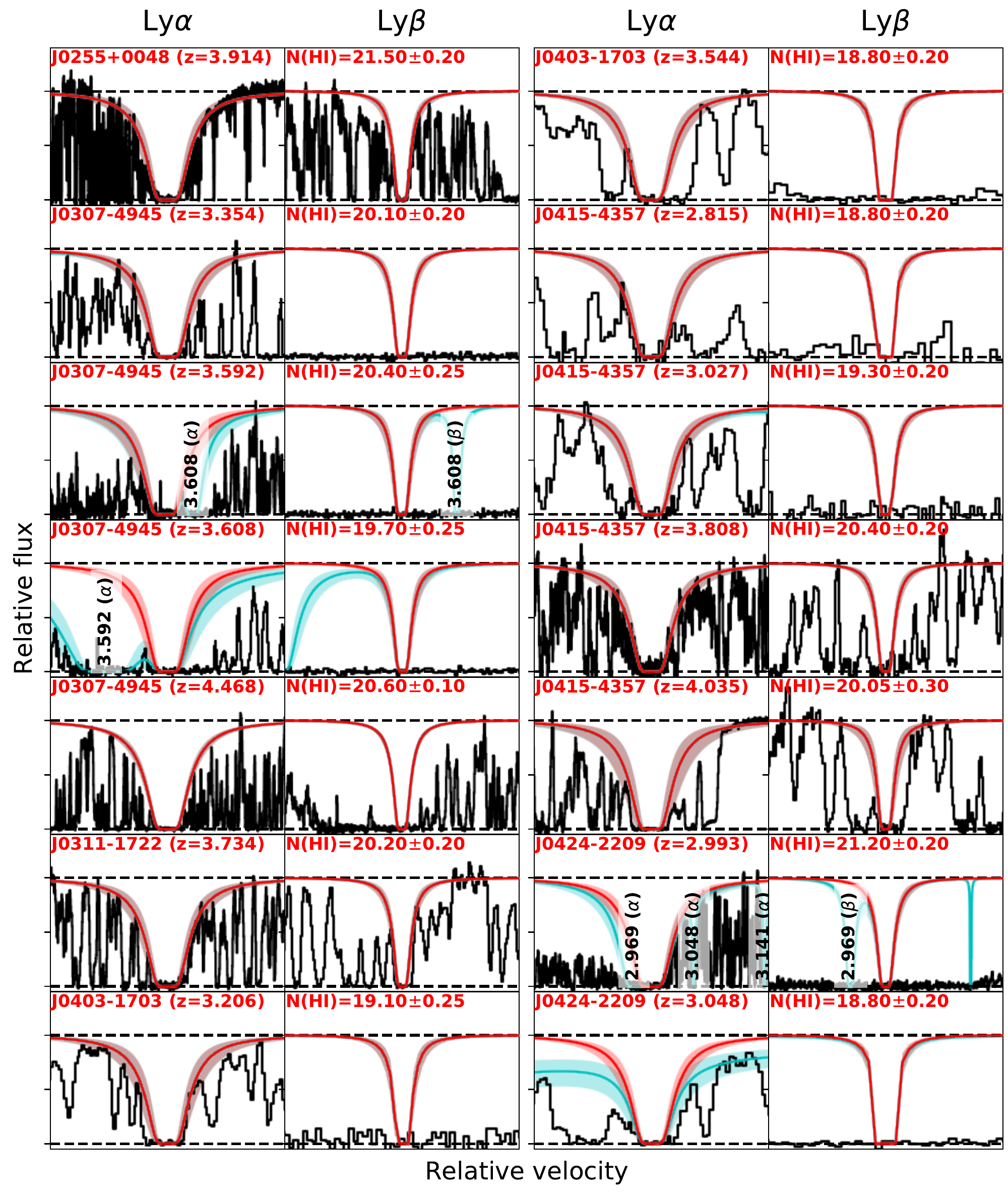}
\end{subfigure}
\caption{(cont'd)}
\end{center}
\end{figure*}

\clearpage

\begin{figure*}
\ContinuedFloat
\begin{center}
\begin{subfigure}{\textwidth}
\includegraphics[width=0.95\textwidth]{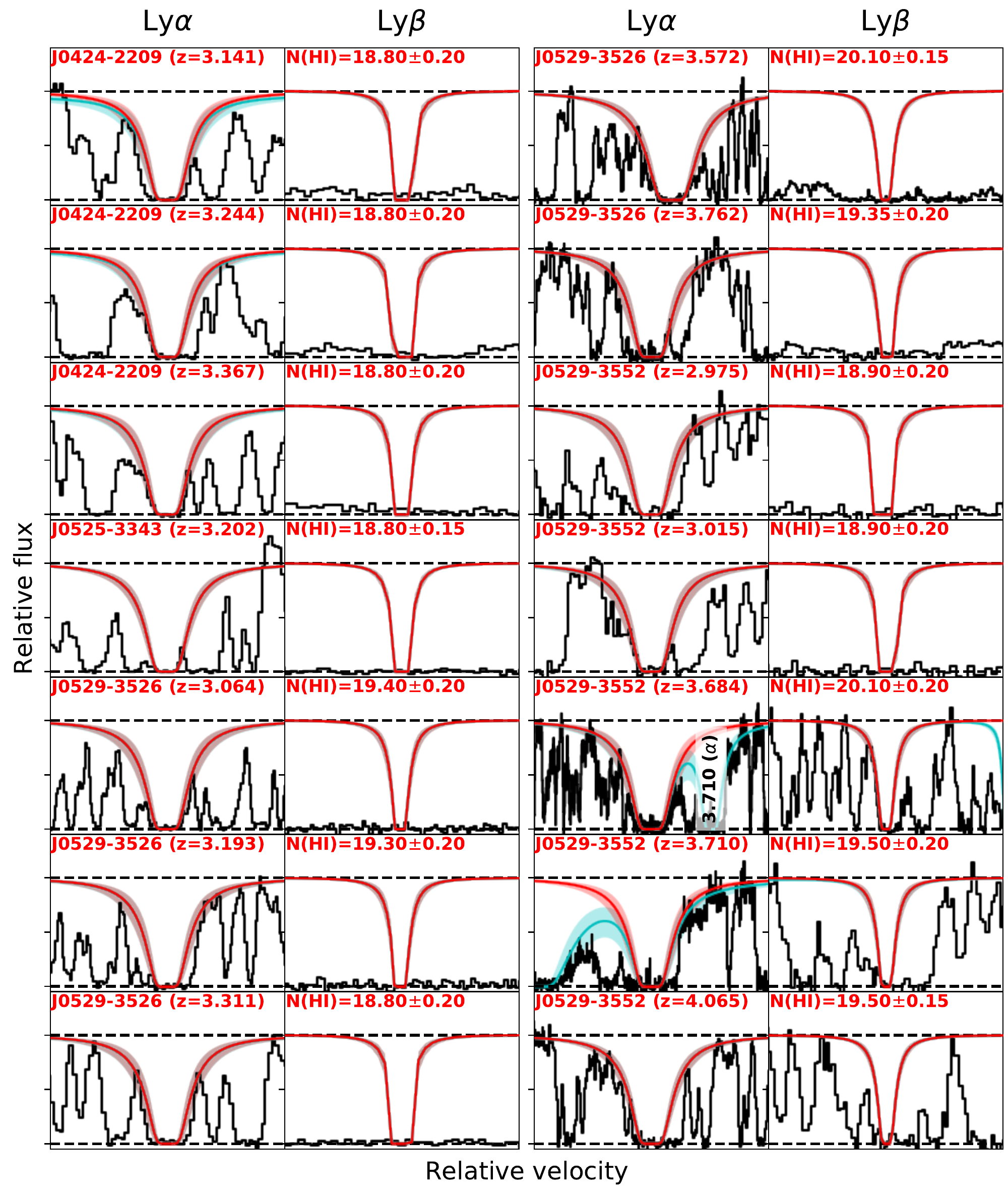}
\end{subfigure}
\caption{(cont'd)}
\end{center}
\end{figure*}

\clearpage

\begin{figure*}
\ContinuedFloat
\begin{center}
\begin{subfigure}{\textwidth}
\includegraphics[width=0.95\textwidth]{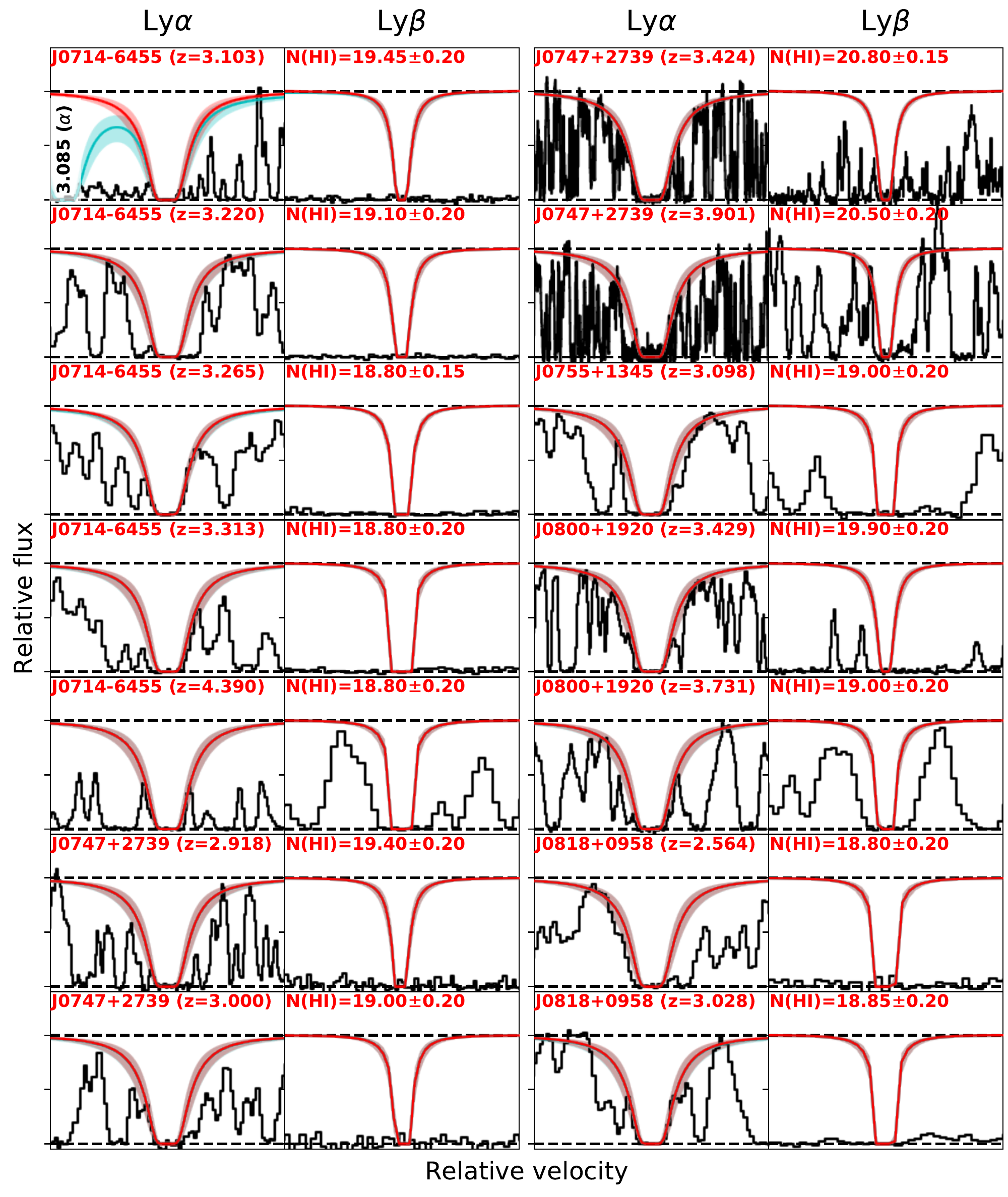}
\end{subfigure}
\caption{(cont'd)}
\end{center}
\end{figure*}

\clearpage

\begin{figure*}
\ContinuedFloat
\begin{center}
\begin{subfigure}{\textwidth}
\includegraphics[width=0.95\textwidth]{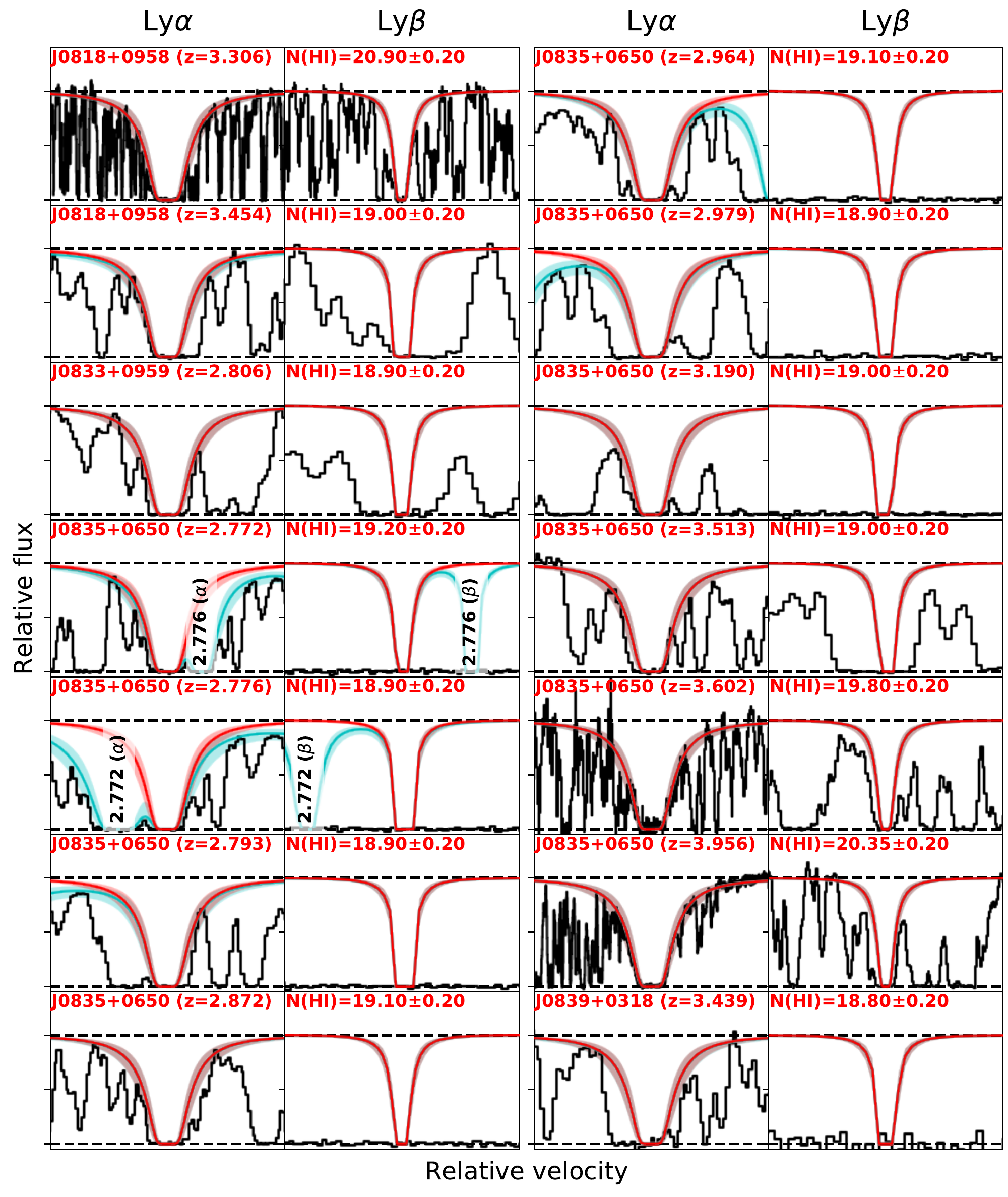}
\end{subfigure}
\caption{(cont'd)}
\end{center}
\end{figure*}

\clearpage

\begin{figure*}
\ContinuedFloat
\begin{center}
\begin{subfigure}{\textwidth}
\includegraphics[width=0.95\textwidth]{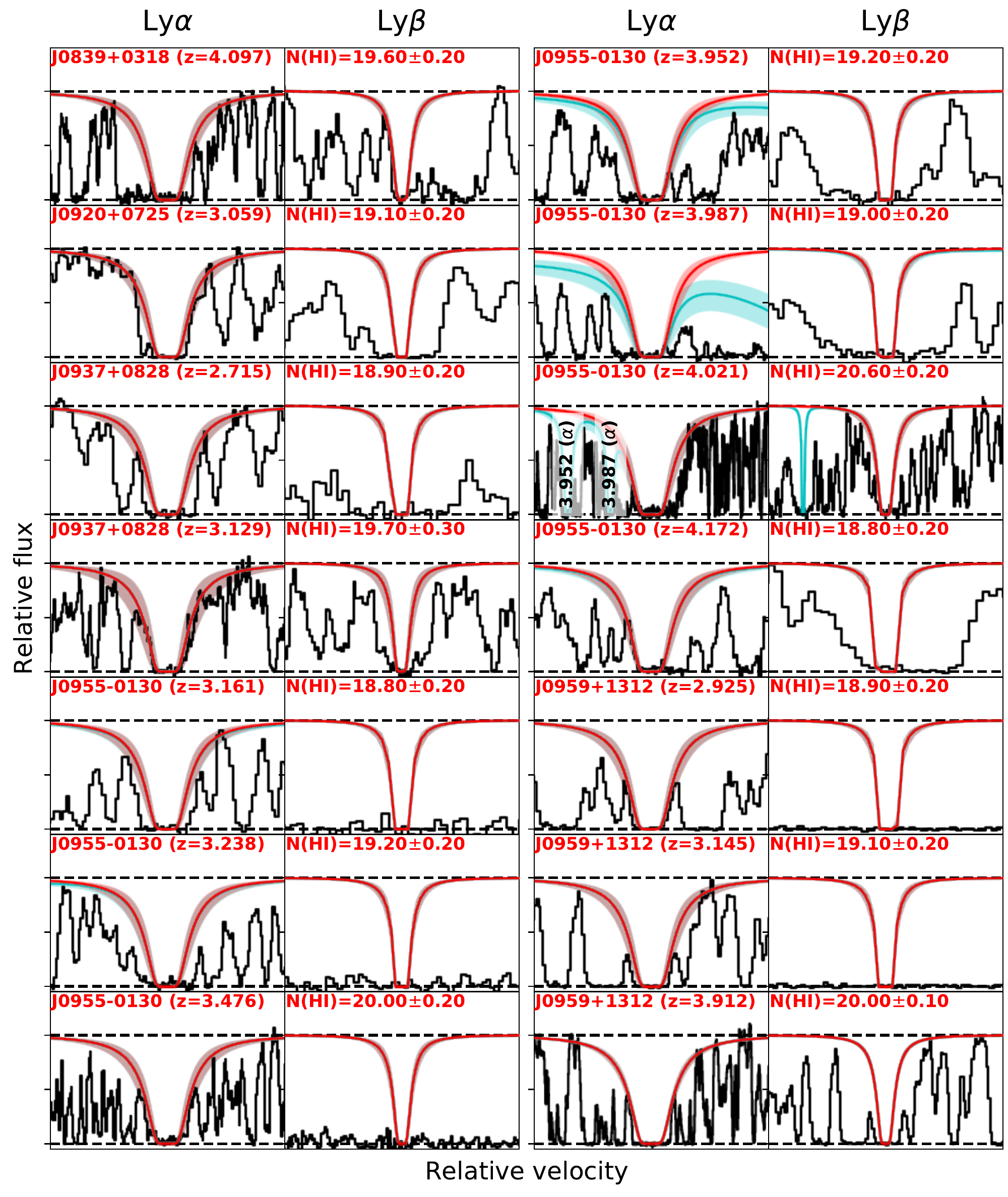}
\end{subfigure}
\caption{(cont'd)}
\end{center}
\end{figure*}

\clearpage

\begin{figure*}
\ContinuedFloat
\begin{center}
\begin{subfigure}{\textwidth}
\includegraphics[width=0.95\textwidth]{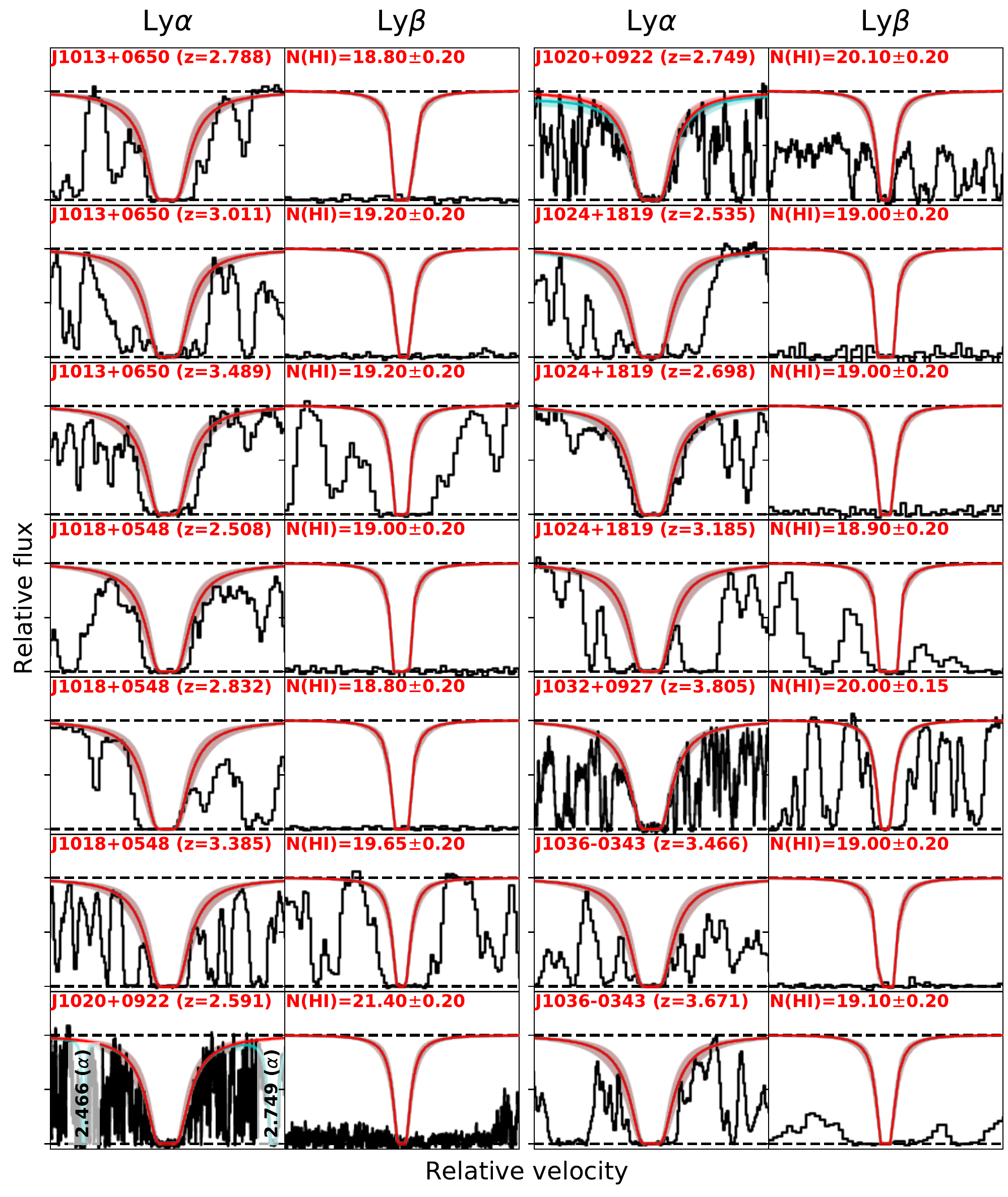}
\end{subfigure}
\caption{(cont'd)}
\end{center}
\end{figure*}

\clearpage

\begin{figure*}
\ContinuedFloat
\begin{center}
\begin{subfigure}{\textwidth}
\includegraphics[width=0.95\textwidth]{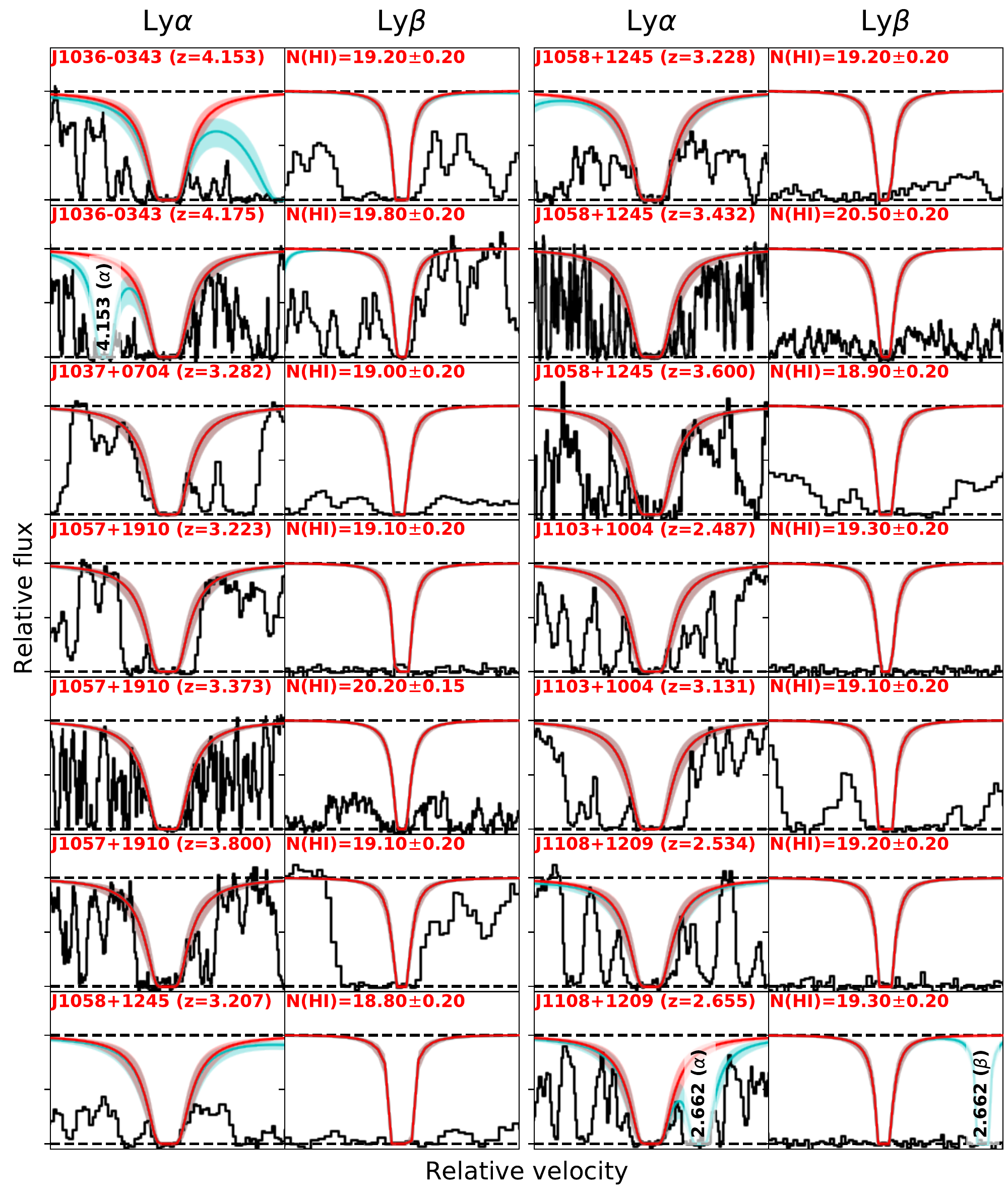}
\end{subfigure}
\caption{(cont'd)}
\end{center}
\end{figure*}

\clearpage

\begin{figure*}
\ContinuedFloat
\begin{center}
\begin{subfigure}{\textwidth}
\includegraphics[width=0.95\textwidth]{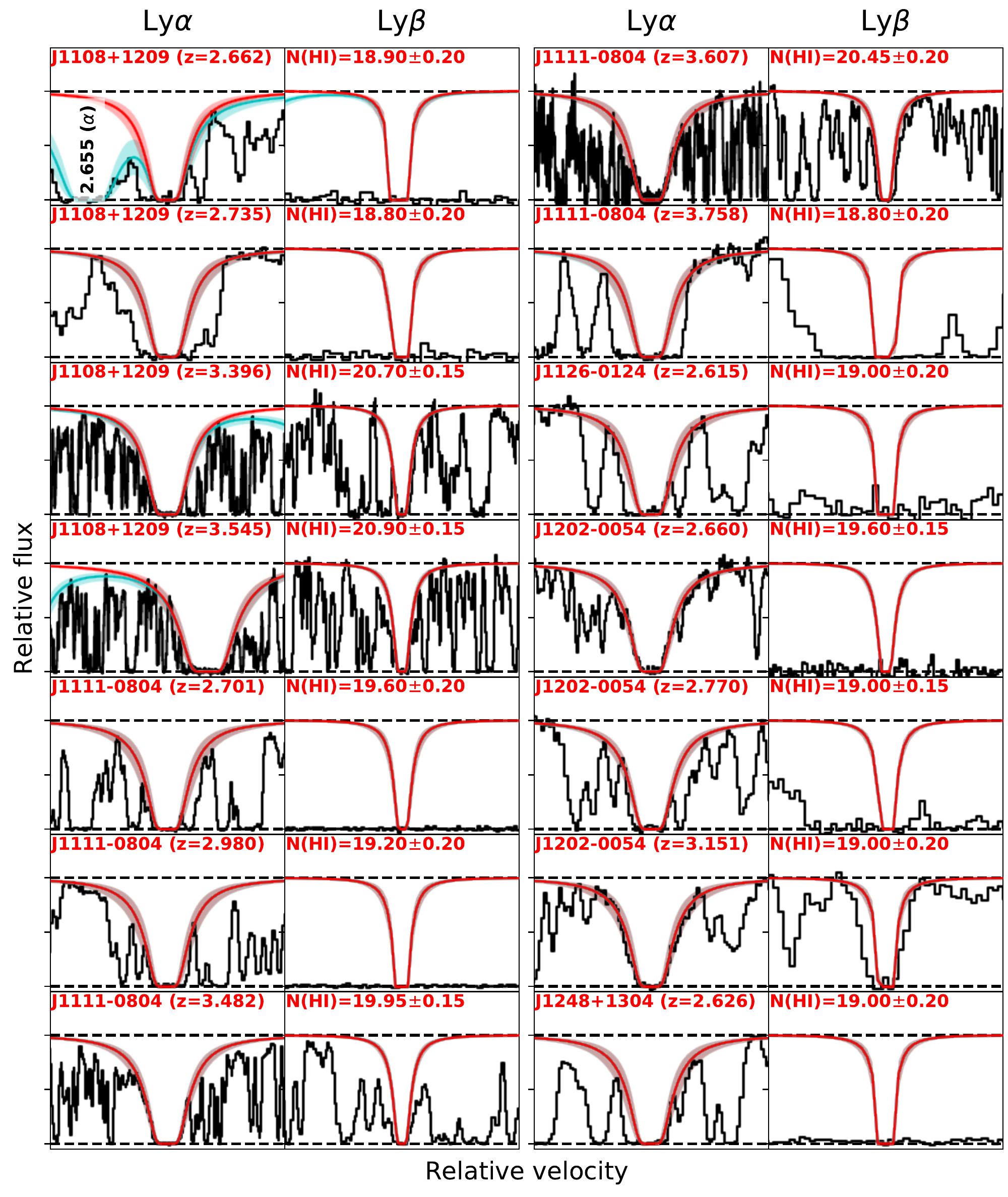}
\end{subfigure}
\caption{(cont'd)}
\end{center}
\end{figure*}

\clearpage

\begin{figure*}
\ContinuedFloat
\begin{center}

\begin{subfigure}{\textwidth}
\includegraphics[width=0.95\textwidth]{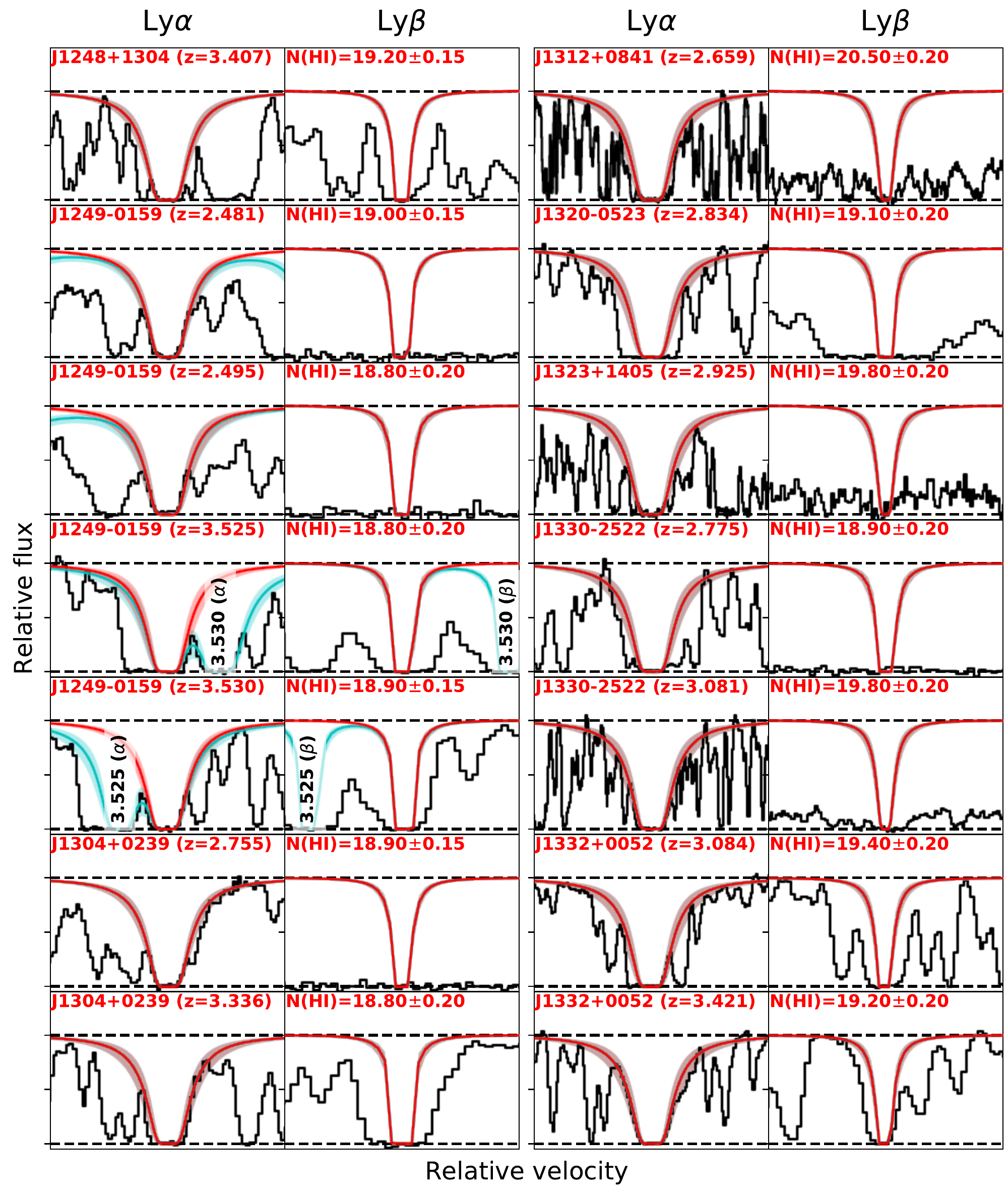}
\end{subfigure}
\caption{(cont'd)}
\end{center}
\end{figure*}

\clearpage

\begin{figure*}
\ContinuedFloat
\begin{center}
\begin{subfigure}{\textwidth}
\includegraphics[width=0.95\textwidth]{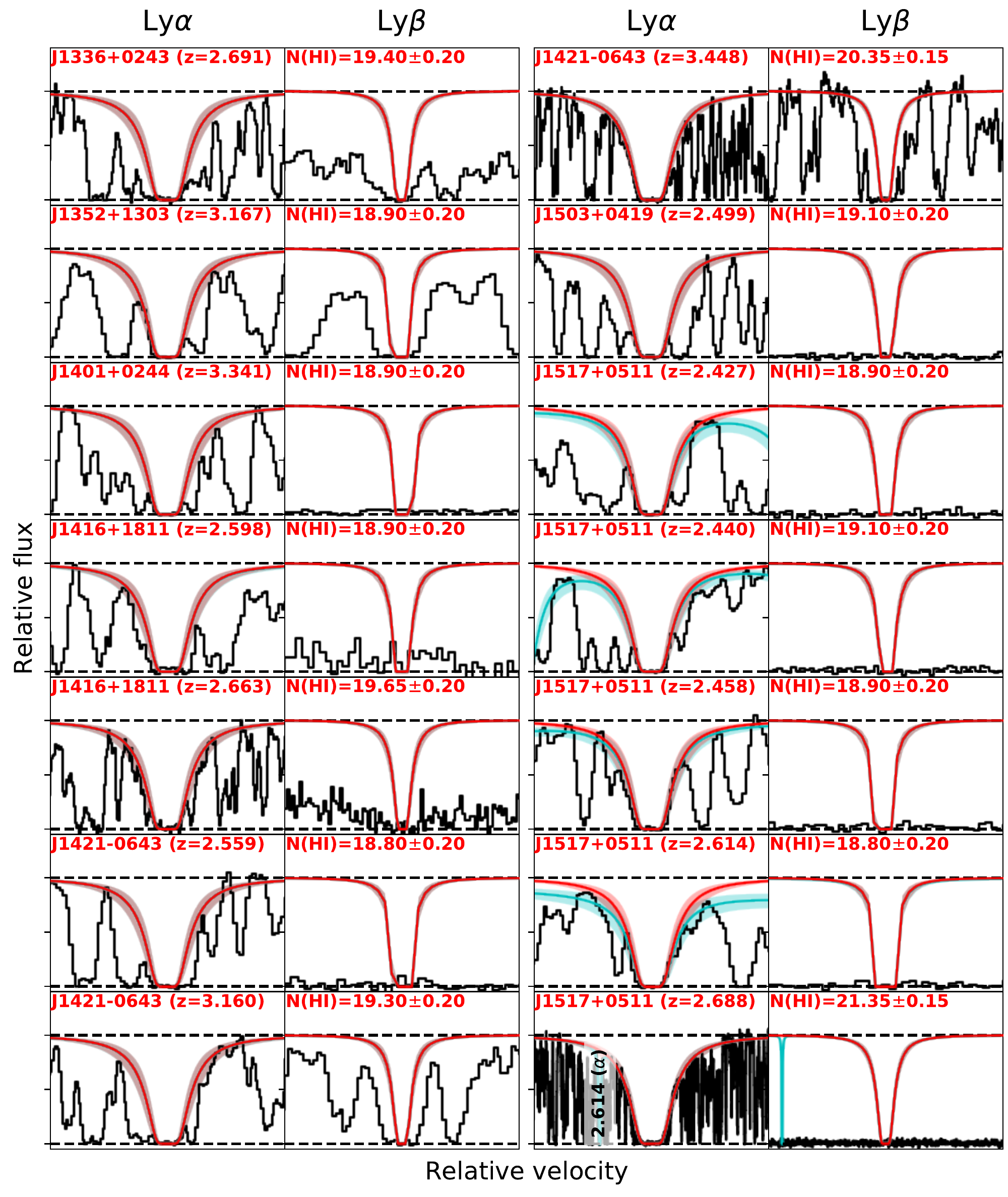}
\end{subfigure}
\caption{(cont'd)}
\end{center}
\end{figure*}

\clearpage

\begin{figure*}
\ContinuedFloat
\begin{center}
\begin{subfigure}{\textwidth}
\includegraphics[width=0.95\textwidth]{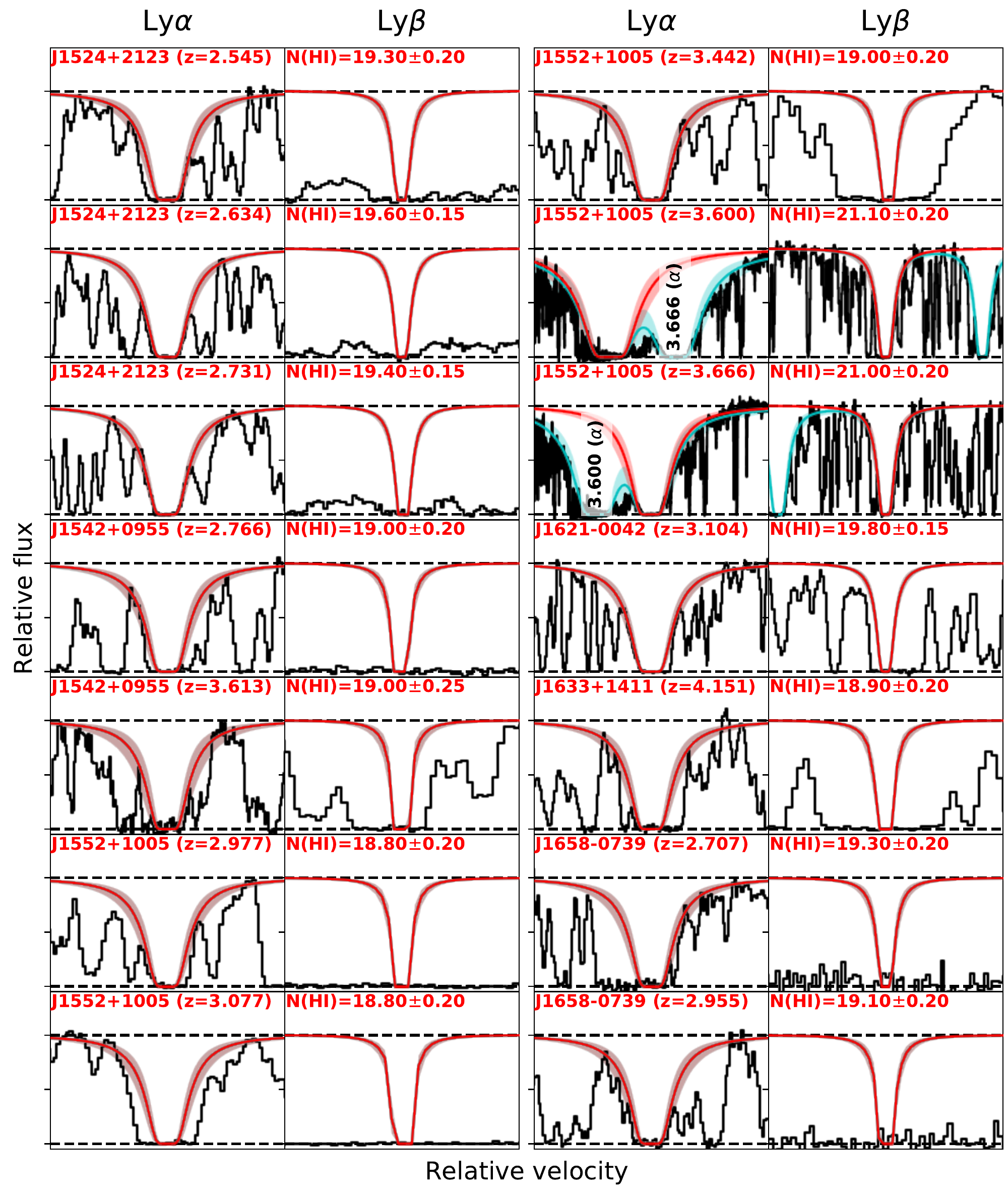}
\end{subfigure}
\caption{(cont'd)}
\end{center}
\end{figure*}

\clearpage

\begin{figure*}
\ContinuedFloat
\begin{center}
\begin{subfigure}{\textwidth}
\includegraphics[width=0.95\textwidth]{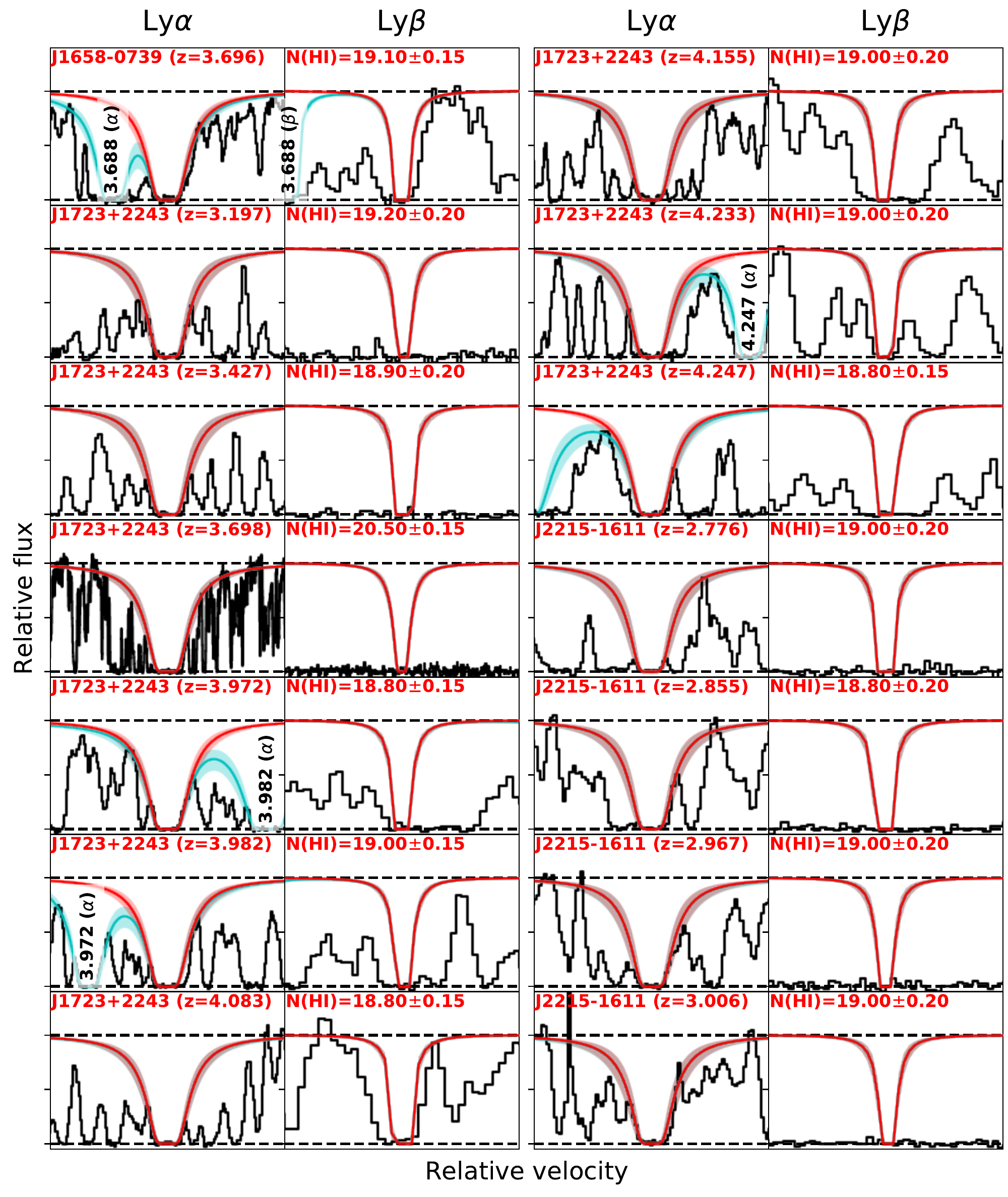}
\end{subfigure}
\caption{(cont'd)}
\end{center}
\end{figure*}

\clearpage

\begin{figure*}
\ContinuedFloat
\begin{center}
\begin{subfigure}{\textwidth}
\includegraphics[width=0.95\textwidth]{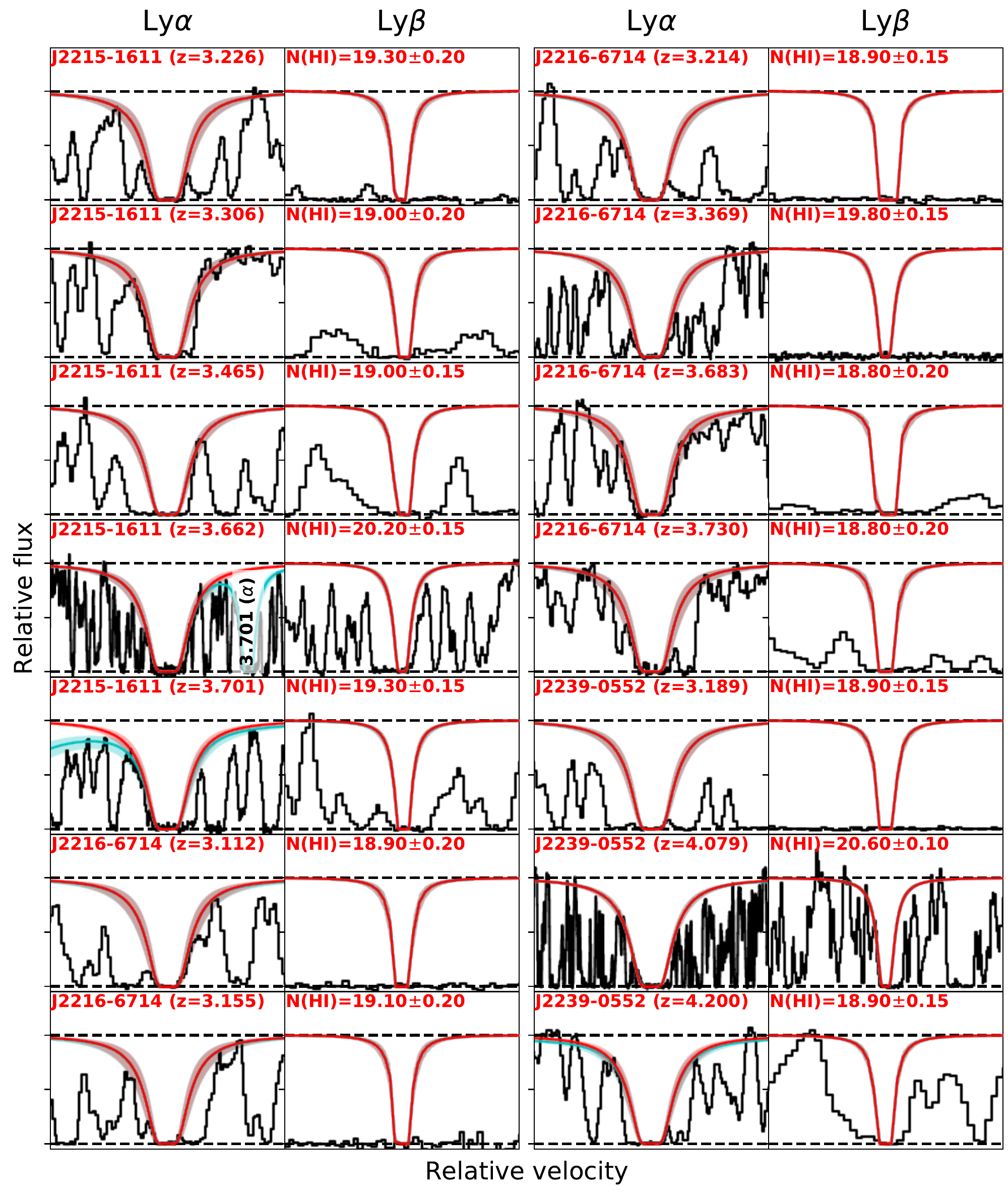}
\end{subfigure}
\caption{(cont'd)}
\end{center}
\end{figure*}

\clearpage

\begin{figure*}
\ContinuedFloat
\begin{center}
\begin{subfigure}{\textwidth}
\includegraphics[width=0.95\textwidth]{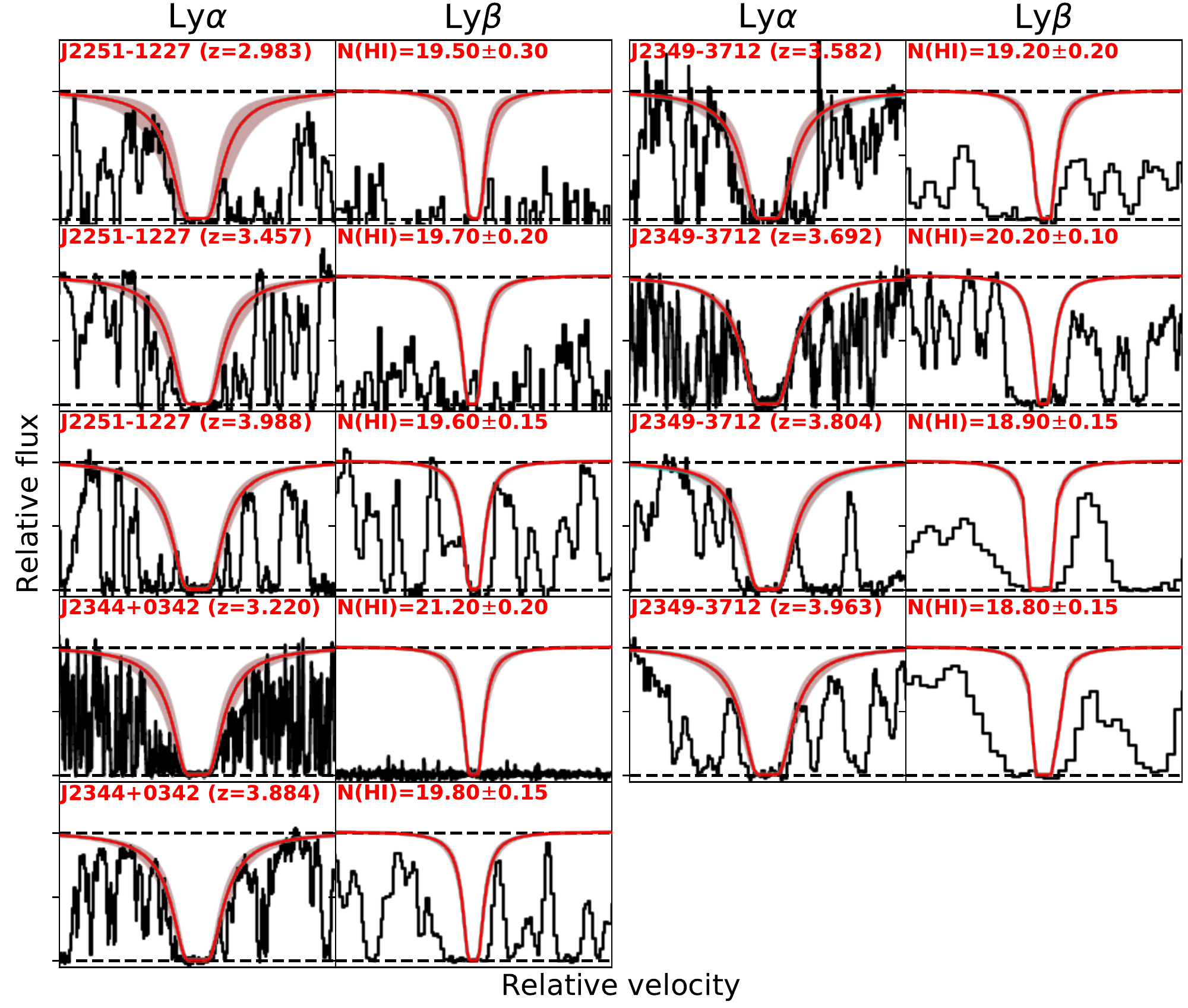}
\end{subfigure}
\caption{(cont'd)}
\end{center}
\end{figure*}

\clearpage

\section{\fNX{} and \omegaDLA{} curves}
\label{sec:AppCurves}
\begin{landscape}
\onecolumn
\begin{tiny}
\begin{center}
\begin{longtable}{l|ccccc|ccccc|ccccc}
\caption{Full sample \fNX{} curves}
\label{tab:fNXFS}\\
\hline
N(\HI{})& & & $2.3\leq$~\zabs{}~$<3.2$& & & & & $3.2\leq$~\zabs{}~$<4.5$& & & & & $2.3\leq$~\zabs{}~$<4.5$& & \\
& \fNX{}$_{,5}$& \fNX{}$_{,17}$& \fNX{}$_{,median}$& \fNX{}$_{,83}$& \fNX{}$_{,95}$& \fNX{}$_{,5}$& \fNX{}$_{,17}$& \fNX{}$_{,median}$& \fNX{}$_{,83}$& \fNX{}$_{,95}$& \fNX{}$_{,5}$& \fNX{}$_{,17}$& \fNX{}$_{,median}$& \fNX{}$_{,83}$& \fNX{}$_{,95}$\\
\hline
\endfirsthead
\multicolumn{16}{l}{{\bfseries \tablename\ \thetable{} -- continued from previous page}}\\
\hline
N(\HI{})& & & $2.3\leq$~\zabs{}~$<3.2$& & & & & $3.2\leq$~\zabs{}~$<4.5$& & & & & $2.3\leq$~\zabs{}~$<4.5$& & \\
& \fNX{}$_{,5}$& \fNX{}$_{,17}$& \fNX{}$_{,median}$& \fNX{}$_{,83}$& \fNX{}$_{,95}$& \fNX{}$_{,5}$& \fNX{}$_{,17}$& \fNX{}$_{,median}$& \fNX{}$_{,83}$& \fNX{}$_{,95}$& \fNX{}$_{,5}$& \fNX{}$_{,17}$& \fNX{}$_{,median}$& \fNX{}$_{,83}$& \fNX{}$_{,95}$\\
\hline
\endhead
\hline \multicolumn{16}{l}{\bfseries Continued on next page}\\
\hline
\endfoot
\endlastfoot
\\

\hline
\hline\multicolumn{16}{c}{SubDLAs}\\
\hline
19.00& 2.88E-20& 3.26E-20& 3.76E-20& 4.33E-20& 4.70E-20& 1.32E-20& 1.48E-20& 1.81E-20& 2.14E-20& 2.35E-20& 2.14E-20& 2.31E-20& 2.59E-20& 2.86E-20& 3.08E-20\\
19.05& 2.51E-20& 2.79E-20& 3.30E-20& 3.74E-20& 4.14E-20& 1.10E-20& 1.28E-20& 1.54E-20& 1.84E-20& 2.02E-20& 1.82E-20& 1.99E-20& 2.24E-20& 2.48E-20& 2.68E-20\\
19.10& 2.09E-20& 2.34E-20& 2.74E-20& 3.19E-20& 3.54E-20& 9.16E-21& 1.05E-20& 1.28E-20& 1.54E-20& 1.70E-20& 1.52E-20& 1.66E-20& 1.88E-20& 2.11E-20& 2.25E-20\\
19.15& 1.69E-20& 1.95E-20& 2.31E-20& 2.66E-20& 2.97E-20& 7.29E-21& 8.46E-21& 1.05E-20& 1.25E-20& 1.40E-20& 1.23E-20& 1.36E-20& 1.55E-20& 1.74E-20& 1.88E-20\\
19.20& 1.35E-20& 1.54E-20& 1.86E-20& 2.18E-20& 2.45E-20& 5.46E-21& 6.76E-21& 8.32E-21& 1.01E-20& 1.14E-20& 9.72E-21& 1.08E-20& 1.24E-20& 1.41E-20& 1.54E-20\\
19.25& 1.06E-20& 1.23E-20& 1.48E-20& 1.76E-20& 1.97E-20& 4.17E-21& 5.10E-21& 6.49E-21& 8.11E-21& 9.27E-21& 7.69E-21& 8.53E-21& 9.93E-21& 1.13E-20& 1.24E-20\\
\end{longtable}
\end{center}
\end{tiny}

\onecolumn
\begin{tiny}
\begin{center}
\begin{longtable}{l|ccccc|ccccc|ccccc}
\caption{Metal-selected sample \fNX{} curves}
\label{tab:fNXMS}\\
\hline
N(\HI{})& & & $2.3\leq$~\zabs{}~$<3.2$& & & & & $3.2\leq$~\zabs{}~$<4.5$& & & & & $2.3\leq$~\zabs{}~$<4.5$& & \\
& \fNX{}$_{,5}$& \fNX{}$_{,17}$& \fNX{}$_{,median}$& \fNX{}$_{,83}$& \fNX{}$_{,95}$& \fNX{}$_{,5}$& \fNX{}$_{,17}$& \fNX{}$_{,median}$& \fNX{}$_{,83}$& \fNX{}$_{,95}$& \fNX{}$_{,5}$& \fNX{}$_{,17}$& \fNX{}$_{,median}$& \fNX{}$_{,83}$& \fNX{}$_{,95}$\\
\hline
\endfirsthead
\multicolumn{16}{l}{{\bfseries \tablename\ \thetable{} -- continued from previous page}}\\
\hline
N(\HI{})& & & $2.3\leq$~\zabs{}~$<3.2$& & & & & $3.2\leq$~\zabs{}~$<4.5$& & & & & $2.3\leq$~\zabs{}~$<4.5$& & \\
& \fNX{}$_{,5}$& \fNX{}$_{,17}$& \fNX{}$_{,median}$& \fNX{}$_{,83}$& \fNX{}$_{,95}$& \fNX{}$_{,5}$& \fNX{}$_{,17}$& \fNX{}$_{,median}$& \fNX{}$_{,83}$& \fNX{}$_{,95}$& \fNX{}$_{,5}$& \fNX{}$_{,17}$& \fNX{}$_{,median}$& \fNX{}$_{,83}$& \fNX{}$_{,95}$\\
\hline
\endhead
\hline \multicolumn{16}{l}{\bfseries Continued on next page}\\
\hline
\endfoot
\endlastfoot
\\

\hline
\multicolumn{16}{c}{SubDLAs}\\
\hline
19.00& 3.76E-21& 5.02E-21& 7.52E-21& 1.00E-20& 1.19E-20& 4.53E-21& 5.77E-21& 7.83E-21& 9.89E-21& 1.15E-20& 5.22E-21& 6.21E-21& 7.71E-21& 9.45E-21& 1.04E-20\\
19.05& 3.35E-21& 4.47E-21& 6.71E-21& 8.94E-21& 1.12E-20& 4.04E-21& 5.14E-21& 6.61E-21& 8.81E-21& 1.03E-20& 4.65E-21& 5.54E-21& 6.87E-21& 8.20E-21& 9.31E-21\\
19.10& 2.49E-21& 3.98E-21& 5.98E-21& 7.97E-21& 9.46E-21& 3.27E-21& 4.25E-21& 5.89E-21& 7.53E-21& 8.83E-21& 3.95E-21& 4.74E-21& 5.92E-21& 7.11E-21& 8.10E-21\\
19.15& 2.22E-21& 3.11E-21& 4.88E-21& 6.66E-21& 8.43E-21& 2.92E-21& 3.50E-21& 4.96E-21& 6.42E-21& 7.58E-21& 3.17E-21& 3.87E-21& 4.93E-21& 6.16E-21& 7.04E-21\\
19.20& 1.98E-21& 2.77E-21& 3.96E-21& 5.93E-21& 7.12E-21& 2.34E-21& 2.86E-21& 4.16E-21& 5.46E-21& 6.24E-21& 2.67E-21& 3.29E-21& 4.08E-21& 5.18E-21& 5.96E-21\\
19.25& 1.41E-21& 2.12E-21& 3.53E-21& 4.94E-21& 5.99E-21& 1.85E-21& 2.32E-21& 3.47E-21& 4.40E-21& 5.33E-21& 2.24E-21& 2.66E-21& 3.49E-21& 4.33E-21& 5.03E-21\\
\end{longtable}
\end{center}
\end{tiny}

\onecolumn
\begin{scriptsize}
\begin{center}
\begin{longtable}{lcc|ccccc|ccccc}
\caption{\lX{} curves}
\label{tab:lX}\\
\hline
$z$& $\Delta$X& $\Delta$z& & & $19.0\leq$~logN(\HI{})~$<20.3$& & & & & $20.3\leq$logN(\HI{})~$<22.0$& & \\
& & & \lX{}$_{,5}$& \lX{}$_{,17}$& \lX{}$_{,median}$& \lX{}$_{,83}$& \lX{}$_{,95}$& \lX{}$_{,5}$& \lX{}$_{,17}$& \lX{}$_{,median}$& \lX{}$_{,83}$& \lX{}$_{,95}$\\
\hline
\endfirsthead
\multicolumn{13}{l}{{\bfseries \tablename\ \thetable{} -- continued from previous page}}\\
\hline
$z$& $\Delta$X& $\Delta$z& & & $19.0\leq$~logN(\HI{})~$<20.3$& & & & & $20.3\leq$logN(\HI{})~$<22.0$& & \\
& & & \lX{}$_{,5}$& \lX{}$_{,17}$& \lX{}$_{,median}$& \lX{}$_{,83}$& \lX{}$_{,95}$& \lX{}$_{,5}$& \lX{}$_{,17}$& \lX{}$_{,median}$& \lX{}$_{,83}$& \lX{}$_{,95}$\\
\hline
\endhead
\hline \multicolumn{13}{l}{\bfseries Continued on next page}\\
\hline
\endfoot
\endlastfoot
\\
\multicolumn{13}{c}{Full Sample}\\
\hline
2.005& 2.5& 0.096& 2.72E-01& 2.93E-01& 3.21E-01& 3.52E-01& 3.75E-01& 6.70E-02& 6.85E-02& 7.08E-02& 7.31E-02& 7.46E-02\\
2.015& 2.5& 0.096& 2.70E-01& 2.90E-01& 3.18E-01& 3.49E-01& 3.71E-01& 6.71E-02& 6.86E-02& 7.08E-02& 7.31E-02& 7.46E-02\\
2.025& 2.5& 0.097& 2.70E-01& 2.90E-01& 3.20E-01& 3.51E-01& 3.73E-01& 6.72E-02& 6.88E-02& 7.11E-02& 7.32E-02& 7.48E-02\\
2.035& 2.5& 0.097& 2.70E-01& 2.88E-01& 3.18E-01& 3.48E-01& 3.70E-01& 6.71E-02& 6.87E-02& 7.10E-02& 7.32E-02& 7.47E-02\\
2.045& 2.5& 0.098& 2.73E-01& 2.93E-01& 3.20E-01& 3.50E-01& 3.72E-01& 6.69E-02& 6.84E-02& 7.07E-02& 7.29E-02& 7.44E-02\\
2.055& 2.5& 0.098& 2.71E-01& 2.91E-01& 3.20E-01& 3.50E-01& 3.72E-01& 6.69E-02& 6.85E-02& 7.07E-02& 7.28E-02& 7.44E-02\\
\end{longtable}
\end{center}
\end{scriptsize}

\onecolumn
\begin{scriptsize}
\begin{center}
\begin{longtable}{l|ccccc|ccccc}
\caption{\omegaDLA{} curves}
\label{tab:omega}\\
\hline
$z$& & & $19.0\leq$~logN(\HI{})~$<20.3$& & & & & $20.3\leq$logN(\HI{})~$<22.0$& & \\
& \omegaDLA{}$_{,5}$& \omegaDLA{}$_{,17}$& \omegaDLA{}$_{,median}$& \omegaDLA{}$_{,83}$& \omegaDLA{}$_{,95}$& \omegaDLA{}$_{,5}$& \omegaDLA{}$_{,17}$& \omegaDLA{}$_{,median}$& \omegaDLA{}$_{,83}$& \omegaDLA{}$_{,95}$\\
\hline
\endfirsthead
\multicolumn{11}{l}{{\bfseries \tablename\ \thetable{} -- continued from previous page}}\\
\hline
$z$& & & $19.0\leq$~logN(\HI{})~$<20.3$& & & & & $20.3\leq$logN(\HI{})~$<22.0$& & \\
& \omegaDLA{}$_{,5}$& \omegaDLA{}$_{,17}$& \omegaDLA{}$_{,median}$& \omegaDLA{}$_{,83}$& \omegaDLA{}$_{,95}$& \omegaDLA{}$_{,5}$& \omegaDLA{}$_{,17}$& \omegaDLA{}$_{,median}$& \omegaDLA{}$_{,83}$& \omegaDLA{}$_{,95}$\\
\hline
\endhead
\hline \multicolumn{11}{l}{\bfseries Continued on next page}\\
\hline
\endfoot
\endlastfoot
\\
\multicolumn{11}{c}{Full Sample}\\
\hline
2.005& 1.37E-04& 1.51E-04& 1.76E-04& 2.01E-04& 2.21E-04& 6.92E-04& 7.20E-04& 7.62E-04& 8.08E-04& 8.41E-04\\
2.015& 1.36E-04& 1.50E-04& 1.74E-04& 1.99E-04& 2.19E-04& 6.93E-04& 7.21E-04& 7.63E-04& 8.08E-04& 8.41E-04\\
2.025& 1.37E-04& 1.52E-04& 1.76E-04& 2.02E-04& 2.21E-04& 6.95E-04& 7.23E-04& 7.65E-04& 8.10E-04& 8.43E-04\\
2.035& 1.36E-04& 1.51E-04& 1.75E-04& 2.00E-04& 2.20E-04& 6.94E-04& 7.22E-04& 7.64E-04& 8.08E-04& 8.41E-04\\
2.045& 1.37E-04& 1.52E-04& 1.76E-04& 2.01E-04& 2.20E-04& 6.86E-04& 7.13E-04& 7.54E-04& 7.98E-04& 8.31E-04\\
2.055& 1.39E-04& 1.54E-04& 1.78E-04& 2.03E-04& 2.23E-04& 6.85E-04& 7.13E-04& 7.54E-04& 7.97E-04& 8.30E-04\\
\end{longtable}
\end{center}
\end{scriptsize}

\end{landscape}

\end{document}